\documentclass{tlp}

\usepackage{latexsym}
\usepackage{url}
\usepackage{aopmath}

\newcommand{\A}{{\mathcal A}}
\newcommand{\T}{{\mathcal T}}

\newcommand{\prt}{\mathbf{output}}
\newcommand{\Cmm}{{\cal M}^{+}}
\newcommand{\cmm}{\mathit{min}^{+}}
\newcommand{\cmM}{\mathit{min}^{+}}
\newcommand{\stb}{\mathit{min}^{+}}

\newcommand{\smpl}{\mathit{\sigma}}
\newcommand{\mmm}{\mathit{min\_mod}}
\newcommand{\stm}{\mathit{stb\_mod}}
\newcommand{\asm}{\mathit{ans\_set}}
\newcommand{\chmm}{\mathit{test\_min}}
\newcommand{\chst}{\mathit{test\_stb}}
\newcommand{\chas}{\mathit{test\_anset}}

\newcommand{\Lit}{\mathit{Lit}}

\newcommand{\At}{\mathit{At}}

\newcommand{\dua}{\overline}

\newcommand{\lla}{\leftarrow}

\newcommand{\n}{\mathbf{not}}

\begin{document}

\submitted{18 November 2004}
\revised{3 June 2005}
\accepted{23 June 2005}

\title[Theory and Practice of Logic Programming]
{Computing minimal models, stable models and answer sets}

\pagerange{\pageref{firstpage}--\pageref{lastpage}}

\author[Z. Lonc and M. Truszczy\'nski]
{Zbigniew Lonc\\
Faculty of Mathematics and Information Science, Warsaw University of
Technology\\
00-661 Warsaw, Poland\\
\email{zblonc@mini.pw.edu.pl}
\and
\and
Miros\l aw Truszczy\'nski,\\
Department of Computer Science, University of Kentucky, Lexington,
KY 40506-0046,
USA\\
\email{mirek@cs.uky.edu}
}
\pagerange{\pageref{firstpage}--\pageref{lastpage}}

\maketitle

\label{firstpage}

\begin{abstract}
We propose and study algorithms to compute minimal models, stable
models and answer sets of $t$-CNF theories, and normal and disjunctive
$t$-programs. We are especially interested in algorithms with {\em
non-trivial} worst-case performance bounds. The bulk of the paper is
concerned with the classes of 2- and 3-CNF theories, and normal and
disjunctive 2- and 3-programs, for which we obtain significantly
stronger results than those implied by our general considerations.
We show that one can find all minimal models of 2-CNF theories and all
answer sets of disjunctive 2-programs in time $O(m 1\mbox{.}4422
\mbox{..}^n)$.
Our main results concern computing stable models of normal 3-programs,
minimal models of 3-CNF theories and answer sets of disjunctive
3-programs. We design algorithms that run in time $O(m1\mbox{.}6701
\mbox{..}^n)$,
in the case of the first problem, and in time $O(mn^2 2\mbox{.}2782
\mbox{..}^n)$, in
the case of the latter two. All these bounds improve by exponential
factors the best algorithms known previously. We also obtain closely
related upper bounds on the number of minimal models, stable models
and answer sets a $t$-CNF theory, a normal $t$-program or a disjunctive
$t$-program may have.
\end{abstract}

\section{Introduction}
\label{intro}

We study the problem of computing {\em minimal} models of CNF theories,
{\em stable} models of normal logic programs and {\em answer sets} of
disjunctive logic programs. We are especially interested in algorithms,
for which we can derive {\em non-trivial} worst-case performance bounds.
Our work builds on studies of algorithms to compute models of
propositional CNF theories \cite{kul99} and improves on our earlier
study of algorithms to compute stable models \cite{lt02b}.

In the paper, by $\At(T)$ and $\At(P)$ we denote the set of atoms
occurring in a theory $T$ or a program $P$, respectively. We represent
models of propositional theories, stable models of normal logic programs
and answer sets of disjunctive logic programs as sets of atoms that are
true in these models and answer sets. When discussing the complexity of
algorithms, we consistently write $n$ for the number of atoms and $m$
for the size (the total number of atom occurrences) of an input theory
$T$ or an input program $P$, even when we do not explicitly mention
them.

Propositional logic with the minimal-model semantics ({propositional
circumscription}) \cite{mca80,li88}, logic programming with stable-model
semantics \cite{gl88} and disjunctive logic programming with the
answer-set semantics \cite{gl90b} are among the most commonly studied
and broadly used knowledge representation formalisms (we refer the
reader to \cite{mt93,bdk97} for a detailed treatment of these logics and
additional pointers to literature). Recently, they have been receiving
much attention due to their role in {\em answer-set programming} (ASP)
--- an emerging declarative programming paradigm. Fast algorithms to
compute minimal models, stable models and answer sets are essential for
the computational effectiveness of propositional circumscription, logic
programming and disjunctive logic programming as answer-set
programming systems.

These computational tasks can be solved by a ``brute-force''
straightforward search. We can compute all minimal models of a CNF
theory $T$ by checking for each set $M\subseteq\At(T)$ whether $M$ is
a minimal model of $T$. To this end, we test first whether $M$ is
a model of $T$. If not, $M$ is not a minimal model of $T$ and we move on
to the next subset of $\At(T)$. Otherwise, we test whether any {\em
proper}
subset of $M$ is a model of $T$. If the answer is yes, then $M$ is not
a minimal model of $T$ and we consider another subset of
$\At(T)$. If the answer is no, then $M$ is a minimal model of $T$.
Testing whether a subset of $\At(T)$ is a model of $T$ can be
accomplished in $O(m)$ steps. Thus, if $|M|=i$, we can verify whether
$M$ is a minimal model of $T$ in time $O(m2^i)$. Checking all sets of
cardinality $i$ requires $O({n\choose i} m2^i)$ steps and checking all
sets --- $O(\sum_{i=0}^n{n \choose i} m2^i) = O(m3^n)$. Thus, this
brute-force approach to compute minimal models works in time
$O(m3^n)$.

To determine whether a set of atoms $M$ is an answer set of a
disjunctive logic program $P$ we need to verify whether $M$ is a minimal
model of the {\em reduct of $P$ with respect to $M$} \cite{gl90b} or,
equivalently, whether $M$ is a minimal model of the propositional theory
that corresponds to the reduct. Thus, a similar argument as before
demonstrates that answer sets of a finite propositional disjunctive
logic program $P$ can also be computed in time $O(m3^n)$.

In the case of stable models we can do better. The task of verifying
whether a set of atoms $M$ is a stable model of a finite propositional
logic program $P$ can be accomplished in time $O(m)$. Consequently,
one can compute all stable models of $P$ in $O(m2^n)$ steps using
an exhaustive search through all subsets of $\At(P)$, and checking for
each of them whether it is a stable model of $P$.

A fundamental question, and the main topic of
interest in this paper, is whether there are algorithms for the three
computational problems discussed here with better worst-case
performance bounds.

We note that researchers proposed several algorithms to compute minimal
models of propositional CNF theories \cite{bp94,nie96}, stable
models of logic programs \cite{sns02} and answer sets of
disjunctive logic programs \cite{eflp00}. Some implementations based on
these
algorithms, for instance, {\em smodels} \cite{sns02} and {\em dlv}
\cite{eflp00}, perform very well in practice. However, so far very little
is known about the worst-case performance of these implementations.

In this paper, we study the three computational problems discussed
earlier. We focus our considerations on {\em $t$-CNF theories} and
{\em $t$-programs}, that is, theories and programs, respectively,
consisting of clauses containing no more than $t$ literals. Such
theories and programs arise often in the context of search problems.
Given a search problem, we often encode its specifications as a 
(disjunctive) DATALOG$^\neg$ program (or, a set of {\em propositional
schemata} --- universally quantified clauses in some function-free 
language \cite{et04a}). Propositional program (CNF theory) corresponding to a
{\em concrete} instance of the search problem is then obtained by 
grounding the (disjunctive) DATALOG$^\neg$ rules (propositional 
schemata) with constants appearing in the descriptions of the instance. 
Since the initial program (set of propositional schemata) is independent 
of particular problem instances, grounding results in propositional
programs (CNF theories) with clauses of bounded length, in other words,
in $t$-programs ($t$-theories) for some, typically quite small, value 
of $t$. In fact in many cases, $t=2$ or $t=3$ (it is so, for instance,
for problems discussed in \cite{mt99,nie99}, once the so called domain
predicates are simplified away).

In our earlier work, we studied the problem of computing stable
models of normal $t$-programs \cite{lt02b}. We obtained some
general results in the case of an arbitrary $t\geq 2$ and were
able to strengthen them for two special cases of $t=2$ and $t=3$.
We presented an algorithm to compute all stable models of a normal
2-program in time $O(m 3^{n/3}) = O(m1\mbox{.}4422\mbox{..}^n)$
and showed its asymptotic optimality. We proposed a similar
algorithm for the class of normal 3-programs and proved that its
running time is $O(m 1\mbox{.}7071\mbox{..}^n)$. Finally, we
applied the techniques developed in this paper to obtain a
non-trivial worst-case performance bound for {\em smodels}, when
input programs are restricted to be $2$-programs.

In this paper, we improve on our results from \cite{lt02b} and extend
them to the problems of computing minimal models of $t$-CNF theories
and answer sets of disjunctive $t$-programs. We present results
concerning the case of an arbitrary $t\geq 2$ but the bulk of the paper
is devoted to 2- and 3-CNF theories and 2- and 3-programs, for which
we significantly strengthen our general results.

First, we show how to find all minimal models of 2-CNF theories and all
answer
sets of disjunctive 2-programs in time $O(m 1\mbox{.}4422\mbox{..}^n)$, 
generalizing
a similar result we obtained earlier for computing stable models of
normal 2-programs. Our main results concern computing stable models of
normal 3-programs, minimal models of 3-CNF theories and answer sets of
disjunctive 3-programs. We design algorithms that run in time
$O(m1\mbox{.}6701\mbox{..}^n)$, for the first problem, and in time $O(mn^2
2\mbox{.}2782\mbox{..}^n)$, for the latter two. These bounds improve by 
exponential
factors
the best algorithms known previously. We also obtain closely related
upper bounds on the number of minimal models, stable models and answer
sets a 2- or 3-theory or program may have.

Our paper is organized as follows. In the next section we state the main
results of the paper. In the remainder of the paper we prove them.
First, in Section \ref{algs}, we present an auxiliary algorithm ${\cmM}$
that, given an arbitrary CNF theory $T$, outputs a family of sets
containing all minimal models of $T$. We also derive some general bounds
on the performance of the algorithm $\cmM$ and on the number of sets
it outputs. In the following section, we adapt the algorithm $\cmM$
to each of the computational tasks of interest to us: finding all
minimal models of CNF theories, stable models of normal programs and
answer sets of disjunctive programs. In Sections \ref{two} and
\ref{three}, we specialize the algorithms from Section \ref{algs2} to
the case of $t$-CNF theories (normal $t$-programs and disjunctive
$t$-programs) for $t=2$ and $3$, respectively. In Section \ref{cov}, we
outline the proof of a main technical lemma, on which all results
concerning 3-CNF theories, 3-programs and disjunctive 3-programs depend.
We present a detailed proof of this result in the appendix. In Section
\ref{ubs}, we discuss the case of an arbitrary $t\geq 2$. In Section
\ref{lbs}, we discuss lower bounds on the numbers of minimal models
(stable models, answer sets) of $t$-CNF theories (normal $t$-programs,
disjunctive $t$-programs). We use these bounds in the last section of
the paper to discuss the optimality of our results and to identify some
open problems for future research.

\section{Main results}
\label{res}

We will now present and discuss the main results of our paper. We start
by stating two theorems that deal with minimal models of 2-CNF theories,
stable models of (normal) 2-programs and answer sets of disjunctive
2-programs. The results concerning stable models of 2-programs were
first presented in \cite{lt02b}. The results about minimal models of
2-CNF theories and answer sets of disjunctive 2-programs are new.

\begin{theorem}
\label{2p1}
There are algorithms to compute all minimal models of 2{-}CNF theories,
stable models of 2-programs and answer sets of disjunctive 2-programs,
respectively, that run in time $O(m 3^{n/3}) = 
O(m1\mbox{.}4422\mbox{..}^n)$.
\end{theorem}

\begin{theorem}
\label{2p2}
Every 2{-}CNF theory (2-program and disjunctive 2-program, respectively)
has at most $3^{n/3} = 1\mbox{.}4422\mbox{..}^n$ minimal models (stable 
models,
answer-sets, respectively).
\end{theorem}

There are 2-CNF theories, 2-programs and disjunctive 2-programs with
$n$ atoms and with $\Omega(3^{n/3})$ minimal models, stable models and
answer sets, respectively. Thus, the bounds provided by Theorems
\ref{2p1} and \ref{2p2} are optimal.

Next, we present results concerning 3-CNF theories, 3-programs and
disjunctive 3-programs. These results constitute the main contribution
of our paper. As in the previous case, we obtain a common upper bound
for the number of minimal models, stable models and answer sets of
3-CNF theories, 3-programs and disjunctive 3-programs, respectively.
Our results improve the bound on the number of stable models of
3-programs from \cite{lt02b} and, to the best of our knowledge, provide
the first non-trivial bounds on the number of minimal models of 3-CNF
theories and answer sets of disjunctive programs.

\begin{theorem}
\label{3p2}
Every 3{-}CNF theory $T$ (every normal 3-program $P$ and every
disjunctive 3-program $P$, respectively) has at most 
$1\mbox{.}6701\mbox{..}^n$
minimal models (stable models, answer-sets, respectively).
\end{theorem}

For 2-CNF theories and 2-programs, the common bound on the number of
minimal models, stable models and answer sets, appeared also as an
exponential factor in formulas estimating the running time of algorithms
to compute the corresponding objects. In contrast, we find that there
is a difference in how fast we can compute stable models of normal
3-programs as opposed to minimal models of 3-CNF theories and answer
sets of disjunctive 3-programs. The reason seems to be that the problem
to check whether a set of atoms is a stable model of a normal program
is in P, while the problems of deciding whether a set of atoms is a
minimal model of a 3-CNF theory or an answer set of a disjunctive
3-program are co-NP complete \cite{cl94,eg95}. For the problem of
computing stable models of normal 3-programs we have the following
result. It constitutes an exponential improvement on the corresponding
result from \cite{lt02b}.

\begin{theorem}
\label{3p1}
There is an algorithm to compute all stable models of normal 3{-}programs
that runs in time $O(m1\mbox{.}6701\mbox{..}^n)$.
\end{theorem}

Our results concerning computing minimal models of 3-CNF theories and
answer sets of disjunctive 3-programs are weaker. Nevertheless, in each
case they provide an exponential improvement over the trivial bound of
$O(m3^n)$ and, to the best of our knowledge, they offer currently the
best asymptotic bounds on the performance of algorithms for these two
problems.

\begin{theorem}
\label{3p1d}
There is an algorithm to compute all minimal models of 3-CNF theories
and answer sets of disjunctive 3-programs, respectively, that runs in
time $O(mn^2 2\mbox{.}2782\mbox{..}^n)$.
\end{theorem}

Proving Theorems \ref{2p1} - \ref{3p1d} is our main objective for the
remainder of this paper.

\section{Technical preliminaries and an auxiliary algorithm}
\label{algs}

We begin by presenting and analyzing an auxiliary algorithm that, given
a CNF theory $T$ computes a superset of the set of all minimal models
of $T$.

In the paper, we consider only CNF theories with no clause containing
multiple occurrences of the same literal, and with no clause containing
both a literal and its dual. The first assumption allows us to treat
clauses interchangingly as disjunctions of their literals or as {\em
sets} of their literals.

Let $T$ be a CNF theory. By $\Lit(T)$ we denote the set of all literals
built of atoms in $\At(T)$. For a literal $\omega$, by $\overline{
\omega}$ we denote the literal that is $\omega$'s dual. That
is, for an atom $a$ we set $\dua{a}=\neg a$ and $\dua{\neg a}=a$.

For a set of literals $L \subseteq \Lit(T)$, we define:
\[
\overline{L}=\{\overline{\omega}\colon \omega\in L\},\ \ L^{+} = \At(T)
\cap L\ \ \mbox{and}\ \ L^{-} = \At(T)\cap \overline{L}\mbox{.}
\]

A set of literals $L$ is {\em consistent} if $L^{+} \cap L^{-}
=\emptyset$. A set of atoms $M \subseteq\At(T)$ is {\em
consistent} with a set of literals $L\subseteq \Lit(T)$, if
$L^{+}\subseteq M$ and $L^{-}\cap M =\emptyset$.

A set of atoms $M \subseteq\At(T)$ is a {\em model} of a theory
$T$ if, for each clause $c\in T$, $c\cap M\not=\emptyset$ or
$c^{-}\cap(\At(T)\setminus M)\not=\emptyset$. A model $M$ of a
theory $T$ is {\em minimal} if no proper subset of $M$ is a model
of $T$.

Let $T$ be a CNF theory and let $L\subseteq\Lit(T)$ be a set of
literals. We define a theory $T_L$ as follows:
\[
T_L = \{c'\colon \mbox{there is $c\in T$ such that $c'=c - \overline{L}$
and $c\cap L = \emptyset$}\}\mbox{.}
\]
Thus, to obtain $T_L$ we remove from $T$ all clauses implied by $L$
(that is, clauses $c\in T$ such that $c\cap L \not= \emptyset$),
and resolving each remaining clause with literals in $L$, that is,
removing from it all literals in $\dua L$. It may happen that $T_L$
contains the empty clause (is contradictory) or is empty (is a
tautology). The theory $T_L$ has the following important properties.

\begin{lemma}
\label{key}
Let $T$ be a CNF theory and $L\subseteq \Lit(T)$. For every
$X\subseteq\At(T)$ that is consistent with $L$:
\begin{enumerate}
\item $X$ is a model of $T$ if and only if $X  - L^{+}$ is a model of
$T_L$.
\item If $X$ is a minimal model of $T$, then $X  - L^{+}$ is a
minimal model of $T_L$.
\item If $L^{+}=\emptyset$ then $X$ is a minimal model of $T$ if and
only if $X$ is a minimal model of $T_L$.
\end{enumerate}
\end{lemma}
\begin{proof} Let $T'$ be
the set of clauses in $T$ that contain a literal from $L$ (in the proof
we view clauses as sets of their literals). Clearly, when constructing
$T_L$, we remove clauses in $T'$ from $T$. Since $X$ is consistent with
$L$, $X$ satisfies all clauses in $T'$. Thus, $X$ is a model of $T$ if
and only if $X$ is a model of $T''=T  - T'$.

Next, we note that every clause $c$ in $T''$ is of the form $c'\cup
c''$, where $c'\in T_L$ and $c''$ consists of literals of the form
$\overline{\omega}$, for some $\omega \in L$. Moreover, every clause
$c'\in T_L$ appears at least once as part of such representation of a
clause $c$ from $T$.

Since $X$ is consistent with $L$, $X$ is a model of a clause $c \in
T''$ if and only if $X$ is a model of $c'$. Since $c'$ contains no
literals from $L$ nor their duals, $X$ is a model of $c'$ if and only
$X  - L^{+}$ is a model of $c'$. It follows that $X$ is a model of
$T''$ (and so, of $T$) if and only if $X  - L^{+}$ is a model of
$T_L$.

To prove part (2) of the assertion, we observe first that $X  -
L^{+}$ is a model of $T_L$ (by part (1)). Let us consider $Y \subseteq
X  - L^{+}$ such that $Y$ is a model of $T_L$. Clearly, $Y\cup L^{+}$
is consistent with $L$ and $Y=(Y\cup L^{+})-L^{+}$. Hence, it follows by
(1) that $Y\cup L^{+}$ is a model of $T$. Since $X$ is consistent with
$L$, $Y\cup L^{+}\subseteq X$. By the minimality of $X$, $Y\cup L^{+}=X$
and, as $Y\cap L^{+}=\emptyset$, we obtain that $Y=X - L^{+}$. Thus, $X -
L^{+}$ is a minimal model of $T_L$.

To prove (3), we only need to show that if $X$ is a minimal model of
$T_L$ then $X$ is a minimal model of $T$ (the other implication follows
from (2)). By (1), $X$ is model of $T$. Let $Y\subseteq X$ be also a
model of $T$. Clearly, $Y$ is consistent with $L$. Thus, again by (1),
$Y$ is a model of $T_L$. By the minimality of $X$, $Y=X$ and $X$ is a
minimal model of $T$.
\end{proof}

Let $T$ be a CNF theory. A family $\mathcal{A}$ of subsets of
$\Lit(T)$ {\em covers} all minimal models of $T$, or is a {\em
cover} for $T$, if $\mathcal{A}\not=\emptyset$, $\emptyset\notin
\mathcal{A}$ and if every minimal model of $T$ is consistent with
at least one set $A\in {\mathcal A}$. A {\em cover function} is a
function which, to every CNF theory $T$ such that
$\At(T)\not=\emptyset$ assigns a cover of $T$.  The family
${\mathcal A}=\{\{a\},\{\overline{a}\}\}$, where $a$ is an atom of
$T$, is an example of a cover for $T$. A function which, to every
CNF theory $T$ such that $\At(T)\not=\emptyset$, assigns $\{
\{a\},\{\overline{a}\}\}$, for some atom $a\in \At(T)$, is an
example of a cover function.

When processing CNF theories, it is often useful to simplify their
structure without changing their logical properties. Let $\sigma$ be a
function assigning to each CNF theory $T$ another CNF theory, $\sigma
(T)$, which is equivalent to $T$, satisfies $\At(\sigma(T))\subseteq
\At(T)$ and which, in some sense, is simpler than $T$. We call each
such function a {\em simplifying} function. For now, we leave the nature
of the simplifications encoded by $\sigma$ open.

We are now in a position to describe the algorithm we promised at the
beginning of this section. That algorithm computes a superset of
the set of all minimal models of an input CNF theory $T$. It is
parameterized with a cover function $\rho$ and a simplifying function
$\sigma$.
That is, different choices for $\rho$ and $\sigma$ specify different
instances of the algorithm. We call this algorithm $\cmm$ and describe
it in Figure \ref{fig1}. To be precise, the notation should explicitly
refer to the functions $\rho$ and $\sigma$ that determine $\stb$. We
omit these references to keep the notation simple. The functions $\rho$
and $\sigma$ will always be clear from the context.

The algorithm $\cmm$ is fundamental to our approach. We derive from it
algorithms for the three main tasks of interest to us: computing
minimal models, stable models and answer sets.

The input parameters of $\cmm$ are CNF theories $T$ and $S$, and a set
of literals $L$. We require that $L \subseteq \Lit(T)$ and $S=
\smpl(T_L)$. We will refer to these two conditions as the {\em
preconditions} for $\cmm$. The input parameter $S$ is determined by the
two other parameters $T$ and $L$ (through the preconditions on $T$,
$S$ and $L$). We choose to specify it explicitly as that simplifies the
description and the analysis of the algorithm.

The output of the algorithm $\cmm(T,S,L)$ is a family $\Cmm(T,L)$ of
sets that contains all minimal models of $T$ that are consistent with
$L$.

The algorithm $\stb$ and the implementations of the cover function
$\rho$ and a simplifying function $\sigma$ that are used in $\stb$
assume a standard linked-list representation of CNF theories.
Specifically, an input CNF theory $T$ is a doubly-linked list of its
clauses, where each clause $c$ in $T$ is a doubly-linked list of its
literals. The total size of such a representation of a CNF theory $T$
is $O(m)$. In addition, for each literal $\omega\in \Lit(T)$, we have
a linked list $C(\omega)$ consisting of all clauses in $T$ that contain
$\omega$ as a literal. More precisely, for each clause $c$ containing
$\omega$, the list $C(\omega)$ contains a pointer to the location of $c$
in $T$ and a pointer to the location of $\omega$ on the list $c$. These
lists can be created from the linked list $T$ in time that is linear in
the size of $T$. Since we assume that clauses do not contain multiple
occurrences of the same literal, we assume that the same property holds
for linked lists that represent them.

\begin{figure}[ht]
\begin{tabbing}
\quad\=\quad\=\quad\=\quad\=\quad\=\quad\=\quad\=\quad\=\quad\=\quad\=
\quad\=\quad\=\\
$\stb(T,S,L)$\\
\>\>\% $T$ and $S$ are CNF theories, $L$ is a set of literals\\
\>\>\% $T$, $S$ and $L$ satisfy the preconditions: $L\subseteq \Lit(T)$, and
$S=\smpl(T_L)$\\
\medskip
1\>\>{\bf if} $S$ does not contain an empty clause {\bf then}\\
2\>\>\>{\bf if} $S=\emptyset$ {\bf then}\\
3\>\>\>\>$M:=L^{{+}}$; $\prt(M)$\\
4\>\>\>{\bf else}\\
5\>\>\>\>${\mathcal A}:=\rho(S)$;\\
6\>\>\>\>{\bf for every} $A\in {\mathcal A}$ {\bf do}\\
7\>\>\>\>\>$\mathit{SA}':= T_{L\cup A}$;\\
8\>\>\>\>\>$\mathit{SA}:=\smpl(\mathit{SA}')$;\\
9\>\>\>\>\>$\stb(T,\mathit{SA},L\cup A)$\\
10\>\>\>\>{\bf end of for}\\
11\>\>\>{\bf end of else} \\
12\>\>{\bf end of} $\stb$.
\end{tabbing}
\caption{Algorithm $\cmm$}
\label{fig1}
\vspace*{-0.15in}
\end{figure}

Clearly, the recursive call in the line (9) is legal as
$SA=\sigma(T_{L\cup A})$. Moreover, since $\rho$ is a cover function,
for every $A\in \rho(S)$, $|A|\geq 1$. Thus, $|\At(SA)|< |\At(S)|$ and
the algorithm terminates. The next lemma establishes the key property
of the output produced by the algorithm $\cmm$.

\begin{lemma}
\label{correct}
Let $\rho$ be a cover function and $\sigma$ be a simplifying function.
For every CNF theory $T$ and a set of literals $L\subseteq \Lit(T)$, if
$X$ is a minimal model of $T$ consistent with $L$, then $X$ is among the
sets returned by $\stb(T,S,L)$, where $S=\sigma(T_L)$.
\end{lemma}
\begin{proof} We prove the assertion proceeding by induction on $|\At(S)|$.
Let us assume that $|\At(S)|=0$ and that $X$ is a minimal model of $T$.
By Lemma \ref{key}, $X  - L^{+}$ is a minimal model of $T_L$ and,
consequently, also of $S$ ($S=\sigma(T_L)$ and so, $S$ and $T_L$ have
the same models). It
follows that $S$ is consistent and, therefore, contains no empty clause.
Since $|\At(S)|=0$, $S= \emptyset$. Consequently, $X  - L^{+}
=\emptyset$ (as $X  - L^{+}$ is a minimal model of $S$). Hence, $X
\subseteq L^{+}$. Furthermore, since $X$ is consistent with $L$, $L^{+}
\subseteq X$. Thus, $X=L^{+}$. Finally, since $S$ is empty, the program
enters line (3) and outputs $X$, as $X=L^{+}$.

For the inductive step, let us assume that $|\At(S)| > 0$ and that
$X$ is a minimal model of $T$ consistent with $L$. By Lemma
\ref{key}, $X  - L^{+}$ is a minimal model of $T_L$ and,
consequently, a minimal model of $S$. Since ${\mathcal A}$, computed in
line (5), is a cover for $S$, there is $A\in {\mathcal A}$ such that
$X  - L^{+}$ is consistent with $A$. Clearly, $\At(L)\cap \At(A)=
\emptyset$. Thus, $X$ is consistent with $L\cup A$. By the induction
hypothesis (the parameters $T$, $\mathit{SA}$ and $L\cup A$ satisfy the
preconditions for the algorithm $\cmm$ and $|\At(S)|> |\At(\mathit{SA}
)|$), the call $\stb(T,\mathit{SA},L\cup A)$, within loop (6), returns
the set $X$. \end{proof}

\begin{corollary}
\label{cor1}
Let $T$ be a CNF theory. The family $\Cmm(T,\emptyset)$ of sets that
are returned by $\stb(T,S,\emptyset)$, where $S=\sigma(T)$, contains all
minimal models of $T$.
\end{corollary}

We will now study the performance of the algorithm $\stb$. We start with
the following observation concerning computing theories $\mathit{SA}'$
in line (7) of the algorithm $\stb$.

\begin{lemma}
\label {rem1}
There is an algorithm which, given a linked-list representation of a
CNF theory $S$ and a set $A$ of literals such that $A\subseteq \Lit(S)$,
constructs a linked-list representation of the theory $\mathit{SA}'$ in
linear time in the size of $S$.
\end{lemma}
\begin{proof} It is clear that with the data structures that we described above,
we can eliminate from the list $S$ all clauses containing literals in
$A$ by traversing lists $C(\omega)$, $\omega\in A$, and by deleting each
clause we encounter in this way. Since $S$ is a doubly-linked list, each
deletion takes constant time, and the overall task takes linear time in
the size of $S$. Similarly, also in linear time in the size of the list
$S$, we can remove literals in the set $\overline{A}$ from each clause
that remains on $S$. \end{proof}

Next, we will analyze the recursive structure of the algorithm $\stb$.
Let $\rho$ be a cover function and $\sigma$ a simplifying function.
For a CNF theory $T$ we define a labeled tree $\T_T$ inductively as
follows. If $T$ contains the empty clause or if $T= \emptyset$ ($T$ is
a tautology), $\T_T$ consists of a single node labeled with $\smpl(T)$.
Otherwise, we form a new node, label it with $\smpl(T)$ and make it the
parent of all trees $\T_{T'_A}$, where $T'=\smpl(T)$, $A\in \rho(T')$.
For every $A\in \rho(T)$ we have $|\At(T'_A)| < |\At(T')|\leq |\At(T)|$.
Thus, the tree $\T_T$ is well defined. We denote the set of leaves of
the tree $\T_T$ by $L(\T_T)$. As the algorithm $\stb$, $\T_T$ depends
on functions $\sigma$ and $\rho$, too. We drop the references to
these functions to keep the notation simple.

It is clear that for every CNF theory $T$ and for every set of literals
$L \subseteq \Lit(T)$, the tree $\T_S$, where $S=T_L$, is precisely the
tree of recursive calls to $\cmm$ made by the top-level call
$\cmm(T,\sigma(S),L)$. In particular, the tree $\T_T$ describes the
structure of the execution of the call $\cmm(T,\sigma(T),\emptyset)$.

We use the tree $\T_T$ to estimate the running time of the algorithm
$\stb$. We say that a cover function $\rho$ is {\em splitting} if for
every theory $T$, such that $|\At(T)|\geq 2$, it returns a cover with
at least two elements. We have the following result.

\begin{lemma}
\label{time}
Let $\rho$ be a splitting cover function and $\sigma$ a simplifying
function. Let $t$ be an integer function such that $t(k)=\Omega(k)$ and
the functions $\rho$ and $\smpl$ can be computed in time $O(t(k))$ on
input theories of size $k$. Then the running time of the algorithm
$\stb$ on input $(T,\sigma(T),\emptyset)$, where $T$ is a CNF theory,
is $O(|L({\cal T}_T)|t(m))$.
\end{lemma}
\begin{proof} Let $T$ be a CNF theory and let $s$ be the number of nodes in
the tree $\T_T$. Then, the number of recursive calls of the algorithm
$\stb$ is also equal to $s$. Clearly, the total time needed for lines
(1)-(6) over all recursive calls to $\stb$ is $O(s\cdot t(m))$ (in the case
of line (6) we only count the time needed to control the loop and not
to execute its content). Indeed, in each recursive call the size of the
theories considered is bounded by $m$, the size of the theory $T$.

We charge each execution of the code in lines (7) and (8) to the
recursive call to $\stb$ that immediately follows. By Lemma \ref{rem1},
line (7) takes time $O(m)$ and line (8) takes time $O(t(m))$ (again, by
the fact that the sizes of theories $\mathit{SA}$ and $\mathit{SA}'$ are
bounded by the size of $T$, that is, by $m$). Thus, the total time
needed for these instructions over all recursive calls to $\stb$ is also
$O(s\cdot t(m))$.

It follows that, the running time of the algorithm $\stb$ is $O(st(m))$.
Since $\rho$ is splitting, every node in the tree $\T_T$, other than
leaves and their parents, has at least two children. Consequently,
$s=O(|L(\T_T)|)$ and the assertion follows.
\end{proof}

We also note that only those recursive calls to $\cmm$ that
correspond to leaves of $\T_T$ produce output. Thus, Corollary
\ref{cor1} imply the following bound on the number of minimal models
of a CNF theory $T$.

\begin{lemma}
\label{ell}
Let $T$ be a CNF theory. The number of minimal models of $T$ is at most
$|L(\T_T)|$. 
\end{lemma}

In order to use Lemmas \ref{time} and \ref{ell} we need a method to
estimate the number of leaves in rooted trees. Let $\T$ be a rooted tree
and let $L(\T)$ be the set of leaves in $\T$. For a node $x$ in $\T$, we
denote by $C(x)$ the set of {\em directed} edges in $\T$ that link $x$
with its children. For a leaf $w$ of $\T$, we denote by $P(w)$ the set
of {\em directed} edges on the unique path in $\T$ from the root of
${\cal T}$ to the leaf $w$. The following observation was shown in
\cite{kul99}.

\begin{lemma}
\label{Kull}
{\rm \cite{kul99}} Let $p$ be a function assigning positive real
numbers to edges of a rooted tree $\T$ such that for every internal
node $x$ in $\T$, $\sum_{e\in C(x)}p(e)=1$. Then,
\[
|L(\T)|\leq\max_{w\in L(\T)} (\prod_{e\in P(w)}p(e))^{-1}\mbox{.}
\]
\end{lemma}

For some particular cover functions, Lemma \ref{Kull} implies
specific bounds on the number of leaves in the tree ${\cal T}_T$.
Let $\mu$ be a function that assigns to every CNF theory $T$ a
real number $\mu(T)$ such that $0\leq \mu(T) \leq |\At(T)|$. We
call each such function a {\em measure}. Given a measure $\mu$, a
simplifying function $\sigma$ is {\em $\mu$-compatible} if for
every CNF theory $S$, $\mu(\smpl(S))\leq \mu(S)$. Similarly, we
say that a cover function $\rho$ is {\em $\mu$-compatible} if for
every CNF theory $S$ such that $\At(S)\not=\emptyset$ and for
every $A\in \rho(S)$, $\mu(S)-\mu(S_A) > 0$. We denote the
quantity $\mu(S)-\mu(S_A)$ by $\Delta(S,S_A)$.

Let $S$ be a CNF theory such that $\At(S)\not=\emptyset$. Let
$\mu$ be a measure and let $\rho$ be a cover function that is
$\mu$-compatible. Since for  every $A\in\rho(A)$,
$\Delta(S,S_A)>0$, there is a unique real number $\tau\geq 1$
satisfying the equation
\begin{equation}
\label{tau}
\sum_{A\in \rho(S)}{\tau^{-\Delta(S,S_A)}} = 1\mbox{.}
\end{equation}
Indeed, for $\tau \geq 1$ the left hand side of the equation (\ref{tau}) 
is a strictly decreasing continuous function of $\tau$. Furthermore, its 
value for $\tau=1$ is at least $1$ (as $\rho(S)\not=\emptyset$) and it
approaches $0$ when $\tau$ tends to infinity. We denote the number
$\tau\geq 1$ satisfying (\ref{tau}) by $\tau_S$ (we drop
references to $\mu$ and $\rho$, on which $\tau_S$ also depends, to
keep the notation simple). We say that $\rho$ is {\em
$\mu$-bounded} by a real number $\tau_0$ if for every CNF theory
$S$ with $|\At(S)|\geq 1$, $\tau_S\leq \tau_0$. We have the
following result --- a corollary to Lemma \ref{Kull}.

\begin{lemma}
\label{thm-ineq}
Let $T$ be a CNF theory and let $\mu$ be a measure. For every
$\mu$-compatible simplifying function $\sigma$ and for every
$\mu$-compatible cover function $\rho$ that is $\mu$-bounded by
$\tau_0$,
\begin{equation}
\label{ineq}
|L(\T_T)|\leq \tau_0^{|\At(T)|}\mbox{.}
\end{equation}
\end{lemma}
\begin{proof} Let $e=(x,y)$ be an edge in $\T_T$ and let $S$ and $S'$ be
CNF theories that label $x$ and $y$, respectively. Since $x$ is not a
leaf in $\T_T$, $\At(S)\not=\emptyset$. Thus, $\rho$ is defined for
$S$. Moreover, by the definition of $\T_T$ it follows that there is an
element $A\in \rho(S)$ such that $S'=\smpl(S_A)$. We define $D(e)=\mu(S)
-\mu(S')$. Since $\smpl$ is $\mu$-compatible, $\mu(S')\leq \mu(S_A)$.
Thus, $D(e)\geq \mu(S)-\mu(S_A)$. Since $\rho$ is $\mu$-compatible, $D(e)
 > 0$.

Let us now set $p(e) = \tau_S^{-\Delta(S,S_A)}$, where $\tau_S$ is the
root of the equation (\ref{tau}). Since, $\rho$ is $\mu$-bounded by
$\tau_0$, we have $p(e)^{-1} = \tau_S^{\Delta(S,S_A)}\leq \tau_0^{\Delta
(S,S_A)} \leq \tau_0^{D(e)}$ (we recall that $\tau_S\geq 1$).

Clearly, for every leaf $w\in \T_T$ we have
\[
\sum_{e\in P(w)} D(e) = \mu(T')-\mu(W)\leq \mu(T)\leq |\At(T)|,
\]
where $T'=\smpl(T)$ and $W$ is the theory that labels $w$. Thus,
\[
(\prod_{e\in P(w)}p(e))^{-1}\leq\prod_{e\in P(w)}\tau_0^{D(e)}=
\tau_0^{\sum_{e\in P(w)}D(e)}\leq \tau_0^{|\At(T)|}\mbox{.}
\]
The function $p$ satisfies the assumptions of Lemma \ref{Kull}.
Consequently, the assertion follows. \end{proof}

Results of this section show that in order to get good performance
bounds for the algorithm $\cmm$ and good bounds on the number of minimal
models of a CNF theory one needs a measure $\mu$ and a splitting cover
function $\rho$ that is $\mu$-compatible and $\mu$-bounded by $\tau_0$,
where $\tau_0$ is as small as possible.

To estimate roots $\tau_S$ of specific equations of type (\ref{tau}),
which we need to do in order to bound all of them from above and
determine $\tau_0$, we will later use the following
straightforward observation.

\begin{lemma}
\label{last}
Let $\mu$ be a measure, $\rho$ a $\mu$-compatible cover function and
$S$ a CNF theory with $|\At(S)|\geq 1$. If for every $A\in \rho(S)$,
$k_{S,A}$ is a positive real such that $\Delta(S,S_A)\geq k_{S,A}$,
then $\tau_S \leq \tau_S'$, where $\tau_S'$ is the root of the equation
\begin{equation}
\label{tau'}
\sum_{A\in \rho(S)}{\tau^{-k_{S,A}}}=1\mbox{.}
\end{equation}
\end{lemma}

\section{Computing minimal models, stable models and answer sets --- a
general case}
\label{algs2}

We will now derive from the algorithm $\cmm$ algorithms for computing
minimal models of CNF theories, stable models of normal programs and
answer sets of disjunctive programs. In this section, we do not assume
any syntactic restrictions.

We start with the problem of computing minimal models of a CNF theory
$T$. Let $\chmm$ be an algorithm which, for a given
CNF theory $T$ and a set of atoms $M\subseteq \At(T)$ returns the
boolean value {\bf true} if $M$ is a minimal model of $T$, and returns
{\bf false}, otherwise.

We now modify the algorithm $\cmm$ by replacing each occurrence of
the command $\prt(M)$ (in line (3)), with the command
\begin{quote}
{\bf if} $\chmm(T,M)$ {\bf then} $\prt(M)$.
\end{quote}
We denote the resulting algorithm by $\mmm$ (we assume
the same preconditions on $\mmm$ as in the case of $\cmm$).
Since all minimal models
of $T$ that are consistent with $L$ belong to $\Cmm(T,L)$ (the output of
$\stb(T,\sigma(T_L),L)$), it is clear that the algorithm $\mmm(T,\sigma
(T_L),L)$ returns all minimal models of $T$ that are consistent with $L$
and nothing else.

Computation of stable models and answer sets of logic programs follows
a similar pattern. First, let us recall that we can associate with
a disjunctive logic program $P$ (therefore, also with every normal logic
program $P$) its propositional counterpart, a CNF theory $T(P)$ consisting
of clauses of $P$ but interpreted in propositional logic and rewritten
into CNF. Specifically, to obtain $T(P)$ we replace each disjunctive
program clause
\[
c_1\vee \ldots \vee c_p \lla a_1,\ldots,a_r,\n(b_1),\ldots,\n(b_s)
\]
in $P$ with a CNF clause
\[
\neg a_1\vee\ldots\vee \neg a_r\vee b_1\vee\ldots\vee b_s\vee c_1\vee
\ldots \vee c_p\mbox{.}
\]
It is well known that stable models (answer sets) of (disjunctive)
logic program $P$ are minimal models of $T(P)$ \cite{mt93}.

Let us assume that $\chst(P,M)$ and $\chas(P,M)$ are algorithms to
check whether a set of atoms $M$ is a stable model and an answer set,
respectively, of a program $P$.

To compute stable models of a logic program $P$ that are
consistent with a set of literals $L$, we first compute the CNF
theory $T(P)$. Next, we run on the triple $T(P)$, $\sigma(T(P)_L)$
and $L$, the algorithm $\cmm$ modified similarly as before (we
note that the triple $(T(P), \sigma(T(P)_L),L)$ satisfies the
preconditions of $\cmm$). Namely, we replace the command $\prt(M)$
(line (3)) with the command
\begin{quote}
{\bf if} $\chst(P,M)$ {\bf then} $\prt(M)$.
\end{quote}
The effect of the change is that we output only those sets in
$\Cmm(T(P),L)$, which are stable models of $P$. Since every stable
model of $P$ is a minimal model of $T(P)$, it is an element of
$\Cmm(T(P),L)$. Thus, this modified algorithm, we will refer to it as
$\stm$, indeed outputs all stable models of $P$ consistent with $L$
and nothing else.

In the same way, we construct an algorithm $\asm$ computing answer
sets of disjunctive programs. The only difference is that we use the
algorithm $\chas$ in place of $\chst$ to decide whether to output a
set.

We will now analyze the performance of the algorithms we just described.
The following observation is evident.

\begin{lemma}
\label{prop}
Let $\rho$ be a splitting cover function and let
\begin{enumerate}
\item the worst-case running time of the algorithm $\stb$ be
$O(t_1(n,m))$, for some integer function $t_1$, and
\item the worst-case running time of the algorithm $\chmm$ ($\chst$ or
$\chas$, depending on the problem) be $O(t_2(n,m))$, for some integer
function $t_2$.
\end{enumerate}
Then the running time of the algorithms $\mmm$, $\stm$ and $\asm$
(in the worst case) is $O(t_1(n,m) {+} \gamma t_2(n,m))$, where
$\gamma= |L(\T_T)|$ or $|L(\T_{T(P)})|$, depending on
the problem, and $T$ and $P$ are an input CNF theory or an input
(disjunctive) program, respectively.
\end{lemma}
\begin{proof} Clearly, the running time of the algorithm $\chmm$ (and,
similarly, $\chst$ and $\chas$) is the sum of the running times of
the algorithm $\stb$ and of all the calls to the algorithm $\mmm$
($\stm$ and $\asm$, respectively). The number of calls to the algorithm
$\mmm$ ($\stm$ and $\asm$, respectively) is equal to the number of nodes
in the tree $\T_T$ ($\T_{T(P)}$, respectively). Since the
cover function $\rho$ is splitting,
that number is $O(\gamma)$. Thus, the assertion follows. \end{proof}


\section{2-CNF theories, 2-programs, disjunctive 2-programs}
\label{two}

The performance of the algorithms $\mmm$, $\stm$ and $\asm$ depends on
the performance of the algorithm $\stb$ and on the performance of the
algorithms $\chmm$, $\chst$ and $\chas$. The performance of the
algorithm $\stb$ depends, in turn, on the performance of the
implementations of the underlying cover function $\rho$ and the
simplifying function $\sigma$.

We note that if $T$ is a 2-CNF theory, then throughout the execution of
the algorithm $\stb$ we only encounter 2-CNF theories (theories $S_A$
are 2-CNF theories if $S$ is a 2-CNF theory). Thus, it is enough to
define and implement a simplifying function $\sigma$ and a cover
function $\rho$ for 2-CNF theories only. We define $\sigma(T)=T$, for
every 2-CNF theory $T$ (we choose the identity function for $\sigma$)
and we have the following result concerning $\rho$.

\begin{lemma}
\label{th-2cnf}
Let a simplifying function $\sigma$ be the identity function. There is a
splitting cover function $\rho$ for 2-CNF theories that can be
implemented to run in time $O(m)$ and such that for every 2-CNF theory
$T$, $|L(\T_T)| \leq 1\mbox{.}4422\mbox{..}^n$.
\end{lemma}
\begin{proof} We recall that we only
consider CNF theories that do not contain clauses with multiple
occurrences of the same literal or occurrences of both a literal and
its dual. Thus, theories we consider here contain no clauses of the
form $\gamma\vee\dua\gamma$, $\gamma\vee\gamma$, where $\gamma$ is a
literal.

To define a splitting function $\rho$ for a 2-CNF theory $S$ with
$\At(S)\not=\emptyset$, we will consider several cases depending on
the properties of $S$. In each of them we assume that situations
covered by the ones considered before are excluded.

\noindent
{\bf Case 1.} $|\At(S)|=1$. Let us assume that $\At(S)=\{x\}$. In this
case, $S=\{x\}$ or $S=\{\dua x\}$ (in this proof, we view clauses as
disjunctions of literals). We define $A_1=\{x\}$ or $A_1=\{\dua
x\}$, respectively. Clearly, $\{A_1\}$ is a cover for $S$. We set
$\rho(S)=\{A_1\}$.

\noindent
{\bf Case 2.} There is a literal $\omega$ such that the unit clause
$\omega$ belongs to $S$. Let $y$ be an atom in $\At(S)$ such that $y$
is not the atom in $\omega$ (such an atom exists, as $|\At(S)|\geq 2$
now). We define $A_1= \{\omega,y \}$ and $A_2= \{\omega,\dua y\}$.
Clearly, $\{A_1,A_2\}$ is a cover for $S$. Indeed, if $M$ is a minimal
model of $S$, then $M$ satisfies $\omega$ and either $M$ satisfies $y$
or $M$ satisfies $\dua y$. We set $\rho(S)=\{A_1,A_2\}$. We also observe
that, since $y$ does not appear in $\omega$, $|A_i|=2$, $i=1,2$.

\noindent
{\bf Case 3.} There is an atom $x$ such that all its occurrences in
clauses of $S$ are negative. Let $y$ be any {\em other} atom in $S$ (as
we argued above, such $y$ exists). We set $A_1=\{\dua x,y
\}$ and $A_2=\{\dua x,\dua y\}$. Let $M$ be a minimal model of $S$. Then
$M$ satisfies $\dua x$ (otherwise, $M'=M  -\{x\}$ would be a model of
$S$, a contradiction with the minimality of $M$). Since $M$ satisfies
either $y$ or $\dua y$, $\{A_1,A_2\}$ is a cover for $S$. We define
$\rho(S)=\{A_1,A_2\}$. Since $x\not= y$, we have $|A_i|=2$, $i=1, 2$.

\noindent
{\bf Case 4.} There is a clause $\dua x \vee \omega$ in $S$. Since we
assume now that Case 3 does not hold, there is also a clause $x \vee
\beta$ in $S$. In this case, we define $A_1=\{x,\omega\}$ and $A_2=
\{\dua x, \beta\}$. It is easy to verify that $\{A_1,A_2\}$ is a cover
for $S$ and we define $\rho(S)=\{A_1,A_2\}$. Since $x$ does not appear
in $\omega$ and $\beta$ (we assume that $S$ does not contain clauses of
the form $\gamma\vee \gamma$ and $\gamma\vee\dua \gamma$, where $\gamma$
is a literal), we also have $|A_i|=2$, $i=1,2$.

\noindent
{\bf Case 5.} All clauses in $S$ are of the form $x\vee y$, where $x$
and $y$ are different atoms.

\noindent
(a) There is an atom, say $x$, that appears in exactly one clause, say
$x\vee y$. Let us define $A_1=\{x,\dua y\}$ and $A_2=\{\dua x, y\}$, and
let $M$ be a minimal model of $S$. If $x\in M$ then $y\notin M$.
Indeed, otherwise $M  -\{x\}$ would be a model of $S$ (as $x$
appears only in the clause $x\vee y$ in $S$). That would contradict the
minimality of $M$. Thus, if $x\in M$, $M$ is consistent with $A_1$. If
$x\notin M$ then, since $M$ is a model of $x\vee y$, $y\in M$. Thus,
in this case, $M$ is consistent with $A_2$. It follows that
$\{A_1,A_2\}$ is a cover for $S$ and we define $\rho(S)=\{A_1,A_2\}$.
Moreover, we also have that $|A_i|=2$, $i=1,2$.

\noindent
(b) There is an atom $x$ that appears in at least three different
clauses,
say $x\vee y_i$, $i=1,2,3$. In this case, we set $A_1=\{x\}$ and
$A_2=\{\dua x, y_1,y_2,y_3\}$. It is evident that $\{A_1,A_2\}$ is a
cover for $S$ and we set $\rho(S)=\{A_1,A_2\}$. We also note that
$|A_1|=1$ and $|A_2|=4$.

\noindent
(c) Every atom appears in exactly two clauses. Let $w$ be an arbitrary
atom in $S$. Let $w\vee u$ and $w\vee v$ be the two clauses in $S$ that
contain $w$. Both $u$ and $v$ also appear in exactly two clauses in $S$.
Let $u\vee u'$ be the clause other than $w\vee u$ (that is, $u'\not=w$)
and $v\vee v'$ be the clause other than $w\vee v$ (that is, $v'\not=w$).
We set $A_1=\{\dua u, u',w\}$, $A_2=\{\dua v, v', w\}$ and $A_3=\{\dua
w,u,v\}$. By the construction, $|A_i|=3$, $i=1,2,3$. Moreover, the
family $\{A_1,A_2,A_3\}$ is a cover for $S$. Indeed, let $M$ be a
minimal model of $S$. If $u,v\in M$, then $w\notin M$, since $M$ is a
minimal model of $S$ and $w\vee u$ and $w\vee v$ are the only clauses in
$S$ containing $w$. If $u\not\in M$, then both $w$ and $u'$ are in $M$.
If $v\not\in M$, then  $w$ and $v'$ are in $M$. Thus, we define
$\rho(S)=\{A_1,A_2,A_3\}$.

Cases 1 - 5 exhaust all possibilities for a 2-CNF theory $S$ such that
$\At(S)\not=\emptyset$. Thus, the definition of $\rho$ is complete.
We argued in each case that $\rho(S)$ is a cover. Therefore $\rho$ is a
cover function. Moreover, in each case when $|\At(S)|\geq 2$, $|\rho(S)|
\geq 2$.  Thus, the cover function $\rho$ is splitting.

The definition of $\rho$ implies a simple algorithm for computing
$\rho(S)$. One needs to identify the first case that applies (it can be
accomplished in linear time in the size of $S$ given a linked-list
representation we discussed in Section \ref{algs}) and output the cover
constructed in that case (it can be done in constant time). Thus, the
overall algorithm can be implemented to run in linear time.

To estimate the number of leaves in the tree ${\cal T}_T$, where
$T$ is a 2-CNF theory, we will use Lemma \ref{thm-ineq}. To this end,
for each 2-CNF theory $S$ we define $\mu(S)= |\At(S)|$. It is evident
that $0\leq \mu(S)\leq |\At(S)|$. Thus, $\mu$ is a measure.

Since for every 2-CNF theory $S$, $\sigma(S)=S$, $\sigma$ is
$\mu$-compatible. Next, it follows directly from the definition of
$\rho$ that for every 2-CNF theory $S$ with $\At(S)\not=\emptyset$ and
for every $A\in \rho(S)$, $\mu(S)-\mu(S_A)>0$. Thus, $\rho$ is
also $\mu$-compatible. We will now show that $\rho$ is $\mu$-bounded by
$1\mbox{.}4422\mbox{..}\ $. In order to do it, for every 2-CNF theory $S$ 
with
$|\At(S)|\geq 1$ we estimate $\tau_S$, the unique positive root of the
equation (\ref{tau}).

Let $S$ be 2-CNF theory such that $|\At(S)|\geq 1$. In Case 1, $\rho(S)$
consists of exactly one set, say $A_1$, and $S_{A_1}=\emptyset$. Thus,
$\mu(S)=1$, $\mu(S_{A_1})=0$ and $\Delta(S,S_{A_1})=1$. The equation
(\ref{tau}) becomes $\tau =1$ and $\tau_S=1$.

In Cases 2 - 4 and 5a, $\rho(S)$ consists of two sets, $A_1$ and $A_2$,
and $\mu(S_{A_i}) \leq \mu(S)-2$, $i=1,2$. In other words,
$\Delta(S,S_{A_i})\geq 2$, $i=1,2$. In each such case, Lemma \ref{last}
implies that $\tau_S \leq \tau_1$, where $\tau_1$ is the root of the
equation $2\tau^{-2}=1$.

In Case 5b, $S$ has two children in $\T_T$, $S_{A_i}$, $i=1,2$. Moreover,
$\mu(S_{A_1}) \leq \mu(S)-1$ and $\mu(S_{A_2}) \leq \mu(S)-4$.
Consequently, $\Delta(S,S_{A_1})\geq 1$ and $\Delta(S,S_{A_2})\geq 4$.
In this case, Lemma \ref{last} implies that $\tau_S \leq \tau_2$, where
$\tau_2$ is the root of the equation $\tau^{-1}+\tau^{-4}=1$.

Finally, in Case 5c, $S$ has three children in $\T_T$,
$S_{A_i}$, $i=1,2,3$, and $\mu(S_{A_i}) \leq \mu(S)-3$, $i=1,2,3$. In
other  words, $\Delta(S,S_{A_i})\geq 3$, $i=1,2,3$. In this case,
$\tau_S \leq \tau_3$, where $\tau_3$ is the root of the equation
$3\tau^{-3}=1$.

Since $\tau_1= 1\mbox{.}4142\mbox{..}\ $, $\tau_2= 1\mbox{.}3802\mbox{..}\ $ 
and $\tau_3=
1\mbox{.}4422\mbox{..}$, it follows by Lemma \ref{thm-ineq} that
$|L(\T_T)|\leq (1\mbox{.}4422\mbox{..})^n$. Thus,
the assertion follows. \end{proof}

\noindent
{\bf Proof of Theorem \ref{2p2}.}
In Section \ref{algs} we noted that for every (disjunctive) 2-program
$P$, stable models (answer sets) of $P$ are minimal models of the
theory $T(P)$. Thus, Theorem \ref{2p2} follows from Lemmas
\ref{th-2cnf} and \ref{ell}.

In addition, Lemmas \ref{th-2cnf}, \ref{rem1} and \ref{time} imply the
following corollary.

\begin{corollary}
\label{cor-2cnf}
There is an implementation of the algorithm $\cmm$ that, for 2-CNF
theories, runs in time $O(m1\mbox{.}4422\mbox{..}^n)$.
\end{corollary}

To derive Theorem \ref{2p1} we will need one more auxiliary fact
concerning testing whether a set of atoms is a minimal model of a 2-CNF
theory.

\begin{proposition}
\label{prop-min}
Let $T$ be a 2-CNF theory and $M\subseteq \At(T)$. There is a
linear-time algorithm to test whether $M$ is a minimal model of $T$.
\end{proposition}
\begin{proof} First, we test whether $M$ is a model of $T$. To this end, we
check whether every clause in $T$ has a literal that is true in $M$ (a
literal of the form $a$, where $a\in M$, or a literal of the form
$\overline{a}$, where $a\notin M$). That task can be accomplished in
linear time in the size of $T$. If $M$ is not a model of $T$, it is not
a minimal model. So, from now on we will assume that $M$ is a model of
$T$.

We now define $L=\{\overline{a}\colon a\in \At(T)  -
M\}$. By Lemma \ref{key}(3), $M$ is a minimal model of $T$
if and only if $M$ is a minimal model of $T_L$.

Let $c$ be a clause of $T_L$. By Lemma \ref{key}(1), $M$ is a model of
$c$. Let us assume that $c$ consists of negated atoms only. Then, $c$
contains a literal $\dua a$, where $a$ is an atom, and $M$ satisfies
$\dua a$. On the other hand, since $c\in T_L$, it follows from the way
$T_L$ is defined that $a\in M$. Thus, we get a contradiction.
Consequently, every clause in $T_L$ is of the form $a$, $a\vee b$ or $a
\vee\overline{b}$. In particular, it follows that $\At(T_L)$ is a model
of $T_L$. By the construction of $T_L$, $\At(T_L)\subseteq M$. Thus,
if $\At(T_L)\not=M$, $M$ is not a minimal model of $T_L$. Therefore,
from now on we can assume that $\At(T_L)=M$.

We form a directed graph $G$ by using atoms in $\At(T_L) (=M)$ as its
vertices and by connecting vertices $a$ and $b$ with a directed edge
$(a,b)$ if and only if $\overline{a}\vee b$ is a clause in $T_L$.
Strongly connected components in the graph $G$ consist of atoms that
must have the same truth value in every model $M'$ of $T_L$ (either all
atoms of a strongly connected component of $G$ are in $M'$ or none of
them is).

By a {\em 0-rank} strongly connected component of $G$ we mean a strongly
connected component $S$ of $G$ such that there is no edge $(a,b)$ in
$G$ with $a\notin S$ and $b\in S$.
One can show that $M$ is a minimal model of $T_L$ if and only if every
0-rank strongly connected component in $G$ contains at least one positive
clause of $T_L$.

Computing strongly connected components and 0-rank strongly
connected components can be done in linear time. Similarly, one can
verify in linear time whether every 0-rank component contains a positive
clause from $T_L$. Thus, the method described in the proof can be
implemented to run in linear time. \end{proof}

\noindent
{\bf Proof of Theorem \ref{2p1}.}
It is well known that testing whether a set of atoms $M$ is a stable
model of a logic program can be done in linear time. Testing whether a
set of atoms $M$ is a minimal model of a 2-CNF theory can be done in
linear time according to Proposition \ref{prop-min}. To test whether a
set of atoms $M$ is an answer set of a disjunctive 2-program one needs
to test whether $M$ is a minimal model of the reduct $P^M$ or,
equivalently, whether $M$ is a minimal model of a 2-CNF theory $T(P^M)$.
Thus, by Proposition \ref{prop-min}, also this task can be accomplished
in linear time. Consequently Theorem \ref{2p1} follows from Lemma
\ref{prop} and Corollary \ref{cor-2cnf}.

\section{3-CNF theories, 3-programs, disjunctive 3-programs}
\label{three}

In this section, we will prove Theorems \ref{3p2}, \ref{3p1} and
\ref{3p1d}. As with 2-CNF theories, the first step is to specify
functions $\sigma$ and $\rho$. Let $T$ be a 3-CNF theory. We define
$\sigma(T)$ to be the theory obtained by eliminating multiple
occurrences of clauses and all 3-clauses that are subsumed by 2-clauses
in $T$.

\begin{lemma}
\label{smpl} Let $T$ be a 3-CNF theory. There is an algorithm that
computes $\smpl(T)$ and runs in linear time in the size of $T$
(assuming a linked-list representation of $T$).
\end{lemma}
\begin{proof} We first create a linked list $Q$ of clauses that contains for
each clause $c$ in $T$ all lists obtained from $c$ by permuting its
elements. With each permutation of $c$, we store a pointer to $c$ on the
list (representing) $T$. The list $Q$ has size at most six
times larger than the size of $T$.

Next, we sort $Q$ lexicographically. This task can be accomplished
in linear time by the radix sort algorithm \cite{ahu74}. Clearly,
if $T$ contains $r\geq 2$ occurrences of a clause $c$, then for
every permutation $c'$ of the literals in $c$, $Q$ will contain a
contiguous segment of $r$ occurrences of $c'$. Conversely, each
contiguous segment of identical elements on the list $Q$ indicates
a clause that appears in $T$ multiple times. Thus, we can identify
all clauses with multiple occurrences in $T$ in a single pass
through $Q$ (we recall that we maintain pointers from clauses in
$Q$ to their original counterparts in $T$) and, then, delete
duplicates.

Similarly, a 3-clause $c\in T$ is subsumed by a 2-clause $d\in T$ if and
only if some permutation $d'$ of $d$ is a prefix of some permutation
$c'$ of $c$ and so, $d'$ is an immediate predecessor of $c'$ on $Q$.
As before, all such 3-clauses can be identified in a single pass
through $Q$ and then deleted from $T$.

Since we maintain $T$ as a doubly-linked list, each deletion can be
performed in linear time. Consequently, the whole process takes linear
time in the size of $T$. \end{proof}

We will next specify an appropriate cover function
$\rho$. Due to the choice of $\sigma$, it suffices to define $\rho(T)$
for every 3-CNF theory $T$ such that $\At(T)\not=\emptyset$ and $T$
contains no 3-clauses subsumed by 2-clauses in $T$, as that is enough to
determine the tree $\T_T$. We have the following result.

\begin{lemma}
\label{mainth}
There is a splitting cover function $\rho$ defined for every 3-CNF
theory $T$ that contains no multiple clauses nor 3-clauses subsumed
by 2-clauses in $T$, which can be implemented to run in linear time and
such that for every 3-CNF theory $T$, $|L(\T_T)| \leq 
1\mbox{.}6701\mbox{..}^n$.
\end{lemma}

We outline a proof of Lemma \ref{mainth} in the next section and
provide full details in the appendix.

\noindent
{\bf Proof of Theorem \ref{3p2}.} Theorem \ref{3p2} follows directly
from Lemmas \ref{ell} and \ref{mainth}.

Next, we note that Lemmas \ref{rem1}, \ref{time} and \ref{mainth} imply
the following corollary.

\begin{corollary}
\label{cor-main}
There is an implementation of the algorithm $\cmm$ that, for 3-CNF
theories, runs in time $O(m1\mbox{.}6701\mbox{..}^n)$.
\end{corollary}

\noindent
{\bf Proof of Theorem \ref{3p1}.}
Since there is a linear-time algorithm to test whether a set of atoms is
a stable model of a logic program, Theorem \ref{3p1} follows from
Corollary \ref{cor-main} and Lemma \ref{prop}.

We will now prove Theorem \ref{3p1d}. We focus on the case of
minimal models of 3-CNF theories. The argument in the case of answer
sets of disjunctive 3-programs is similar. We start with a simple result
on testing minimality.

\begin{proposition}
\label{simp}
Let $f$ be an integer function and $t$ an integer such that $t\geq 2$.
If there is an algorithm that decides in time $O(f(m,n))$ whether a
$t$-CNF theory $T$ is satisfiable, then there is an algorithm that
decides in time $O(|M|f(m+1,n))$ whether a set $M\subseteq \At(T)$ is
a minimal model of a $t$-CNF theory $T$.
\end{proposition}
\begin{proof} Let $M=\{a_1,\ldots,a_k\}$. We define $L=\{\dua x\colon
x\in \At(T)  - M\}$ and observe that $M$ is a minimal model of $T$
if and only if $M$ is a model of $T$ and none of $t$-CNF theories
$T_L\cup \{\dua a_i\}$, $i=1,\ldots,k$, is satisfiable. Thus, to
decide if $M$ is a minimal model of $T$, we first check if $M$ is
a model of $T$ (in time $O(m)$) and then apply the algorithm
checking satisfiability of $k=|M|$ $t$-CNF theories $T_L\cup
\{\dua a_i\}$ of size $m+1$ each (in time $O(f(m+1,n))$ each).
\end{proof}

Satisfiability of 3-CNF theories can be decided in time
$O(m 1\mbox{.}481^n)$ \cite{dgh02}. Thus, by Proposition
\ref{simp}, there is an an algorithm to decide whether a set
$M\subseteq \At(T)$ is a minimal model of a $3$-CNF theory $T$
that runs in time $O(|M|m\cdot 1\mbox{.}481^{|M|})$.

\noindent
{\bf Proof of Theorem \ref{3p1d}.}
We can assume that $n\geq 4$. Let $\beta$ be a real number such that
$0\mbox{.}6 \leq \beta<1$ (we will specify $\beta$ later). We will now 
estimate
the running time of the algorithm $\mmm$, in which the procedure $\chmm$
is an implementation of the method described above. By the proof of
Lemma \ref{prop}, this running time is the sum of the running time of
the algorithm $\cmm$ and of the total time $t_{\mathit{min}}$ needed to
execute all calls to $\chmm$ throughout the execution of $\mmm$. By
Corollary \ref{cor-main}, the first component is $O(m 
1\mbox{.}6701\mbox{..}^n)$.

To estimate $t_{\mathit{min}}$, we split
$\Cmm(T,\emptyset)$ into two parts:
\[
{\cal M}_1=\{M\in \Cmm(T,\emptyset)\colon |M| \geq \beta n\}\ \mbox{and}\
{\cal M}_2=\{M\in \Cmm(T,\emptyset)\colon |M| < \beta n\}\mbox{.}
\]
Clearly, the total time $t_{\mathit{min}}$ needed to execute all
calls to $\chmm$ throughout the execution of $\cmm$ is:
\[
t_{\mathit{min}} = O(\sum_{M\in{\cal M}_1} m |M| (1\mbox{.}481)^{|M|} +
\sum_{M\in{\cal M}_2} m |M| (1\mbox{.}481)^{|M|})\mbox{.}
\]
We have
\begin{eqnarray*}
\sum_{M\in{\cal M}_1} m |M| (1\mbox{.}481)^{|M|} &\leq& \sum_{i\geq \beta n}
m{n\choose i} i (1\mbox{.}481)^{i}\\
&\leq& m(n+1-\lceil \beta n\rceil)\lceil \beta n\rceil{n\choose {\lceil
\beta n\rceil}} (1\mbox{.}481)^{\lceil \beta n\rceil}\\
&\leq& \beta mn^2{n\choose {\lceil\beta n\rceil}}(1\mbox{.}481)^{\lceil 
\beta n\rceil}\mbox{.}
\end{eqnarray*}
The second inequality follows from that fact that for every $i$, $i \geq
0\mbox{.}6n$,
\[
m{n\choose i} i (1\mbox{.}481)^{i} \geq m{n\choose i+1} (i+1)
(1\mbox{.}481)^{i+1},
\]
and from the observation that the number of terms in the
sum is $n+1-\lceil \beta n\rceil$. The last inequality follows by an
easy calculation from the assumptions that $n\geq 4$ and $\beta\geq
0\mbox{.}6$.

To estimate the second term, we note that $|{\cal M}_2|\leq
|\Cmm(T,\emptyset)| \leq (1\mbox{.}6701\mbox{..})^n$ and, for
every $M\in {\cal M}_2$, $|M| < \beta n$. Thus,
\[
\sum_{M\in{\cal M}_2} m |M| (1\mbox{.}481)^{|M|} \leq (\beta m n) 
(1\mbox{.}6701\mbox{..})^n
(1\mbox{.}481)^{\beta n}\mbox{.}
\]
Let us choose $\beta$ to be the smallest $\beta' \geq 0.6$ such that
${n\choose {\lceil\beta' n\rceil}}=O(1\mbox{.}6701\mbox{..}^n)$.
One can verify that $\beta = 0\mbox{.}7907\mbox{..}\ $. For this
$\beta$, we have
\[
t_{\mathit{min}}= O(mn^2 (1\mbox{.}6701\mbox{..}  (1\mbox{.}481)^\beta)^n) =
O(mn^2 2\mbox{.}2782\mbox{..}^n),
\]
which completes the proof of Theorem \ref{3p1d}.

\section{An outline of the proof of Lemma \ref{mainth}}
\label{cov}

To prove Lemma \ref{mainth}, we will follow a similar approach to that
we used in the proof of Lemma \ref{th-2cnf}. In the proof we define a
cover function $\rho$ and verify that it satisfies all the requirements
of the lemma. Because of our choice of the simplifying function, it is
enough to define $\rho$ for every 3-CNF theory $S$ such that $\At(S)\not
=\emptyset$ and $S$ contains no multiple occurrences of clauses nor
3-clauses subsumed by 2-clauses of $S$.

To estimate the number of leaves in the tree ${\cal T}_T$, we will
introduce a measure $\mu$ and show that $\smpl$ is $\mu$-compatible and
that $\rho$ is $\mu$-compatible and $\mu$-bounded by 
$1\mbox{.}6701\mbox{..}\ $.
Specifically, for a 3-CNF theory $S$, we define $\mu(S)= n(S) -\alpha
k(S)$, where $n(S)=|\At(S)|$, $k(S)$ is the maximum number of 2-clauses
in $S$ with pairwise disjoint sets of atoms, and $\alpha$ is a constant
such that $0< \alpha< 1$. We will specify the constant $\alpha$ later.
At this point we only note that for every $\alpha\in [0,1]$ and for
every 3-CNF theory $S$, $0\leq \mu(S)\leq|At(S)|$. Thus, $\mu$ is indeed
a measure (no matter what $\alpha\in[0,1]$ we choose).

Let $S$ be a 3-CNF theory such that $|\At(S)|\geq 1$. If $S'=\smpl(S)$
then $|\At(S)|\geq |\At(S')|$. Moreover, since the sets of 2-clauses in
$S$ and $S'$ are the same, $k(S)=k(S')$. Thus, $\mu(S)\geq \mu(S')$
and, consequently, $\smpl$ is $\mu$-compatible.

We will now outline the construction of $\rho(S)$,
where $S$ is a 3-CNF theory such that $\At(S)\not=\emptyset$ and $S$
contains no multiple occurrences of clauses nor 3-clauses subsumed by
2-clauses in $S$. We consider several cases depending on the structure
of $S$. We design the cases so that every such 3-CNF theory $S$ falls
into exactly one of them. Moreover, we design these cases so that one
can decide which case applies to $S$ in linear time in the size of $S$.
In each case, for a 3-CNF theory $S$ we describe a specific cover for
$S$ and use it as the value of $\rho(S)$. In each case, it is clear
that $\rho(S)$ can be output in constant time. Consequently, it follows
that computing $\rho(S)$ can be accomplished in linear time. Moreover,
in each case when $|\At(S)|\geq 2$, we have that $|\rho(S)|\geq 2$. That
implies that $\rho$ is splitting.

In each case of the definition of $\rho(S)$ and for each $A\in \rho(S)$,
we will determine a {\em positive} real number $k_{A,S}$ (in general,
depending on
$\alpha$) such that $\Delta(S,S_A)=\mu(S)-\mu(S_A)\geq k_{A,S}$. We will
use these values to state the equation (\ref{tau'}), whose root
$\tau_S'$ provides, by Lemma \ref{last}, an upper bound for $\tau_S$,
the root of the equation (\ref{tau}). The fact that the numbers $k_{A,S}$
are positive implies that the cover function $\rho$ is $\mu$-compatible.

Once we describe all the cases, we will then finally select $\alpha$. We
will do it so that in each case considered when defining $\rho(S)$, the
values $k_{A,S}$ are positive and $\tau_S' \leq 1\mbox{.}6701\mbox{..}\ $. 
That
property shows that $\rho$ is $\mu$-bounded by $1\mbox{.}6701\mbox{..}\ $.

The properties of the cover function $\rho$ and of the function $\smpl$
together with Lemma \ref{thm-ineq} imply now Lemma \ref{mainth}.

In Cases 1 - 3 of the definition of the function $\rho$ we deal with
three simple situations when $|\At(S)|=1$, when $S$ contains a 1-clause,
and when some atom appears only negated in clauses of $S$.

Cases 4 - 10 cover situations when $S$ contains a 2-clause. Case 4
covers the situation when there is a pair of 2-clauses in $S$ with a
common atom. Therefore in the remaining cases, we assume that the sets
of atoms of 2-clauses in $S$ are pairwise disjoint. That assumption
makes it easier to analyze $\mu(S)-\mu(S_A)$ and obtain bounds
$k_{S,A}$. Indeed, under that assumption, $k(S)$ is simply equal
to the number of 2-clauses in $S$.

Cases 5 and 6 together cover the situations when there is a
2-clause and a 3-clause in $S$ such that the 2-clause has a
literal, which is dual to a literal occurring in the 3-clause, and
when the set of atoms of the 2-clause is a subset of the set of
atoms of the 3-clause. That allows us to assume in all subsequent
cases that for every atom $a$ occurring in a 2-clause $c$ in $S$,
all occurrences of $a$ in clauses of $S$ are positive. Indeed,
they cannot be all negative by Case 3. Moreover, by Cases 1 and 4,
all clauses in $S$ that contain $a$ and are different from $c$ are
3-clauses. If any of them contains a negated occurrence of $a$,
Case 5 or 6 would apply.

In Case 7 we assume that there is an atom of a 2-clause that does not
belong to any other clause in $S$. In Case 8 we consider the situation
when some atom of a 2-clause appears as a literal in exactly one
3-clause. Finally, in Cases 9 and 10 we assume that there is an atom of
a 2-clause, which belongs to at least 2 different 3-clauses.

It is easy to verify that Cases 1 - 10 exhaust all possibilities when
there is a 2-clause or a 1-clause in $S$. Therefore, from now on we
assume that all clauses in $S$ are 3-clauses. For an atom $a\in \At(S)$,
we denote by $T(a)$ the theory consisting of the clauses of $S$, in
which $a$ is one of the literals.

In Case 11 we consider theories $S$ such that, for some atom $a\in
\At(S)$, there are two clauses in $T(a)$ such that one of them contains
a literal, which is dual to a literal in the other one. Thus, in the
remaining cases we assume that, for each atom $a\in \At(S)$, the theory
$T(a)$ does not contain dual literals. We denote by $\Gamma(a)$ an
undirected graph whose vertices are the literals different from $a$
occurring in the clauses of $T(a)$ and a pair of literals $\beta\gamma$
is an edge in $\Gamma(a)$ if $a\vee\beta\vee\gamma$ is a clause in
$T(a)$. We call the number of neighbors of a vertex in a graph
$\Gamma(a)$ the {\em degree} of the vertex.

In Cases 12 - 20 we assume that there is an atom $a\in \At(S)$, for
which the graph $\Gamma(a)$ has some specified structural properties.
In Case 12 we assume that for some atom $a$, there is a vertex in the
graph $\Gamma(a)$ with at least $5$ neighbors. Case 13 covers the
situation when the maximum degree of a vertex in some graph $\Gamma(a)$
is $3$ or $4$. In Case 14 we assume that there is an atom $a$ such that
$\Gamma(a)$ has at least $4$ independent edges. In Cases 15 - 20 we
consider the theories $S$ such that the graph $\Gamma(a)$, for some
$a\in \At(S)$, has no vertices of degree $3$ or more and is not
isomorphic to any of the following three graphs: $C_3\cup P_1$ (a graph
whose components are a triangle and a single edge), $P_3\cup P_1$
(a graph whose components are a 3-edge path and a single edge) and
$3K_2$ (a graph whose components are three single edges).

Finally, in Case 21 we assume that, for all atoms $a\in \At(S)$, the
graphs $\Gamma(a)$ are isomorphic to one of the graphs $C_3\cup P_1$,
$P_3\cup P_1$, $3K_2$. First we consider the case when some atom of
$\At(S)$ occurs in $S$ negated. Next we assume that all occurrences of
atoms in $S$ are positive.

Let us now explain the choice of a particular value of $\alpha$ in the
definition of the measure $\mu(S)=|\At(S)|-\alpha k(S)$. Our goal is to
get as good an upper bound for the number of leaves in $\T_T$ as we
can. We choose the value of $\alpha$ so that the maximum $\tau_0$ of the
solutions of the equation (\ref{tau'}) over all cases considered in the
definition of $\rho$ be as small as possible.

It turns out that the Cases 9(iii) and 14 are, in a sense,
``extremal''. In Case 9(iii) of the definition of $\rho(S)$, the
equation (\ref{tau'}) specializes to
\begin{equation}
\label{eq11}
\tau^{-1+\alpha}+\tau^{-4+3\alpha}+2\tau^{-6+4\alpha}+\tau^{-8+5\alpha}=1\mbox{.}
\end{equation}
In Case 14, the equation (\ref{tau'}) becomes
\begin{equation}
\label{eq22} \tau^{-1}+\tau^{-1-4\alpha}=1\mbox{.}
\end{equation}

It is easy to verify that the value $\tau_1=\tau_1(\alpha)$ of the
positive root of the equation (\ref{eq11}) satisfies the inequality
$\tau_1>1$ and grows, when $\alpha$ grows from $0$ to $1$. On the other
hand, the value $\tau_2=\tau_2(\alpha)$ of the positive root of the
equation (\ref{eq22}) satisfies the inequality $\tau_2>1$ and decreases,
when $\alpha$ grows from $0$ to $1$. The larger of the roots $\tau_1,
\tau_2$ is minimized when $\tau_1=\tau_2$. This equality happens to be
achieved for $\alpha=0\mbox{.}1950\mbox{..}\ $. For this value of $\alpha$, 
we have
$\tau_1=\tau_2=1\mbox{.}6701\mbox{..}\ $. Moreover, it can be checked by 
direct
computations that in all remaining cases, if 
$\alpha=0\mbox{.}1950\mbox{..}$, then
the values $k_{S,A}$ are positive and the roots of the equation
(\ref{tau'})
are smaller than $1\mbox{.}67$. Thus, for $\alpha=0\mbox{.}1950\mbox{..}\ $, 
$\rho$ is
$\mu$-compatible and $\mu$-bounded by $1\mbox{.}6701\mbox{..}\ $.

\section{The case of an arbitrary $t\geq 2$}
\label{ubs}

In this section, we briefly discuss computing minimal models of $t$-CNF
theories, stable models of normal $t$ programs and answer sets of
disjunctive $t$-programs for an arbitrary $t\geq 2$. First, we recall
the following result from \cite{lt02b}. When stating it, by $\alpha_t$
we denote the unique positive root of the equation $1 + \tau+ \tau^2+
\ldots+\tau^{t-1} =\tau^t$. It is easy to see that $\alpha_t\leq
2-1/2^t$ and one can show
numerically that $\alpha_2=1\mbox{.}6180\mbox{..}\ $, $\alpha_3=
1\mbox{.}8393\mbox{..}\ $, $\alpha_4 = 1\mbox{.}9275\mbox{..}\ $ and
$\alpha_5 = 1\mbox{.}9659\mbox{..}\ $.

\begin{theorem}[\cite{lt02b}]
\label{stb-tcnf} 
Let $t$ be an integer, $t\geq
2$. There is an algorithm computing all stable models of
$t${-}programs that runs in time $O(m \alpha_t^n)$.
\end{theorem}

Theorem \ref{stb-tcnf} can also be derived for the results we presented
in this paper. Let $T$ be a CNF theory that contains a $t$-clause
$\beta_1\vee\ldots\vee\beta_t$. It is easy to see that the family of
sets $\{A_1,\ldots,A_t\}$, where $A_i=\{\dua\beta_1,\ldots,\dua
\beta_{i-1}, \beta_i\}$, $1\leq i\leq t$, is a cover for $T$. By
exploiting that observation, we can show that there is a splitting
cover function $\rho$ such that for every $t$-CNF theory $T$,
\begin{enumerate}
\item $\rho(T)$ can be computed in time $O(m)$, and
\item $|L(\T_T)| \leq \alpha_t^n$ (we use the identity function for
the simplifying function $\sigma$ and we use the measure $\mu(S)=|\At(S
)|$ to derive that bound).
\end{enumerate}
It follows that the corresponding implementation of the algorithm $\cmm$
runs in $O(m\alpha_t^n)$ steps.
Moreover, the family $\Cmm(T,\emptyset)$ that is returned by $\cmm(T,T,
\emptyset)$ satisfies $|\Cmm(T,\emptyset)| \leq \alpha_t^n$. Reasoning
as in other places in the paper, it is easy to derive Theorem
\ref{stb-tcnf} from these observations.

Since $\Cmm(T,\emptyset)$ contains all minimal models of $T$, we also
get the following result.

\begin{theorem}
\label{num-tcnf}
Every t{-}CNF theory $T$ (every normal t-program $P$ and
every disjunctive t-program $P$,
respectively) has at most $\alpha_t^n\leq (2-1/2^t)^n$ minimal
models (stable models, answer-sets, respectively).
\end{theorem}

Next, we will construct algorithms for computing all minimal models of
$t$-CNF theories and all answer sets of disjunctive $t$-programs in the
case of an arbitrary $t\geq 2$. Not surprisingly, in the case of $t=2$
and $t=3$ these results are weaker than those we obtained earlier in
the paper. Our approach does not depend on general results developed
earlier in Sections \ref{algs} and \ref{algs2}. It exploits instead
recent results on deciding satisfiability of $t$-CNF theories
\cite{dgh02}.

\begin{theorem}
\label{mmast}
There is an algorithm to compute all minimal models of $t$-CNF theories
and answer sets of disjunctive $t$-programs, respectively, that runs in
time $O(q(m)(3-2/(t+1))^n)$, for some polynomial $q$.
\end{theorem}
\begin{proof} There is a polynomial $q'$ such that the satisfiability of
$t$-CNF theories can be decided in time $O(q'(m) (2-2/(t+1))^n)$
\cite{dgh02}. Thus, by Proposition \ref{simp}, there is an an
algorithm to decide whether a set $M\subseteq \At(T)$ is a minimal model
of a $t$-CNF theory $T$, which runs in time
\[
O(|M|q'(m+1)(2-2/(t+1))^{|M|}) = O(q(m)(2-2/(t+1))^{|M|}),
\]
where $q(m)=mq'(m+1)$. Using this algorithm as a minimality-testing
procedure in the straightforward algorithm to compute all minimal models
of a $t$-CNF theory $T$, which generates all subsets of $\At(T)$ and
tests each of them for being a minimal model of $T$, yields a method to
compute all minimal models of a $t$-CNF theory that runs in time:
\[
O(\sum_{i=0}^n {n \choose i} q(m) (2-2/(t+1))^i) = 
O(q(m)(3-2/(t+1))^n\mbox{.}
\]
The argument in the case of answer sets is similar. \end{proof}

We note that the method we used to derive Theorem \ref{3p1d} can be
generalized to an arbitrary $t\geq 3$ (using the observations made after
Theorem \ref{stb-tcnf}). However, the bound $|\Cmm(T,\emptyset)|\leq
\alpha_t^n$ is too week to yield algorithms faster than the algorithm
described in the proof of Theorem \ref{mmast}. The only exception is the
case of $t=4$, where by generalizing the proof of Theorem \ref{3p1d} we
can derive an algorithm constructing all minimal models of 4-CNF theories
(answer sets of disjunctive 4-programs) in $O(m 2\mbox{.}5994\mbox{..}^n)$ 
steps.
Since the improvement over the bound of $O(q(m)2\mbox{.}6^n)$ implied by
Theorem \ref{mmast} is so small, we omit the details of the derivation
of the $O(m 2\mbox{.}5994\mbox{..}^n)$ bound.

\section{Lower bounds}
\label{lbs}

In this section, we derive lower bounds on the number of minimal models,
stable models or answer sets a $t$-CNF theory, a $t$-program and a
disjunctive $t$-program may have. We will also derive lower bounds
on the running time of algorithms for computing {\em all} minimal
models, stable models or answer sets of $t$-CNF theories and
$t$-programs.


All examples we construct have a similar structure. Let $X$ be a set of
atoms and let $t$ be an integer such that $2\leq t \leq |X|$. By
$E_{t,X}$ we denote a $t$-CNF theory consisting of all clauses of
the form $a_{1}\vee \ldots\vee a_{t}$, where atoms $a_i$ belong to
$X$ and are pairwise distinct. By $E^d_{t,X}$, we mean a disjunctive
$t$-program consisting of the same clauses, but treated as
disjunctive-program clauses. Finally, by $E^p_{t,X}$ we mean a
$t$-program consisting of all program clauses of the form $a_{t}\lla
\n(a_{1}), \ldots, \n(a_{t-1})$, where, as before, all atoms $a_i$
belong to $X$ and are pairwise distinct. The sizes of $E_{t,X}$ and
$E^d_{t,X}$ are the same and are equal to $t{|X| \choose t}$. The size
of $E^p_{t,X}$ is $t^2{|X| \choose t}$.

Minimal models of $E_{t,X}$, stable models of $E^p_{t,X}$ and answer
sets of $E^d_{t,X}$ coincide. In fact, in each case, they are precisely
$(|X|-t+1)$-element subsets of $X$. Thus, $E_{t,X}$, $E^d_{t,X}$ and
$E^p_{t,X}$ have $|X| \choose t-1$ minimal models, stable models and
answer sets, respectively.

Let $k$ be a positive integer and let us consider $k(2t-1)$ {\em
distinct} elements $x_{i,j}$, where $i=1,\ldots,k$ and $j=1,\ldots,
2t-1$. We define $X_i=\{x_{i,1},\ldots,x_{i,{2t-1}}\}$ and set
\[
F_{t,k} =\bigcup_{i=1}^k E_{t,X_i},\ \ F^p_{t,k} =\bigcup_{i=1}^k
E^p_{t,X_i}\ \ \mbox{and}\ \ F^d_{t,k} =\bigcup_{i=1}^k E^d_{t,X_i}\mbox{.}
\]
Clearly, $F_{t,k}$ is a $t$-CNF theory, $F^p_{t,k}$ is a normal
$t$-program and $F^d_{t,k}$ is a disjunctive $t$-program. Moreover,
since each of these theories (programs) is the {\em disjoint} union of
$k$ {\em isomorphic} components $E_{t,X_i}$ ($E^d_{t,X_i}$, $E^p_{t,X_
i}$, respectively), we have the following simple observations:
\begin{enumerate}
\item $|\At(F_{t,k})| = |\At(F^p_{t,k})| = |\At(F^d_{t,k})| = k(2t-1)$.
\item The size of $F_{t,k}$ and $F^d_{t,k}$ is $kt{2t-1 \choose t}$; the
size of $F^p_{t,k}$ is $kt^2{2t-1 \choose t}$.
\item $F_{t,k}$, $F^p_{t,k}$ and $F^d_{t,k}$ have ${2t{-}1 \choose
t}^k$ minimal models, stable models and answer sets, respectively, and
each of these models or answer sets has $kt$ elements.
\end{enumerate}

Let us define $\mu_t= {2t{-}1 \choose t}^{1/{2t{-}1}}$. The
observations (1)-(3) imply the following result.

\begin{theorem}
\label{lb}
Let $t$ be an integer, $t\geq 2$. There are positive constants
$d_t$, $D_t$ and $D'_t$ such that for every $n\geq 2t{-}1$ there is a
$t$-CNF theory $T$ (a $t${-}program $P$ or a disjunctive $t$-program $Q$,
respectively) with $n$ atoms and such that
\begin{enumerate}
\item The size $m$ of $T$ ($P$ and $Q$, respectively) satisfies $m \leq
d_t n$
\item The number of minimal models of $T$ (stable models of $P$ or
answer sets of $Q$, respectively) is at least $D_t \mu_t^n$ and the sum
of their cardinalities is at least $D'_t n \mu_t^n$.
\end{enumerate}
\end{theorem}
\begin{proof} We will prove the assertion only in the case of CNF theories.
The arguments for $t$-programs and disjunctive $t$-programs are similar.

We will show that $d_t = 2{2t-1 \choose t}$, $D_t= \mu_t^{-(2t-1)}$
and $D'_t=D_t/4$ have the required properties.

Let $n\geq 2t-1$. We select $k$ to be the largest integer such
that $k(2t-1)\leq n$. Clearly, $k\geq 1$. We select a set $X$ of
$n-k(2t-1)$ atoms, all of them different from atoms $x_{i,j}$ that
appear in the theory $F_{t,k}$. Finally, we define $T =
F_{t,k}\cup E_{t,X}$.

Clearly, $T$ contains $n$ atoms. Moreover, the size of $T$, $m$,
satisfies $m = m'+m''$, where $m'$ and $m''$ are the sizes of
$F_{t,k}$ and $E_{t,X}$, respectively. Since $k\geq 1$, $m'' < m'$.
Thus, $m\leq 2m'=2kt{2t-1 \choose t}$ (Observation 2). Since
$kt\leq k(2t-1)\leq n$, we obtain $m\leq nd_t$.

The number of minimal models of $T$ is greater than or equal to the number
of minimal models of $F_{t,k}$ (since the sets of atoms of $F_{t,k}$ and
$E_{t,X}$ are disjoint, every minimal model of $F_{t,k}$ extends to a
minimal model of $T$). By Observation 3, the latter  number is
$\mu_t^{k(2t-1)}$. We now have
\[
\mu_t^{k(2t-1)} \geq  \mu_t^{n-(2t-1)} = \mu_t^{-{(2t-1)}}\mu_t^n =
D_t\mu_t^n\mbox{.}
\]

Each of the minimal models has size at least $kt$ (again, by the
fact that the sets of atoms of $F_{t,k}$ and $E_{t,X}$ are
disjoint). Since $kt=\lfloor n/(2t-1)\rfloor t\geq n/4$, the total
size of all minimal models of $T$ is at least $D_t'n\mu_t^n$.
\end{proof}

As a corollary to Theorem \ref{lb}, we obtain the following result.

\begin{corollary}
\label{clb}
Let $t$ be an integer, $t\geq 2$.
\begin{enumerate}
\item There is a $t$-CNF theory (a $t$-program, a disjunctive $t$-program)
with $n$ atoms and $\Omega(\mu_t^n)$ minimal models (stable models,
answer sets, respectively).
\item Every algorithm computing all minimal models of $t$-CNF theories
(stable models of $t${-}programs, answer sets of disjunctive
$t$-programs, respectively) requires in the worst case at least
$\Omega(n\mu_t^n)$ steps.
\item Let $0< \alpha < \mu_t$. For every polynomial $f$, there is no
algorithm for computing all minimal models of $t$-CNF theories
(stable models of $t${-}programs, answer sets of disjunctive
$t$-programs, respectively) with worst{-}case performance of
$O(f(m)\alpha^n)$.
\end{enumerate}
\end{corollary}
The lower bound given by Corollary \ref{clb}(1) specializes to
(approximately) $\Omega(1\mbox{.}4422\mbox{..}^n)$ and
$\Omega(1\mbox{.}5848\mbox{..}^n)$, for $t=2$ and $3$, respectively. 
Similarly, the
lower bound given by Corollary \ref{clb}(2) specializes to
(approximately) $\Omega(n1\mbox{.}4422\mbox{..}^n)$ and
$\Omega(n1\mbox{.}5848\mbox{..}^n)$, for $t=2$ and $3$, respectively.


\section{Discussion}
\label{disc}

The algorithms we presented in the case of 2-CNF theories, and normal and
disjunctive 2-programs have worst-case performance of 
$O(m1\mbox{.}4422\mbox{..}^n)$.
The algorithm we designed for the task of computing stable models of normal
3-programs runs in time $O(m1\mbox{.}6701\mbox{..}^n)$. Finally, our 
algorithms
for computing minimal models of 3-CNF theories and answer sets of
disjunctive logic programs run in time $O(mn^2 2\mbox{.}2782\mbox{..}^n)$. 
All these
bounds improve by exponential factors over the corresponding
straightforward ones.

The key question is whether still better algorithms are possible.
In this context, we note that our algorithms developed for the
case of 2-CNF theories and 2-programs are optimal, as long as we
are interested in {\em all} minimal models, stable models and
answer sets, respectively. However, we can compute a {\em single}
minimal model of a 2-CNF theory $T$ or decide that $T$ is
unsatisfiable in {\em polynomial} time. Indeed, it is well known
that we can compute a model $M$ of $T$ or decide that $T$ is
unsatisfiable in polynomial time. In the latter case, no minimal
models exist. In the former one, by the proof of Proposition
\ref{simp}, $M$ is a minimal model of $T$ if and only if theories
$T_L\cup\{\dua a\}$, where $L=\{\dua b\colon b \in \At(T)  - M\}$
and $a\in M$, are all unsatisfiable.
%
Thus, by means of polynomially many satisfiability checks we
either determine that $M$ is a minimal model of $T$ or find $a\in
M$ such that $M-\{a\}$ is a model of $T$. In contrast, deciding
whether a 2-program has a stable model and whether a disjunctive
2-program has an answer set is NP-complete. Thus, it is unlikely
that there are polynomial-time algorithms to compute a single
stable model (answer set) of a (disjunctive) 2-program or decide
that none exist. Whether our bound of
$O(m1\mbox{.}4422\mbox{..}^n)$ can be improved by an exponential
factor if we are interested in computing a single stable model or
a single answer set, rather than all of them, is an open problem.

The worst-case behavior of our algorithms designed for the case of
3-CNF theories and 3-programs does not match the lower bound of
$O(n1\mbox{.}5848\mbox{..}^n)$ implied by Corollary \ref{clb}. Thus,
there is still room for improvement, even when we want to compute {\em
all} minimal models, stable models and answer sets. In fact, we
conjecture that exponentially faster algorithms exist.

In the case of 3-CNF theories, reasoning similarly as in the case of
2-CNF theories, and using the proof of Proposition \ref{simp} and the
algorithm from \cite{dgh02}, shows that in time $O(p(m) 1\mbox{.}481^n)$,
where $p$ is a polynomial, one can compute {\em one} minimal model of
a 3-CNF theory $T$ or determine that $T$ is unsatisfiable.
This is a significantly better bound than $O(mn^22\mbox{.}2782\mbox{..}
^n)$ that we obtained for computing {\em all} minimal models. We do not
know however, whether the bound $O(p(m) 1\mbox{.}481^n)$ is optimal.
Furthermore, we do not know whether an exponential improvement over the
bound of $O(mn^22\mbox{.}2782\mbox{..}^n)$ is possible if we want to compute 
a single
answer set of a disjunctive
3-program or determine that none exists.
Similarly, we do not know whether one can compute a single stable
model of a 3-program or determine that none exists in time
exponentially lower than $O(m1\mbox{.}6701\mbox{..}^n)$.

In some cases, our bound in Theorem \ref{3p1d} can be improved.
Let $\cal F$ be the class of all CNF theories consisting of clauses
of the form $a_1\vee\ldots\vee a_p$ or $a\vee \dua b$, where $a_1,
\ldots, a_p$, $a$ and $b$ are atoms. Similarly, let $\cal G$ be the
class of all disjunctive programs with clauses of the form $a_1\vee
\ldots\vee a_p \lla \n(b_1),\ldots,\n(b_r)$ or $a\lla b,\n(b_1),\ldots,
\n(b_r)$, where $a_1,\ldots, a_p$, $b_1,\ldots, b_r$, $a$ and $b$ are
atoms. Checking whether a set $M$ is a minimal model of a theory from
${\cal F}$ or an answer set of a program from $\cal G$ is in the class
P (the task can be accomplished in linear time by extending the argument
we used to establish Proposition \ref{prop-min}). Thus, by Lemma
\ref{prop}, we obtain the following result.

\begin{theorem}
\label{cor8}
There is an algorithm to compute
minimal models of 3-CNF theories in $\cal F$ (answer sets of
disjunctive 3-programs in $\cal G$, respectively), that runs in
time $O(m 1\mbox{.}6701\mbox{..} ^n)$.
\end{theorem}

Finally, we stress that our intention in this work was to better
understand the complexity of problems to compute stable models of
programs, minimal models of CNF theories and answer sets of disjunctive
programs. Whether our theoretical results and algorithmic techniques
we developed here will have any significant practical implications is
a question for future research.

\section*{Acknowledgments}
\vspace*{-0.1in}

An extended abstract of this paper was published in Proceedings of the 
International Conference on Logic Programming, ICLP{-}2003. This research 
was supported by the National Science Foundation under Grants No. 0097278 
and IIS-0325063, and by the Polish Academy of Sciences and Warsaw 
University of Technology.


\newpage
\section*{Appendix}

We complete here the proof of Lemma \ref{mainth}, outlined in
Section \ref{cov}. We defined there a measure $\mu$ by setting
\[
\mu(S)=n(S)-\alpha k(S),
\]
where $n(S)$ is the number of atoms in $S$, $k(S)$ is the maximum number
of 2-clauses in $S$ with mutually disjoint sets of atoms, and $\alpha=
0\mbox{.}1950\mbox{..}\ $.

We introduced in Section \ref{cov} a simplifying function $\sigma$,
which eliminates from a 3-CNF theory $T$ repeated occurrences of the
equivalent clauses and also all 3-clauses subsumed by 2-clauses in $T$.
We showed there that $\sigma$ is $\mu$-compatible and that $\sigma(T)$
can be computed in linear time in the size of $T$.

To complete the proof, we still need to define a cover function $\rho$
postulated in Lemma \ref{mainth}. As we argued in Section \ref{cov}, it
is enough to
define $\rho(S)$ when $S$ is a 3-CNF theory with at least one atom and
such that $S$ contains no multiple occurrences of clauses nor 3-clauses
subsumed by a 2-clause in $S$. In addition, we will assume that no
clause in $S$ contains multiple occurrences of a literal or a pair
of dual literals (given a linked-list representation of $S$, this
condition can be enforced, if necessary, in linear time in the size of
$S$).

For each such theory $S$ we will define $\rho(S)$. To this end, we
consider 21 cases that we specified in Section \ref{cov}. When
discussing a case, we assume that none of the previously considered
cases applies.

In each case, we describe $\rho(S)$. Then, for each $A\in \rho(S)$,
we find a positive real number $k_{A,S}$ such that $\Delta(S,S_A)=
\mu(S)-\mu(S_A) \geq k_{A,S}$. To this end, we find lower bounds for
$n(S)-n(S_A)$ and $k(S_A)-k(S)$. Clearly, $|\At(A)| \leq n(S)-n(S_A)$
and we use $|\At(A)|$ to estimate $n(S)-n(S_A)$ from below. In each case
 we consider, the cardinality of $\At(A)$ is equal to the number of
literals we list as members of $A$. In Cases 2 - 5, we provide explicit
proofs of that claim. In all other cases arguments are similar and we
only outline them or omit them altogether. As concerns $k(S_A)-k(S)$, we
provide lower bounds and arguments to justify them in each case we
consider. For the most part, the proofs take advantage of the fact that
once we settle Case 4, we can assume that all 2-clauses in $S$ have
pairwise disjoint sets of atoms.

Establishing positive bounds $k_{A,S}$ for $\Delta(S,S_A)$ shows that
$\rho$ is $\mu$-compatible. It also yields a specific instance of the
equation (\ref{tau'}). Assuming $\alpha=0\mbox{.}1950\mbox{..}\ $, we find 
the root
$\tau_S'$ of the equation (\ref{tau'}). In each case, $\tau_S'\leq
1\mbox{.}6701\mbox{..}\ $. By Lemma \ref{thm-ineq}, that implies that $\rho$ 
is
$\mu$-bounded by $1\mbox{.}6701\mbox{..}\ $ and completes the proof of Lemma
\ref{mainth}.

In what follows, we consistently use $c$, possibly with indices,
to denote clauses. We use Greek alphabet letters to denote
literals and Latin alphabet letters to denote atoms. Throughout
the proof, we strictly adhere to the assumption that $b$, $d$,
$e$, $g$, $\ell$ and $o$ denote atoms appearing in literals
$\beta$, $\delta$, $\epsilon$, $\gamma$, $\lambda$ and $\omega$,
respectively, and we extend this notation to the case when we use
these letters together with indices. Finally, we consistently view
clauses as disjunctions of their literals.

We denote by $I$ a largest set of 2-clauses in $S$ with pairwise
disjoint sets of atoms.


\noindent
{\bf Case 1.} $|\At(S)|=1$, say $\At(S)=\{b\}$.

In this case, $S=\{\beta\}$ or $S=\{\beta,\dua \beta\}$. Clearly,
\[
\A=\{\{\beta\}\}
\]
is a cover for $S$ and we set $\rho(S)=\A$. It is easy to see that
$\At(S_A)=0$. Moreover, since neither $S$ nor $S_A$ contain 2-clauses,
$\Delta(S,S_A)=1$. The equation (\ref{tau'}) becomes
\[
\tau=1
\]
and its root, $\tau_S'$ satisfies $\tau_S' =1$.

\noindent
{\bf Case 2.} There is a 1-clause in $S$.

Let $\omega$ be the literal of a 1-clause in $S$. Since $|At(S)|\geq
2$,
there is an atom $y\in \At(S)$ such that $o\not=y$. It is evident
that
\[
{\A} =\{\{\omega,y\},\{\omega,\dua y\}\}
\]
is a cover for $S$ and we set $\rho(S)=\A$.

For every $A\in \A$, the theory $S_A$ contains all 2-clauses of $I$
that do not contain $o$ or $y$. Thus $k(S_A)\geq k(S)-2$. It follows that
for $A\in \A$,
\[
\Delta(S,S_A) \geq 2 - 2\alpha\mbox{.}
\]
Moreover, since $\alpha<1$, for every $A\in \A$, $\Delta(S,S_A) > 0$.
The equation (\ref{tau'}) becomes
\[
2\tau^{-2+2\alpha}=1\mbox{.}
\]
and $\tau_S'\leq 1\mbox{.}54$.

\noindent
{\bf Case 3.} There is an atom $a$ that appears negated in every clause
in $S$.

Clearly, $a$ does not belong to any minimal model of $S$ or, in
other words, every {\em minimal} model of $S$ is consistent with
$\{\dua a\}$. Let $y$ be an atom in $\At(S)$ such that $y\not=a$. It
follows that
\[
{\A} =\{\{\dua a,y\},\{\dua a,\dua y\}\}
\]
is a cover for $S$ and we set $\rho(S) =\A$. Reasoning as
Case 2, we obtain that for every $A\in \A$,
\[
\Delta(S,S_A) \geq 2 - 2\alpha > 0\mbox{.}
\]
Moreover, the equation (\ref{tau'}) becomes
\[
2\tau^{-2+2\alpha}=1
\]
and $\tau_S'\leq 1\mbox{.}54$.

\noindent
{\bf Case 4.} There are two 2-clauses in $S$, which contain a common
atom.

Because of the assumption we adopt for Case 4, there is a clause,
say $c_1$ such that $c_1\notin I$. Since $I$ is a largest set of
2-clauses in $S$ with pairwise disjoint sets of atoms, there is a
clause, say $c_2$, in $I$ such that $c_1$ and $c_2$ have a common
atom.

\noindent
{\bf Subcase (i).} $c_1=\dua\omega \vee \gamma$ and $c_2=\omega \vee
\beta$ (that is, the common atom appears in $c_1$ and $c_2$ ``in the
opposite ways'').

We note that every model consistent with $\omega$ is also consistent
with $\gamma$, as it satisfies the clause $c_1$. Similarly, every model
consistent with $\dua\omega$ is also consistent with $\beta$. Thus,
\[
\A =\{\{\omega, \gamma\},\{\dua\omega, \beta\}\}\mbox{.}
\]
is a cover for $S$ and we define $\rho(S)={\A}$.

Clearly, $|\At(\{\omega, \gamma\})|=2$ (otherwise, the 2-clause $c_1$
would contain a multiple occurrence of an atom or a pair of dual
literals). Similarly, $|\At(\{\dua\omega, \beta\})|=2$, as well.

The theory $S_{\{\omega,\gamma\}}$ contains all 2-clauses of $I$ except
for $c_2$ and, possibly, a clause in $I$ containing the atom of $\gamma$.
Thus, $k(S_{\{\omega,\gamma\}})\geq k(S)-2$. The theory $S_{\{
\dua\omega,\beta\}}$, on the other hand, contains all 2-clauses of $I$
except for $c_2$. Hence, $k(S_{\{\dua\omega,\beta\}})\geq k(S)-1$. It
follows that
\[
\Delta(S,S_A)\geq \left\{
            \begin{array}{ll}
              2-2\alpha & \mbox{if $A=\{\omega,\gamma\}$}\\
              2-\alpha & \mbox{if $A=\{\dua\omega, \beta\}\mbox{.}$}
            \end{array}
         \right.
\]
Since $\alpha<1$, $2-2\alpha>0$ and $2-\alpha>0$. Thus, for every $A\in
{\A}$, $\Delta(S,S_A)>0$. Moreover, with $2-2\alpha$ and $2-\alpha$
as numbers $k_{A,S}$, $A\in \A$, the equation (\ref{tau'}) becomes
\[
\tau^{-2+2\alpha}+\tau^{-2+\alpha}=1\mbox{.}
\]
For $\alpha=0\mbox{.}1950\mbox{..}\ $ its root, $\tau_S'$, satisfies 
$\tau_S'\leq 1\mbox{.}51$.

\noindent
{\bf Comment.} From now on we will not explicitly state the numbers
$k_{A,S}$. We will specify them implicitly in inequalities bounding
$\Delta(S,S_A)$ from below. In each case, it will be straightforward to
see that the numbers are positive, due to the fact that $\alpha<1$.
In each case, we will present an instance of the equation (\ref{tau'}),
implied
by the bounds established in the case, as well as the root of the
equation, computed under the assumption that $\alpha=0\mbox{.}1950\mbox{..}\ 
$.

\noindent
{\bf Subcase (ii).} $c_1=\omega \vee \beta$, $c_2=\omega \vee \gamma$
(that is, the common atom to $c_1$ and $c_2$ appears in $c_1$ and
$c_2$ ``in the same way'').

Every model consistent with $\dua\omega$ is consistent with $\beta$
and $\gamma$. Thus, the following family
\[
{\A} =\{\{\omega\}, \{\dua\omega, \beta, \gamma\}\}
\]
is a cover for $S$. We use it as the value of $\rho(S)$.

Since $c_1$ and $c_2$ are 2-clauses of $S$, $o\not= b, g$. Moreover, as
$c_1\notin I$ and $c_2\in I$, $c_1\not= c_2$
and, consequently, $\beta\not= \gamma$. We can also assume that $\beta
\not=\dua\gamma$ (if that was not the case,
Subcase (i) would  apply). Thus, $b\not=g$ and $|\At(\{\dua
\omega, \beta, \gamma\})|=3$.

The theory $S_{\{\omega\}}$ contains all 2-clauses of $I$ except for
$c_2$. Thus $k(S_{\{\omega\}})\geq k(S)-1$. The theory $S_{\{\dua\omega,
\beta,\gamma\}}$ contains all 2-clauses of $I-\{c_2\}$ that do not
contain $g$ (the atom of $\gamma$). Thus
$k(S_{\{\dua\omega,\beta,\gamma\}})\geq k(S)-2$. It follows that
for every $A\in \A$, we have
\[
\Delta(S,S_A)\geq
\left\{
     \begin{array}{ll}
    1-\alpha & \mbox{if $A=\{\omega\}$}\\
    3-2\alpha & \mbox{if $A=\{\dua\omega, \beta, \gamma\}\mbox{.}$}
     \end{array}
\right.
\]
The equation (\ref{tau'}) becomes
\[
\tau^{-1+\alpha}+\tau^{-3+2\alpha}=1
\]
and $\tau_S' \leq 1\mbox{.}58\mbox{..}\ $ (we recall that we take 
$\alpha=0\mbox{.}1950\mbox{..}\ $).

\noindent
{\bf Comment.}
There are no other possibilities in Case 4. From now on we will assume
that the set of {\em all} 2-clauses in $S$ consists of clauses which do
not have common atoms. Thus, $k(S)$ is simply equal to the number of
{\em all} 2-clauses in $S$. Moreover, when simplifying $S$ with respect
to a literal $\omega$ (removing clauses subsumed by $\omega$ and
eliminating $\dua\omega$ from other clauses of $S$ as part of the
computation of $S_A$), at most one 2-clause will be eliminated in the
process and $k(S)$ will decrease at most by 1. Throughout the proof, we
denote that set of 2-clauses of $S$ by $I$.

\noindent
{\bf Case 5.} There are clauses $c_1=\omega\vee \beta\vee\gamma$ and
$c_2=\dua\beta\vee\delta$ in $S$ such that $d\notin\At(c_1)$.

\noindent
{\bf Subcase (i).} Neither $o$ nor $g$ is an atom of a
2-clause in $S$.

The family
\[
\A =\{\{\beta,\delta\}, \{\dua\beta\}\}
\]
is a cover for $S$ (indeed, if a model of $S$ is not consistent with
$\dua\beta$, it is consistent with $\beta$ and, due to clause $c_2$,
with $\delta$). We define $\rho(S)=\A$.

Since $\dua \beta \vee \delta$ is a clause in $S$, $b\not=d$ and,
consequently, $|\At(\{\beta,\delta\})|=2$. Moreover, all 2-clauses
in $I  -\{c_2\}$ are 2-clauses of $S_{\{\beta,\delta\}}$ and
$S_{\{\dua\beta\}}$. Thus, $k(S_{\{\beta,\delta\}})\geq k(S)-1$. We also
have that $c_3=\omega\vee \gamma$ is a 2-clause in $S_{\{\dua\beta\}}$.
Since $S$ contains no 3-clauses subsumed by 2-clauses in $S$, $c_3\notin
S$. By the assumption adopted for this subcase, no atom of $c_3$ belongs
to any 2-clause in $I-\{c_2\}$. Thus, $k(S_{\{\dua\beta\}})\geq k(S)$
and, for every $A\in \A$,
\[
\Delta(S,S_A)\geq \left\{
            \begin{array}{ll}
              2-\alpha & \mbox{if $A=\{\beta,\delta\}$}\\
              1 & \mbox{if $A=\{\dua\beta\}\mbox{.}$}
            \end{array}
         \right.
\]
The equation (\ref{tau'}) becomes
\[
\tau^{-2+\alpha}+\tau=1
\]
and $\tau_S' \leq 1\mbox{.}67$.

\noindent
{\bf Subcase (ii).} There is a 2-clause $c_3\in I$ of the form
$c_3=\dua\omega\vee\epsilon$ or $c_3=\dua\gamma\vee\epsilon$. We will
assume $c_3=\dua\omega\vee\epsilon$. The other case is symmetric.

Every model of $S$ consistent with $\beta$ is consistent with
$\delta$ (clause $c_2$). Every model consistent with
$\{\dua\beta,\omega\}$ is consistent with $\epsilon$ (clause
$c_3$). Finally, every model consistent with
$\{\dua\beta,\dua\omega\}$ is consistent with $\gamma$ (clause
$c_1$). Therefore, the family
\[
\A =\{\{\beta,\delta\},\{\dua\beta,\omega,\epsilon\},
\{\dua\beta,\dua\omega,\gamma\}\}
\]
is a cover for $S$. We define $\rho(S)=\A$.

Since $\dua \beta \vee \delta$ is a clause in $S$, $b\not=d$. Thus,
$|\At(\{\beta,\delta\})|=2$. Similarly, since $c_1$ is a 3-clause in
$S$, the atoms $b$, $o$ and $g$ are pairwise
distinct and $\At(\{\dua\beta,\dua\omega,\gamma\})|=3$. Since
$d\notin\At(c_1)$, $d\not=o$. We already noted that $b\not=o$. Thus,
$c_2\not=c_3$. Consequently, $\At(c_2)\cap \At(c_3)=\emptyset$ (by the
assumption made after Case 4) and $e\not=b$. Finally, $o\not=e$, as $o$
and $e$ appear together in $c_3$. Thus, $|\At(\{\dua\beta,\omega,
\epsilon\})|=3$.

All 2-clauses of $S$ except for $c_2$ are 2-clauses in $S_{\{\beta,
\delta\}}$ and it follows that $k(S_{\{\beta,\delta\}})\geq k(S)-1$.
All 2-clauses of $S$ except for $c_2$ and $c_3$ are 2-clauses in
$S_{\{\dua\beta,\omega,\epsilon\}}$. Hence, $k(S_{\{\dua\beta,\omega,
\epsilon\}})\geq k(S)-2$. Finally, all 2-clauses in $I-\{c_2,c_3\}$
that do not contain $g$ are 2-clauses in $S_{\{\dua\beta,\dua\omega,
\gamma\}}$ and, consequently, $k(S_{\{\dua\beta,\dua\omega,\gamma\}})
\geq k(S)-3$. Hence, for every $A\in \A$,
\[
\Delta(S,S_A)\geq \left\{
           \begin{array}{ll}
         2-\alpha & \mbox{if $A=\{\beta,\delta\}$}\\
         3-2\alpha & \mbox{if $A=\{\dua\beta,\omega,\epsilon\}$}\\
         3-3\alpha & \mbox{if $A=\{\dua\beta,\dua\omega,\gamma\}\mbox{.}$}
       \end{array} \right.
\]
These bounds yield the following instance of the equation (\ref{tau'}):
\[
\tau^{-2+\alpha}+\tau^{-3+2\alpha}+\tau^{-3+3\alpha}=1\mbox{.}
\]
Its root $\tau_S'$ satisfies $\tau_S'\leq 1\mbox{.}64$.

\noindent
{\bf Subcase (iii).} There is a 2-clause $c_3\in I$ of the form
$c_3=\omega\vee\epsilon$ and $g$ does not belong to any 2-clause
in $I$ (or the symmetric situation with the roles of $\omega$ and
$\gamma$ interchanged).

Every model of $S$ consistent with $\beta$ is consistent with $\delta$
(due to clause $c_2$). Moreover, every model consistent with $\{\dua
\beta,\dua\omega\}$ is consistent with $\{\gamma,\epsilon\}$
(clauses $c_1$ and $c_3$). Therefore, the family
\[
\A =\{\{\beta,\delta\}, \{\dua\beta,\omega\}, \{\dua\beta,\dua\omega,
\gamma,\epsilon\}\}
\]
is a cover for $S$ and we take it as the value of $\rho(S)$.

Arguing as in the previous subcase, we show that $e\not=
o,b$. Moreover, we observe that $e\not=g$,
as we assumed that $g$ does not belong to any 2-clause in $I$.
Thus, the cardinality of the last set in $\A$ is 4 (it is easy to see
that each of the other two sets has 2 elements).

All 2-clauses of $S$ except for $c_2$ are still 2-clauses in
$S_{\{\beta,\delta\}}$. Thus, $k(S_{\{\beta,\delta\}})\geq k(S)-1$.
Since the atom of $\gamma$ does not belong to any 2-clause in $I$, all
2-clauses of $S$ except for $c_2$ and $c_3$ are 2-clauses in
$S_{\{\dua\beta,\omega\}}$ and in $S_{\{\dua\beta,\dua\omega,\gamma,
\epsilon\}}$. Thus, $k(S_{\{\dua\beta,\omega\}})\geq k(S)-2$ and
$k(S_{\{\dua\beta,\dua\omega,\gamma,\epsilon\}})\geq k(S)-2$. Hence,
for every $A\in \A$,
\[
\Delta(S,S_A)\geq \left\{
            \begin{array}{ll}
              2-\alpha & \mbox{if $A=\{\beta,\delta\}$}\\
              2-2\alpha & \mbox{if $A=\{\dua\beta,\omega\}$}\\
              4-2\alpha & \mbox{if $A=\{\dua\beta,\dua\omega,\gamma,
                    \epsilon\}\mbox{.}$}
            \end{array}
         \right.
\]
The equation (\ref{tau'}) becomes
\[
\tau^{-2+\alpha}+\tau^{-2+2\alpha}+\tau^{-4+2\alpha}=1,
\]
and $\tau_S'\leq 1\mbox{.}67$.

\noindent
{\bf Subcase (iv).} There are 2-clauses $c_3,c_4\in I$ of the form
$c_3=\gamma\vee\epsilon$ and $c_4=\omega\vee\lambda$.

Every model consistent with $\{\dua\omega,\gamma\}$ is consistent with
$\lambda$ (clause $c_4$). Moreover, every model consistent with $\{\dua
\omega,\dua\gamma\}$ is consistent with $\{\lambda,\epsilon,\beta,
\delta\}$ (clauses $c_4$, $c_3$, $c_1$ and $c_2$). Therefore the family
\[
\A =\{\{\omega\},\{\dua\omega,\gamma,\lambda\}, \{\dua\omega,
\dua\gamma,\lambda,\epsilon,\beta,\delta\}\}
\]
is a cover for $S$ and we choose it to define $\rho(S)$.

Since $d\notin\At(c_1)$, $d\not=o,g$. Since $b, o$ and $g$ are the atoms
of the 3-clause $c_1$, $b\not=o,g$. Thus, $c_2\not=c_3$ and
$c_2\not=c_4$.
Let us assume that $c_3=c_4$. Since $\gamma\not=\omega$, it follows that
$\epsilon=\omega$. Consequently, $c_3$ subsumes $c_1$, a contradiction.
Thus, $c_3\not=c_4$ and so, $c_2,c_3$ and $c_4$ are pairwise different.
By Case 4, it follows that $c_2,c_3$ and $c_4$ have pairwise disjoint
sets of atoms. Consequently, $|\At(\{\dua\omega,\gamma,\lambda \})|=3$
and $|\At(\{\dua \omega, \dua\gamma,\lambda,\epsilon,\beta,\delta\}\})|
=6$.

All 2-clauses of $S$ except for $c_4$ are 2-clauses in $S_{\{\omega\}}$.
Thus, $k(S_{\{\omega\}})\geq k(S)-1$. All 2-clauses of $S$ except for
$c_3$ and $c_4$ are 2-clauses in $S_{\{\dua\omega,\gamma,\lambda\}}$.
Consequently, $k(S_{\{\dua\omega,\gamma,\lambda\}})\geq k(S)-2$. Finally,
all 2-clauses of $S$ except for $c_2$, $c_3$ and $c_4$ are 2-clauses in
$S_{\{\dua\omega,\dua\gamma,\lambda,\epsilon,\beta,\delta\}}$ and so
$k(S_{\{\dua\omega,\dua\gamma,\lambda,\epsilon,\beta,\delta\}})\geq
k(S)-3$. Hence, for every  $A\in \A$,
\[
\Delta(S,S_A)\geq \left\{
            \begin{array}{ll}
              1-\alpha & \mbox{if $A=\{\omega\}$}\\
              3-2\alpha & \mbox{if $A=\{\dua\omega,\gamma,\lambda\}$}\\
              6-3\alpha & \mbox{if $A=\{\dua\omega,\dua\gamma,\lambda,
                    \epsilon,\beta,\delta\}\mbox{.}$}
            \end{array}
         \right.
\]
For these bounds, the equation (\ref{tau'}) becomes
\[
\tau^{-1+\alpha}+\tau^{-3+2\alpha}+\tau^{-6+3\alpha}=1,
\]
and $\tau_S'\leq 1\mbox{.}66$ (assuming $\alpha=0\mbox{.}1950\mbox{..}\ $).

\noindent
{\bf Case 6.} There are clauses $c_1,c_2\in S$ such that $c_1$ is a
2-clause, $c_2$ is a 3-clause and $\At(c_1) \subseteq\At(c_2)$.

Since $c_1$ does not subsume $c_2$, we can assume that
$c_1=\omega\vee\gamma$ and $c_2=\omega'\vee\dua\gamma\vee\beta$,
where ${\omega}'=\omega$ or ${\omega}'=\dua\omega$.

\noindent
{\bf Subcase (i).} The atom $b$ does not occur in any 2-clause.

Every model consistent with $\dua\gamma$ is consistent with $\omega$
(clause $c_1$). Therefore the family
\[
\A =\{\{\dua\gamma,\omega\},\{\gamma\}\}
\]
is a cover for $S$ and we set $\rho(S)=\A$.

All 2-clauses of $S$ except for $c_1$ are 2-clauses in the theories
$S_{\{\dua\gamma, \omega\}}$ and $S_{\{\gamma\}}$. It follows that
$k(S_{\{\dua\gamma,\omega\}})\geq k(S)-1$. Moreover, $c_3=\omega'
\vee\beta$ is a 2-clause in $S_{\{\gamma\}}$. Since $o$
appears in a 2-clause $c_1$, $o$ does not appear in any other
2-clause in $S$. Thus, $c_3$ is different from all 2-clauses in
$S_{\{\gamma\}}$ that belong to $I  -\{c_1\}$. By the assumption
we adopted in this subcase, $b$ does not belong to any 2-clause
in $I  -\{c_1\}$ either. Thus, $k(S_{\{\gamma\}})\geq k(S)$. It
follows that for every $A\in \A$,
\[
\Delta(S,S_A)\geq \left\{
         \begin{array}{ll}
      2-\alpha & \mbox{if $A=\{\dua\gamma,\omega\}$}\\
          1 & \mbox{if $A=\{\gamma\}\mbox{.}$}
         \end{array}
    \right.
\]
In this case, we obtain the following instance of (\ref{tau'}):
\[
\tau^{-2+\alpha}+\tau=1\mbox{.}
\]
Its root $\tau_S'$ satisfies $\tau_S'\leq 1\mbox{.}67$.

\noindent
{\bf Subcase (ii).} The atom $b$ belongs to some 2-clause in $S$.

Let $c_3$ be a 2-clause in $S$ that contains $b$. Then $c_3=\dua
\beta\vee\epsilon$ or $c_3=\beta\vee\epsilon$, for some literal
$\epsilon$, where $e\not=b$. We note that $e\not=o,g$ (as all
2-clauses in $S$ have pairwise disjoint
sets of atoms). Thus, if $c_3 =\dua\beta\vee\epsilon$, Case 5 would
apply to $c_3$ and $c_2$, a contradiction. It follows that $c_3=\beta
\vee\epsilon$.

Every model consistent with $\{\dua\beta,\dua\gamma\}$ is consistent
with $\{\epsilon,\omega\}$ (clauses $c_3$ and $c_1$), and every model
consistent with $\{\dua\beta,\gamma\}$ is consistent with $\{\epsilon,
{\omega}'\}$ (clauses $c_3$ and $c_2$). Therefore the family
\[
\A =\{\{\beta\},\{\dua\beta,\dua\gamma,\epsilon,\omega\},
\mbox{$\{\dua\beta,\gamma,\epsilon,{\omega'}\}$}\}
\]
is a cover for $S$ and we use it as the value of $\rho(S)$ in this
case.

Every 2-clause of $I  -\{c_3\}$ is a 2-clause of $S_{\{\beta\}}$.
Thus, $k(S_{\{\beta\}})\geq k(S)-1$. Moreover, every 2-clause of
$I  - \{c_1,c_3\}$ is a 2-clause in both $S_{\{\dua\beta,\dua
\gamma,\epsilon,\omega\}}$ and $S_{\{\dua\beta,\gamma,\epsilon,
{\omega}'\}}$. Hence, $k(S_{\{\dua\beta,\dua\gamma,\epsilon,\omega\}})
\geq k(S)-2$ and $k(S_{\{\dua\beta,\gamma,\epsilon,{\omega'}\}})\geq
k(S)-2$. It follows that for every $A\in \A$,
\[
\Delta(S,S_A)\geq \left\{
            \begin{array}{ll}
              1-\alpha & \mbox{if $A=\{\beta\}$}\\
              4-2\alpha & \mbox{if $A=\{\dua\beta,\dua\gamma,\epsilon,
         \omega\},\{\dua\beta,\gamma,\epsilon,{\omega'}\}\mbox{.}$}
            \end{array}
         \right.
\]
The equation (\ref{tau'}) becomes
\[
\tau^{-1+\alpha}+2\tau^{-4+2\alpha}=1\mbox{.}
\]
Its root is $\tau_S'\leq 1\mbox{.}65$.

\noindent
{\bf Comment.} Let $\beta\vee \omega$ be a 2-clause in $S$. Let $c$ be a
clause in $S$ such that $b\in \At(c)$. If $c$ is a 1-clause, Case 2
applies. If $c$ is a 2-clause, Case 4 applies. Thus, we can assume that
$c$ is a 3-clause. If $\dua\beta$ is a literal of $c$ and $o\notin
\At(c)$, then Case 5 applies. If $o\in \At(c)$, then Case 6 applies.
Thus, we can assume that every clause $c\in S$ such that $b\in\At(c)$
contains $\beta$ as its literal. If $\beta=\dua b$, Case 3 applies. Thus,
from now on we can assume that atoms of 2-clauses of $S$ have only
positive occurrences in $S$.

\noindent
{\bf Case 7.} There is an atom $a$ that occurs in a 2-clause of $S$,
say $c_1=a \vee b$, and in no other clause of $S$.

Every {\em minimal} model consistent with $a$ is consistent with $\dua b$.
Indeed, if a model $M$ of $S$ contains both $a$ and $b$ then $M  -
\{a\}$ is a model of $S$, too, as $c_1$ is the only clause in $S$ that
contains $a$. Moreover, every model consistent with $\dua a$ is
consistent with $b$ (clause $c_1$). Hence, the family
\[
{\A} =\{\{a,\dua b\},\{\dua a,b\}\}
\]
is a cover for $S$ and we define $\rho(S)=\A$.

Clearly, 2-clauses in $I  -\{c_1\}$ are 2-clauses in both
$S_{\{a,\dua b\}}$ and $S_{\{\dua a,b\}}$. Thus, $k(S_{\{a,\dua b\}})
\geq k(S)-1$ and $k(S_{\{\dua a,b\}})\geq k(S)-1$. Hence, for every $A
\in {\A}$,
\[
\Delta(S,S_A)\geq 2-\alpha > 0\mbox{.}
\]
The equation (\ref{tau'}) becomes:
\[
2\tau^{-2+\alpha}=1
\]
and we have $\tau_S'\leq 1\mbox{.}47$.

\noindent
{\bf Case 8.} There is an atom $a$ that occurs in a 2-clause of $S$,
say $c_1=a\vee b$, and there is exactly one other clause in $S$, say
$c_2$, such that $a\in \At(c_2)$.

Since 2-clauses in $S$ do not have atoms in common, $c_2$ is a 3-clause.
We will assume that $c_2= a\vee\gamma\vee\delta$.

Every model consistent with $\dua a$ is consistent with $b$
(clause $c_1$). Every {\em minimal} model consistent with $\{ a,b\}$ is
consistent with $\{\dua\gamma,\dua\delta\}$. Indeed, if a model $M$ is
consistent with $\{ a,b,\gamma\}$ or $\{ a,b,\delta\}$ then $M  -
\{a\}$ is a model of $S$, too, (as $a$ belongs to two clauses $c_1$ and
$c_2$ only). Hence, the family
\[
{\A} =\{\{\dua a,b\},\{a,\dua b\},\{a,b,\gamma,\delta\}\}
\]
is a cover for $S$ and we take it as $\rho(S)$ in this case.

All 2-clauses in $I  -\{c_1\}$ are 2-clauses in both $S_{\{\dua a,
b\}}$ and $S_{\{a,\dua b\}}$, which implies $k(S_{\{\dua a,b\}})\geq
k(S)-1$ and $k(S_{\{a,\dua b\}})\geq k(S)-1$. Moreover, all 2-clauses
in $I$ except for $c_1$ and 2-clauses in $I$ that contain $g$
and $d$ (there are at most two such 2-clauses) are 2-clauses in
$S_{\{a,b,\gamma,\delta\}}$. Thus, $k(S_{\{a,b,\gamma,\delta\}})\geq
k(S)-3$. Consequently,
\[
\Delta(S,S_A)\geq \left\{
            \begin{array}{ll}
              2-\alpha & \mbox{if $A=\{\dua a,b\},\{a,\dua b\}$}\\
              4-3\alpha & \mbox{if $A=\{a,b,\gamma,\delta\}\mbox{.}$}
            \end{array}
         \right.
\]
The equation (\ref{tau'}) becomes
\[
2\tau^{-2+\alpha}+\tau^{-4+3\alpha}=1
\]
and $\tau_S'\leq 1\mbox{.}65$.

\noindent
{\bf Case 9.} For some atom $a$ that occurs in a 2-clause, say
$c_1$, there are two other clauses $c_2$ and $c_3$ such that $\At(c_2)
\cap\At(c_3)=\{a\}$.

Since 2-clauses in $S$ do not have atoms in common, $c_2$ and $c_3$ are
3-clauses. Throughout Case 9, we assume that $c_1=a\vee b$, $c_2=a\vee
\gamma\vee \delta$ and $c_3=a\vee\epsilon\vee\lambda$.

We have $b\notin\At(c_i)$, $i=2,3$, as otherwise Case 6 would apply.
Thus, since $\At(c_2)\cap\At(c_3)=\{a\}$, it follows that all atoms $a$,
$b$, $g$, $d$, $e$ and $\ell$ are pairwise different.

\noindent
{\bf Subcase (i).} The atoms $g$, $d$, $e$ and $\ell$ do not belong to
any 2-clause in $S$.

Clearly, the family
\[
{\A}=\{\{ a\},\{\dua a,b\}\}
\]
is a cover for $S$ and we set $\rho(S)=\A$.

All 2-clauses in $I  -\{c_1\}$ are 2-clauses in $S_{\{a\}}$ and
$S_{\{\dua a,b\}}$. It follows that $k(S_{\{a\}})\geq k(S)-1$. Moreover,
the clauses $c_4=\gamma\vee \delta$ and $c_5=\epsilon\vee \lambda$ are
2-clauses in $S_{\{\dua a,b\}}$. Since no 3-clause in $S$ is subsumed
by a 2-clause in $S$, $c_4,c_5 \notin S$. By the assumption we adopted
for the current subcase, the atoms of $c_4$ and $c_5$ do not appear in
any 2-clause of $I-\{c_1\}$. Thus, $k(S_{\{\dua a,b\}})\geq k(S)+1$.
It follows that for every $A\in \A$,
\[
\Delta(S,S_A)\geq \left\{
            \begin{array}{ll}
              1-\alpha & \mbox{if $A=\{ a\}$}\\
              2+\alpha & \mbox{if $A=\{\dua a,b\}\mbox{.}$}
            \end{array}
         \right.
\]
The corresponding instance of the equation (\ref{tau'}) is
\[
\tau^{-1+\alpha}+\tau^{-2-\alpha}=1
\]
and $\tau_S'\leq 1\mbox{.}66$, for $\alpha=0\mbox{.}1950\mbox{..}\ $.

\noindent
{\bf Subcase (ii).} The atoms $g$ and $d$ or the atoms $e$ and $\ell$
do not belong to 2-clauses in $S$.

We will assume that $e$ and $\ell$ do not belong to 2-clauses in $S$
(the other case is symmetric). Furthermore, we can assume that $g$ or
$d$, say $g$, is an atom of a 2-clause (otherwise, Case 9(i) would
apply). Let $c_4=g\vee h$ be a 2-clause in $S$ containing $g$ (we note
that the other literal in the clause must be an atom). We can assume
that $d\not=h$ as, otherwise, $\At(c_4)\subseteq\At(c_2)$ and Case 6
would apply. Finally, we note that $c_2=a\vee g\vee\delta$ (since there
is a 2-clause in $S$ containing $g$, $g$ appears positively in every
clause in $S$).

Every model of $S$ consistent with $\dua a$ is consistent with $b$
(clause $c_1$) and every model consistent with $\{\dua a,\dua g\}$
is consistent with $\{b,h,\delta\}$ (clauses $c_1$, $c_4$ and $c_2$).
Hence, the family
\[
{\A} =\{\{a\},\{\dua a,g,b\},\{\dua a,\dua g,b,h,\delta\}\}
\]
is a cover for $S$ and we set $\rho(S)=\A$.

All 2-clauses of $I  -\{c_1\}$ are 2-clauses in $S_{\{a\}}$ and
so, $k(S_{\{a\}})\geq k(S)-1$. Similarly, all 2-clauses of $I  -
\{c_1,c_4\}$ are 2-clauses of $S_{\{\dua a,g,b\}}$. In addition,
$c_5= \epsilon\vee\lambda$ is a 2-clause in $S_{\{\dua a,g,b\}}$.
Reasoning as before we see that $c_5$ is not a 2-clause of $S$.
Moreover, by the assumption we adopted earlier in this subcase,
its atoms do not occur in 2-clauses of $I-\{c_1,c_4\}$. Thus,
$k(S_{\{\dua a,g,b\}}) \geq k(S)-1$. Finally, every clause in $I  -
\{c_1,c_4\}$ that does not contain $d$ is a 2-clause of $S_{\{\dua a,
\dua g,b,h,\delta\}}$. In
addition, the clause $c_5= \epsilon\vee\lambda$ is a 2-clause in
$S_{\{\dua a,\dua g,b,h,\delta\}}$. As before, $c_5$ is not a 2-clause
of $S$ and it does not have atoms in common with other 2-clauses of
$S_{\{\dua a,\dua g,b,h,\delta \}}$. Thus, $k(S_{\{\dua a,\dua g,b,h,
\delta\}})\geq k(S)-2$. Hence,
\[
\Delta(S,S_A)\geq \left\{
            \begin{array}{ll}
              1-\alpha & \mbox{if $A=\{ a\}$}\\
              3-\alpha & \mbox{if $A=\{\dua a,g,b\}$}\\
              5-2\alpha & \mbox{if $A=\{\dua a,\dua g,b,h,\delta\}\mbox{.}$}
            \end{array}
         \right.
\]
The bounds listed above are positive. The equation (\ref{tau'}) becomes
\[
\tau^{-1+\alpha}+\tau^{-3+\alpha}+\tau^{-5+2\alpha}=1
\]
and for $\alpha=0\mbox{.}1950\mbox{..}\ $, $\tau_S'\leq 1\mbox{.}67$.

\noindent
{\bf Subcase (iii).} At least one of the atoms $g$ and $d$ belongs
to a 2-clause, say $c_4$, such that $\At(c_4)\cap \At(c_3)=\emptyset$,
and at least one of the atoms $e$ and $\ell$  belongs to a 2-clause, say
$c_5$, such that $\At(c_5)\cap \At(c_2)=\emptyset$.

Without loss of generality, we will assume that $g$ and $e$ have
the postulated property, and that $c_4=g\vee h$ and $c_5=e\vee f$, for
some atoms $h$ and $f$. We can assume that $d\not=h$ and $\ell\not=f$
(otherwise, Case 6 would apply). Since $h\notin \At(c_3)$ and $f\notin
\At(c_2)$, the atoms $a,b,g,h,e,f, d$ and $\ell$ are pairwise different.

Let $M$ be a model of $S$. If $M$ is consistent with $\{\dua a,g,e\}$,
then $M$ is consistent with $b$ (clause $c_1$). If $M$ is consistent
with $\{\dua a,g,\dua e\}$ then $M$ is consistent with $\{b,f,\lambda\}$
(clauses $c_1$, $c_5$ and $c_3$). If $M$ is consistent with $\{\dua a,
\dua g,e\}$ then it is consistent with $\{b,h,\delta\}$ (clauses $c_1$,
$c_4$ and $c_2$). Finally, if $M$ is consistent with $\{\dua a,\dua g,
\dua e\}$, then $M$ is consistent with $\{b,h,f,\delta,\lambda\}$
(clauses $c_1$, $c_4$, $c_5$, $c_2$ and $c_3$). Hence, the family
\[
{\A} =\{\{a\},\{\dua a,g,e,b\},\{\dua a,g,\dua e,b,f,\lambda\},
\{\dua a,\dua g,e,b,h,\delta\},\{\dua a,\dua g,\dua e,b,h,f,\delta,
\lambda\}\}
\]
is a cover for $S$ and we define $\rho(S)=\A$.

All 2-clauses in $I  -\{c_1\}$ are 2-clauses in $S_{\{a\}}$ and so,
$k(S_{\{a\}})\geq k(S)-1$. All 2-clauses in $I  -\{c_1,c_4,c_5\}$
are 2-clauses in $S_{\{\dua a,g,e,b\}}$ and so, $k(S_{\{\dua a,g,e,b\}})
\geq k(S)-3$. Similarly, 2-clauses in $I  -\{c_1,c_4,c_5\}$ that do not
contain $\ell$ (that condition excludes at most one such 2-clause)
are 2-clauses in $S_{\{\dua a,g,\dua e,b,f,\lambda\}}$. Thus, $k(S_{
\{\dua a,g,\dua e,b,f,\lambda\}})\geq k(S)-4$. All 2-clauses of $I
  -\{c_1,c_4,c_5\}$ that do not contain $d$ are 2-clauses in $S_{\{\dua
a,\dua g,e,b,h,\delta\}}$. Thus, $k(S_{\{\dua a,\dua g,e,b ,h,\delta
\}})\geq k(S)-4$. Finally, all 2-clauses of $I  -\{c_1, c_4,c_5\}$
that do not contain $d$ and $\ell$ (that condition excludes at most
two 2-clauses) are 2-clauses in $S_{\{\dua a,\dua g,\dua e,b, h,f,
\delta,\lambda\}}$. Therefore, $k(S_{\{\dua a,\dua g,\dua e,b,h,f,
\delta,\lambda\}})\geq k(S)-5$. Hence,
\[
\Delta(S,S_A)\geq \left\{
            \begin{array}{ll}
              1-\alpha & \mbox{if $A=\{ a\}$}\\
              4-3\alpha & \mbox{if $A=\{\dua a,g,e,b\}$}\\
              6-4\alpha & \mbox{if $A=\{\dua a,g,\dua e,b,f,\lambda\},
                \{\dua a,\dua g,e,b,h,\delta\}$}\\
              8-5\alpha & \mbox{if $A=\{\dua a,\dua g,\dua e,b,h,f,
                \delta,\lambda\}\mbox{.}$}
            \end{array}
         \right.
\]
The equation (\ref{tau'}) becomes
\[
\tau^{-1+\alpha}+\tau^{-4+3\alpha}+2\tau^{-6+4\alpha}+\tau^{-8+5\alpha}=1,
\]
and $\tau_S'=1\mbox{.}6701\mbox{..}\ $.

\noindent
{\bf Comment.}
We can assume now that there is a 2-clause, say $c_4$, in $S$ such that
$|\At(c_4)\cap\{g,d\}|=1$ and $|\At(c_4)\cap\{e,\ell\}|=1$. Otherwise,
one of the subcases (i)-(iii) would apply. Without loss of generality,
we will assume that $c_4=g\vee e$.

\noindent
{\bf Subcase (iv).} The atoms $d$ and $\ell$ do not belong to any
2-clauses in $S$.

Every model of $S$ consistent with $\{\dua a,g\}$ is consistent with
$b$ (clause $c_1$). Moreover, every model of $S$ consistent with $\{\dua
a,\dua g\}$ is consistent with $\{ b,\delta,e\}$ (clauses $c_1$, $c_2$
and $c_4$). Hence, the family
\[
{\A} =\{\{a\},\{\dua a,g,b\},\{\dua a,\dua g,b,\delta,e\}\}
\]
is a cover for $S$ and we set $\rho(S)=\A$.

All 2-clauses in $I  -\{c_1\}$ are 2-clauses in $S_{\{a\}}$. Thus,
$k(S_{\{a\}})\geq k(S)-1$. All 2-clauses in $I  -\{c_1,c_4\}$ are
2-clauses in $S_{\{\dua a,g,b\}}$. Moreover, $c_5=e\vee\lambda$ is also
a 2-clause in $S_{\{\dua a,g,b\}}$. Since $\ell$ does not belong to
any 2-clause in $S$, $c_5$ is not a clause of $S$ and has no atoms in
common with any 2-clause in $I-\{c_1,c_4\}$. Thus, $k(S_{\{\dua a,g,b\}
})\geq k(S)-1$. Finally, $c_1$ and $c_4$ are the only 2-clauses of $S$,
which are not 2-clauses of $S_{\{\dua a,\dua g,b,\delta,e\}}$.
Consequently, $k(S_{\{\dua a,\dua g,b,\delta,e\}})\geq k(S)-2$. Hence,
\[
\Delta(S,S_A)\geq \left\{
            \begin{array}{ll}
              1-\alpha & \mbox{if $A=\{ a\}$}\\
              3-\alpha & \mbox{if $A=\{\dua a,g,b\}$}\\
              5-2\alpha & \mbox{if $A=\{\dua a,\dua g,b,\delta,e\}\mbox{.}$}
            \end{array}
         \right.
\]
The equation (\ref{tau'}) becomes
\[
\tau^{-1+\alpha}+\tau^{-3+\alpha}+\tau^{-5+2\alpha}=1
\]
and $\tau_S'\leq 1\mbox{.}67$.

\noindent
{\bf Subcase (v).} Exactly one of the literals $d$ and $\ell$ belongs
to a 2-clause in $S$.

Without loss of generality we will assume that $S$ contains a
2-clause $c_5=d\vee j$ (and consequently, $\ell$ does not belong to a
2-clause in $S$). We note that $c_2=a\vee g\vee d$ and $c_3=a\vee e\vee
\lambda$.

Let $M$ be a model of $S$. If $M$ is consistent with $\{\dua a,d,e\}$,
then it is consistent with $\{b\}$ (clause $c_1$). If $M$ is consistent
with $\{\dua a,d,\dua e\}$, it is also consistent with $\{ b,g,\lambda
\}$ (clauses $c_1$, $c_4$ and $c_3$). If $M$ is consistent with $\{\dua
a,\dua d,e\}$, it is also consistent with $\{b,j,g\}$ (clauses $c_1$,
$c_5$ and $c_2$). Finally, if $M$ is consistent with $\{\dua a,\dua d,
\dua e\}$, it is consistent with $\{ b,j,g,\lambda\}$ (clauses $c_1$,
$c_5$, $c_2$ and $c_3$). Hence, the family
\[
{\A} =\{\{a\},\{\dua a,d,e,b\},\{\dua a,d,\dua e,b,g,\lambda\},
\{\dua a,\dua d,e,b,j,g\},\{\dua a,\dua d,\dua e,b,j,g,\lambda\}\}
\]
is a cover for $S$ and we define $\rho(S)=\A$.

All 2-clauses in $I  -\{c_1\}$ are 2-clauses in $S_{\{a\}}$. Thus,
$k(S_{\{a\}})\geq k(S)-1$. Moreover, all 2-clauses in $I
  -\{c_1,c_4,c_5\}$ are 2-clauses in $S_{\{\dua a,d,e,b\}}$,
$S_{\{\dua a,d,\dua e,b,g,\lambda\}}$, $S_{\{\dua a,\dua d,e,b,j,g\}}$
and $S_{\{\dua a,\dua d,\dua e,b,j,g,\lambda\}}$. Hence,
\[
\Delta(S,S_A)\geq \left\{
            \begin{array}{ll}
              1-\alpha & \mbox{if $A=\{ a\}$}\\
              4-3\alpha & \mbox{if $A=\{\dua a,d,e,b\}$}\\
              6-3\alpha & \mbox{if $A=\{\dua a,d,\dua e,b,g,\lambda\},
               \{\dua a,\dua d,e,b,j,g\}$}\\
              7-3\alpha & \mbox{if $A=\{\dua a,\dua d,\dua 
e,b,j,g,\lambda\}\mbox{.}$}
            \end{array}
         \right.
\]
The equation (\ref{tau'}) becomes
\[
\tau^{-1+\alpha}+\tau^{-4+3\alpha}+2\tau^{-6+3\alpha}+\tau^{-7+3\alpha}=1\mbox{.}
\]
Its root $\tau_S'$ satisfies $\tau_S'\leq 1\mbox{.}67$.

\noindent
{\bf Subcase (vi).} Both atoms $d$ and $\ell$ belong to 2-clauses of
$S$.

We can assume that $d$ and $\ell$ form a 2-clause $c_5=d\vee \ell$ as,
otherwise, Case 9(iii) would apply. We note that $c_2=a\vee g\vee d$
and $c_3=a\vee e\vee \ell$.

We assume first that $c_1$, $c_2$ and $c_3$ are the only clauses
containing $a$. We note that
\[
{\A'}=\{\{\dua a\},\{ a,\dua b\},\{ a,b,\dua g,\dua d\},\{ a,b,
\dua e,\dua \ell\}\}
\]
is a cover for $S$. Indeed, let $M$ be a minimal model of $S$
inconsistent with every set of $\A'$. It follows that $a\in M$.
As $M$ is inconsistent with $\{ a,\dua b\}$, $b\in M$. Furthermore,
since $M$ is inconsistent with $\{ a,b,\dua g,\dua d\}$, $g\in M$ or
$d\in M$. Similarly, $e\in M$ or $\ell\in M$. Since $a$ belongs to the
clauses $c_1$, $c_2$ and $c_3$ only, $M - \{a\}$ is a model of $S$,
contradicting the minimality of $M$.

Let $M$ be a model of $S$. If $M$ is consistent with $\dua a$, it is
consistent with $b$, as well (clause $c_1$). If $M$ is consistent
with $\{a,b,\dua g,\dua d\}$, it is consistent with $\{ e,\ell\}$
(clauses $c_4$ and $c_5$).
Lastly, if $M$ is consistent with $\{a,b,\dua e,\dua \ell\}$, it is
consistent with $\{ g,d\}$ (clauses $c_4$ and $c_5$). Since $\A'$ is
a cover for $S$, it follows that the family
\[
{\A}=\{\{\dua a,b\},\{ a,\dua b\},\{ a,b,\dua g,\dua d,e,\ell\},
\{a,b,\dua e,\dua \ell,g,d\}\}
\]
is a cover for $S$, as well. We set $\rho(S)=\A$.

All 2-clauses in $I  -\{c_1\}$ are 2-clauses in $S_{\{\dua a,b\}}$
and in $S_{\{ a,\dua b\}}$. Thus, $k(S_{\{\dua a,b\}})\geq k(S)-1$ and
$k(S_{\{ a,\dua b\}})\geq k(S)-1$. Moreover, all 2-clauses of $I  -
\{c_1,c_4,c_5\}$ are 2-clauses in $S_{\{ a,b,\dua g,\dua d,e,\ell\}}$
and $S_{\{ a,b,\dua e,\dua \ell,g,d\}}$. Consequently, $k(S_{\{ a,b,\dua
g,\dua d,e,\ell\}})\geq k(S)-3$ and $k(S_{\{ a,b,\dua e,\dua \ell,g,d\}})
\geq k(S)-3$. Hence,
\[
\Delta(S,S_A)\geq \left\{
            \begin{array}{ll}
              2-\alpha & \mbox{if $A=\{\dua a,b\},\{ a,\dua b\}$}\\
              6-3\alpha & \mbox{if $A=\{ a,b,\dua g,\dua d,e,\ell\},
          \{a,b,\dua e,\dua \ell,g,d\}\mbox{.}$}
            \end{array}
         \right.
\]
The equation (\ref{tau'}) becomes
\[
2\tau^{-2+\alpha}+2\tau^{-6+3\alpha}=1
\]
and its root satisfies $\tau_S'\leq 1\mbox{.}61$.

\noindent
{\bf Comment.} To complete Case 9(vi) (and, in the same time, Case
9), we still need to consider a situation when $S$ contains a 3-clause
$c$, different from $c_2$ and $c_3$ and such that $a$ is an atom of $c$
(since $a$ appears in a 2-clause, it appears in $c$ positively). If $a$
is the only atom common to $c$ and $c_2$ then, replacing the clause
$c_3$ with $c$, we obtain a situation where Case 9(ii) or 9(iii)
applies.

We can assume then that $c$ and $c_2$ have at least two atoms in common.
If $c$ and $c_2$ have 3 atoms in common then, since each of these common
atoms belongs to a 2-clause, it appears positively in every clause in
$S$. Consequently, $c=c_2$, a contradiction. It follows that $c$ and
$c_2$ have exactly 2-atoms in common. Replacing $c_3$ by $c$ we get a
situation where Case 10(ii) applies. We will consider it below. At that
point, Case 9 will be closed.

\noindent
{\bf Case 10.} For some atom $a$ that occurs in a 2-clause, say $c_1$,
there are two other clauses $c_2$ and $c_3$ that contain $a$ and
$|\At(c_2)\cap\At(c_3)|\geq 2$.

As before, $c_2$ and $c_3$ are 3-clauses. Throughout Case 10, we will
assume that $c_1=a\vee b$, $c_2=a\vee\gamma\vee\delta$ and $c_3=a\vee
\epsilon\vee\lambda$. Without loss of generality, we can assume that
$g=e$, that is, that $\gamma$ and $\epsilon$ have the same atom.

{\bf Subcase (i).} $\gamma=\dua\epsilon$.


As the atom $g$ occurs negatively in $S$ (in $c_2$ or $c_3$), $g$ does
not belong to any 2-clause.

A model of $S$ consistent with $\{\dua a,\gamma\}$ is also consistent
with $\{ b,\lambda\}$ (clauses $c_1$ and $c_3$). Moreover, a model of
$S$ consistent with $\{\dua a,\dua\gamma\} (=\{\dua a,\epsilon\})$ is
consistent with $\{ b,\delta\}$ (clauses $c_1$ and $c_2$). Hence, the
family
\[
{\A} =\{\{a\},\{\dua a,\gamma,b,\lambda\},\{\dua a,\epsilon,b,
\delta\}\}
\]
is a cover for $S$ and we set $\rho(S)=\A$.

All 2-clauses of $I  -\{c_1\}$ are 2-clauses of $S_{\{a\}}$ and so,
$k(S_{\{a\}})\geq k(S)-1$. All 2-clauses in $I  -\{c_1\}$ that do not
contain $\ell$ (the atom of $\lambda$) are 2-clauses in $S_{\{\dua a,
\gamma,b,\lambda\}}$. Thus, $k(S_{\{\dua a,\gamma,b,\lambda\}})\geq
k(S)-2$. Similarly, all 2-clauses of $I -  \{c_1\}$ that do not contain
$d$ (the atom of $\delta$) are 2-clauses in $S_{\{\dua a,\epsilon,b,
\delta\}}$ and so, $k(S_{\{\dua a,\epsilon,b,\delta\}})\geq k(S)-2$.
Hence,
\[
\Delta(S,S_A)\geq \left\{
            \begin{array}{ll}
              1-\alpha & \mbox{if $A=\{ a\}$}\\
              4-2\alpha & \mbox{if $A=\{\dua a,\gamma,b,\lambda\},
          \{\dua a,\epsilon,b,\delta\}\mbox{.}$}
            \end{array}
         \right.
\]
The equation (\ref{tau'}) becomes
\[
\tau^{-1+\alpha}+2\tau^{-4+2\alpha}=1
\]
and $\tau_S'\leq 1\mbox{.}65$.

\noindent
{\bf Comment.}
We will assume in the remaining subcases of Case 10 that $|\At(c_2)\cap
\At(c_3)|=2$, as otherwise, Case 10(i) would apply. Indeed, let us assume
that $c_2$ and $c_3$ have three atoms in common. Since $c_2\not=c_3$,
there is a literal in $c_2$ whose dual appears in $c_3$, precisely the
situation covered by Case 10(i).

\noindent
{\bf Subcase (ii).} $\gamma=\epsilon$ and $\gamma$ belongs to a 2-clause.

Clearly, in this subcase $\gamma$ is an atom, that is $\gamma=g$. Let
the 2-clause containing $g$ be $c_4=g\vee h$.

Every model of $S$ consistent with $\{\dua a,g\}$ is consistent with
$\{b\}$ (clause $c_1$). Moreover, every model of $S$ consistent with
$\{\dua a,\dua g\}$ is consistent with $\{ b,h,\delta,\lambda\}$
(clauses $c_1$, $c_4$, $c_2$ and $c_3$). Hence, the family
\[
{\A} =\{\{a\},\{\dua a,g,b\},\{\dua a,\dua g,b,h,\delta,\lambda\}\}
\]
is a cover for $S$ and we set $\rho(S)=\A$.

All 2-clauses of $I  - \{c_1\}$ are 2-clauses of $S_{\{a\}}$. Thus, $k(
S_{\{a\}})\geq k(S)-1$. All 2-clauses of $I  - \{c_1,c_4\}$ are
2-clauses in $S_{\{\dua a,g,b\}}$ and so, $k(S_{\{\dua a,g,b\}})\geq
k(S)-2$. Finally, all 2-clauses of $I  - \{c_1,c_4\}$ that do not
contain $d$ and $\ell$ are 2-clauses in $S_{\{\dua a,\dua g,b,h,\delta,
\lambda\}}$ and so, $k(S_{\{\dua a,\dua g,b,h,\delta,\lambda\}})\geq
k(S)-4$. Hence,
\[
\Delta(S,S_A)\geq \left\{
            \begin{array}{ll}
              1-\alpha & \mbox{if $A=\{ a\}$}\\
              3-2\alpha & \mbox{if $A=\{\dua a,g,b\}$}\\
              6-4\alpha & \mbox{if $A=\{\dua a,\dua g,b,h,\delta,
          \lambda\}\mbox{.}$}
            \end{array}
         \right.
\]
The equation (\ref{tau'}) becomes
\[
\tau^{-1+\alpha}+\tau^{-3+2\alpha}+\tau^{-6+4\alpha}=1
\]
and $\tau_S'\leq 1\mbox{.}67$.

\noindent
{\bf Subcase (iii).} $\gamma=\epsilon$, $\gamma$ does not belong to any
2-clause, and at most one of $d$ and $\ell$ belongs to a 2-clause.

Every model of $S$ consistent with $\{\dua a,\gamma\}$ is consistent
with $\{b\}$ (clause $c_1$). Moreover, every model consistent with
$\{\dua a,\dua\gamma\}$ is consistent with $\{b,\delta,\lambda\}$
(clauses $c_1$, $c_2$ and $c_3$). Hence, the family
\[
{\A} =\{\{a\},\{\dua a,\gamma,b\},\{\dua a,\dua\gamma,b,\delta,
\lambda\}\}
\]
is a cover and we set $\rho(S)=\A$.

All 2-clauses in $I  - \{c_1\}$ are 2-clauses of $S_{\{a\}}$ and $S_{\{
\dua a,\gamma,b\}}$. Thus, $k(S_{\{a\}})\geq k(S)-1$ and $k(S_{\{\dua a,
\gamma,b\}})\geq k(S)-1$. Moreover, all 2-clauses of $I  - \{c_1\}$ that
do not contain $d$ or $\ell$ (by our assumption, this condition excludes
at most one clause) are 2-clauses in $S_{\{\dua a,\dua\gamma,
b,\delta,\lambda\}}$ and so, $k(S_{\{\dua a,\dua\gamma,b,\delta,\lambda
\}})\geq k(S)-2$. Hence,
\[
\Delta(S,S_A)\geq \left\{
            \begin{array}{ll}
              1-\alpha & \mbox{if $A=\{ a\}$}\\
              3-\alpha & \mbox{if $A=\{\dua a,\gamma,b\}$}\\
              5-2\alpha & \mbox{if $A=\{\dua a,\dua\gamma,b,\delta,
          \lambda\}\mbox{.}$}
            \end{array}
         \right.
\]
the equation (\ref{tau'}) becomes
\[
\tau^{-1+\alpha}+\tau^{-3+\alpha}+\tau^{-5+2\alpha}=1
\]
and $\tau_S'\leq 1\mbox{.}67$.

\noindent
{\bf Subcase (iv).} $\gamma=\epsilon$, $\gamma$ does not belong to any
2-clause and $d$ and $\ell$ belong to 2-clauses of $S$ (possibly to the
same 2-clause).

Since $\delta$ and $\lambda$ belong to 2-clauses, $\delta=d$ and
$\lambda=\ell$.

First, we assume that $c_1$, $c_2$ and $c_3$ are the only clauses that
contain $a$. Clearly, the collection
\[
{\A'} =\{\{a,\dua b\},\{a,b\},\{\dua a,\gamma\},\{\dua a,\dua\gamma
\}\}
\]
is a cover for $S$.

Every minimal model $M$ of $S$ consistent with $\{a,b\}$ is consistent
with $\{\dua\gamma\}$ (otherwise, $M-\{ a\}$ would be a model of $S$,
too). Every model consistent with $\{\dua a,\gamma\}$ is consistent
with $\{b\}$ (clause $c_1$). Finally, every model consistent with
$\{\dua a,\dua\gamma\}$ is consistent with $\{ b,d,\ell\}$ (clauses
$c_1$, $c_2$ and $c_3$). Hence, the family
\[
{\A} =\{\{a,\dua b\},\{a,b,\dua\gamma\},\{\dua a,\gamma,b\},\{\dua
a,\dua\gamma,b,d,\ell\}\}
\]
is a cover for $S$ and we set $\rho(S)=\A$.

All 2-clauses of $I  -\{c_1\}$ are 2-clauses of $S_{\{a,\dua b\}}$,
$S_{\{a,b, \dua\gamma\}}$ and $S_{\{\dua a,\gamma,b\}}$. Thus,
$k(S_{\{a,\dua b\}})\geq
k(S)-1$, $k(S_{\{a,b,\dua\gamma\}})\geq k(S)-1$ and $k(S_{\{\dua
a,\gamma,b\}})\geq k(S)-1$. Moreover, all 2-clauses of $I  -\{c_1\}$
that do not contain $d$ and $\ell$ are 2-clauses in $S_{\{\dua a,\dua
\gamma,b,d,\ell\}}$ and so, $k(S_{\{\dua a,\dua\gamma,b,d,\ell\}})\geq
k(S)-3$. Hence,
\[
\Delta(S,S_A)\geq \left\{
            \begin{array}{ll}
              2-\alpha & \mbox{if $A=\{a,\dua b\}$}\\
              3-\alpha & \mbox{if $A=\{a,b,\dua\gamma\},\{\dua a,
        \gamma,b\}$}\\
              5-3\alpha & \mbox{if $A=\{\dua 
a,\dua\gamma,b,d,\ell\}\mbox{.}$}
            \end{array}
         \right.
\]
The equation (\ref{tau'}) becomes
\[
\tau^{-2+\alpha}+2\tau^{-3+\alpha}+\tau^{-5+3\alpha}=1
\]
and $\tau_S'\leq 1\mbox{.}66$.

Let us suppose now that $c_1$, $c_2$ and $c_3$ are not the only clauses
that contain $a$. Let $c$ be a clause in $S$ containing $a$ and different
from $c_2$ and $c_3$. Since $a$ appears in a 2-clause, $c$ is a
3-clause.

If $\At(c)\cap \At(c_2)=\{a\}$ or $\At(c)\cap \At(c_3)=\{a\}$, then
Case 9 applies (which, we finally settled with Case 10(ii)). Thus, we
can assume that $c$ has two common atoms with $c_2$ and two common atoms
with $c_3$. Since $d\not=\ell$ (otherwise, $c_2=c_3$), there are two
possibilities: (1) $d$ and $\ell$ are atoms of $c$ and, since $d$ and
$\ell$ appear in 2-clauses, $c=a\vee d\vee\ell$, and (2) $g$ is the atom
of $c$, which means that $\gamma$ is a literal in $c$ (otherwise, Case
10(i) would apply). The first possibility is covered by Case 10(ii),
which applies to $c_2$ and $c$. Thus, we can assume that the second
possibility holds. Let $f$ be the atom of $c$ other than $a$ and $g$.
If $f$ does not belong to a 2-clause, then Case 10(iii) would apply
to $c_2$ and $c$.
Hence, we can assume that $f$ belongs to a 2-clause and, consequently,
it appears positively in all clauses in $S$. In particular, $c=a\vee
\gamma \vee f$. Since $c\not=c_2,c_3$, $f\not=d,\ell$.

Every model of $S$ consistent with $\{\dua a,\gamma\}$ is consistent
with $b$ (clause $c_1$). Moreover, every model of $S$ consistent with
$\{\dua a,\dua\gamma\}$ is consistent with $\{ b,d,\ell,f\}$ (clauses
$c_1$, $c_2$, $c_3$ and $c$). Hence, the family
\[
{\A} =\{\{a\},\{\dua a,\gamma,b\},\{\dua a,\dua\gamma,b,d,\ell,
f\}\}
\]
is a cover for $S$ and we set $\rho(S)=\A$.

All 2-clauses in $I-\{c_1\}$ are 2-clauses in $S_{\{a\}}$ and $S_{\{
\dua a,\gamma,b\}}$. Thus, $k(S_{\{a\}})\geq k(S)-1$ and $k(S_{\{\dua
a,\gamma,b\}})\geq k(S)-1$. Moreover, all 2-clauses of $I-\{c_1\}$ that
do not contain $d$, $\ell$ and $f$ are 2-clauses of $S_{\{\dua a,\dua
\gamma,b,d,\ell,f\}}$ and so, $k(S_{\{\dua a,\dua\gamma,b,d,\ell,f\}})
\geq k(S)-4$. Hence,
\[
\Delta(S,S_A)\geq \left\{
          \begin{array}{ll}
      1-\alpha & \mbox{if $A=\{a\}$}\\
      3-\alpha & \mbox{if $A=\{\dua a,\gamma,b\}$}\\
      6-4\alpha & \mbox{if $A=\{\dua a,\dua\gamma,b,d,\ell,f\}\mbox{.}$}
      \end{array} \right.
\]
The equation (\ref{tau'}) becomes
\[
\tau^{-1+\alpha}+\tau^{-3+\alpha}+\tau^{-6+4\alpha}=1
\]
and $\tau_S'\leq 1\mbox{.}64$.

\noindent
{\bf Comment.} From now on we will assume that $S$ does not contain
2-clauses, that is, $k(S)=0$. For an atom $a$, we will denote by $T(a)$
the set of 3-clauses in $S$ with positive occurrences of $a$.

\noindent
{\bf Case 11.} There is an atom $a$ and two 3-clauses $c_1$ and $c_2$
in $T(a)$ such that $c_1$ contains a literal which is dual to some
literal occurring in $c_2$.

Without losing generality, we may assume that $c_1=a\vee \beta\vee
\gamma$ and $c_2=a\vee\dua\beta\vee\delta$.

\noindent
{\bf Subcase (i).} $\gamma=\delta$.

Every model consistent with $\dua a$ is consistent with $\gamma$. Hence,
the family
\[
{\A} =\{\{a\},\{\dua a,\gamma\}\}
\]
is a cover for $S$ and we set $\rho(S)=\A$. Since $k(S_A)\geq 0$, for
$A\in \A$,
\[
\Delta(S,S_A)\geq \left\{
            \begin{array}{ll}
              1 & \mbox{if $A=\{ a\}$}\\
              2 & \mbox{if $A=\{\dua a,\gamma\}\mbox{.}$}
            \end{array}
         \right.
\]
The equation (\ref{tau'}) becomes
\[
\tau^{-1}+\tau^{-2}=1
\]
and $\tau_S'\leq 1\mbox{.}62$.

\noindent
{\bf Subcase (ii).} There is a 3-clause $c_3$ in $T(a)-\{c_1,c_2\}$
such that $b\in\At(c_3)$.

Without losing generality we can assume that $\beta$ is a literal of
$c_3$. Consequently, $c_3=a\vee
\beta\vee \epsilon$ and, since $c_3\not=c_1$, $\epsilon\not=\gamma$.
Moreover, if $\epsilon=\dua\gamma$, Case 11(i) would apply to $c_1$ and
$c_3$. Thus, $\epsilon\not=\dua\gamma$ and $e\not= g$.

Every model of $S$ consistent with $\{\dua a,\beta\}$ is consistent
with $\delta$ (clause $c_2$). Moreover, every model of $S$ consistent
with $\{\dua a,\dua\beta\}$ is consistent with $\{ \gamma,\epsilon\}$
(clauses $c_1$ and $c_3$). Hence, the family
\[
{\A} =\{\{a\},\{\dua a,\beta,\delta\},\{\dua a,\dua\beta,\gamma,
\epsilon\}\}
\]
is a cover for $S$ and we define $\rho(S)=\A$. It also follows
that
\[
\Delta(S,S_A)\geq \left\{
            \begin{array}{ll}
              1 & \mbox{if $A=\{ a\}$}\\
              3 & \mbox{if $A=\{\dua a,\beta,\delta\}$}\\
              4 & \mbox{if $A=\{\dua a,\dua\beta,\gamma,\epsilon\}\mbox{.}$}
            \end{array}
         \right.
\]
The equation (\ref{tau'}) becomes
\[
\tau^{-1}+\tau^{-3}+\tau^{-4}=1
\]
and $\tau_S'\leq 1\mbox{.}62$.

\noindent
{\bf Comment.} From now on we can assume that no clause in $T(a)-
\{c_1,c_2\}$ contains $b$.

\noindent
{\bf Subcase (iii).} There is a 3-clause $c_3$ in $T(a)-\{c_1,c_2\}$
that contains neither $d$ nor $g$.

Let $c_3=a\vee\epsilon\vee\lambda$ (we note that $c_3$ does not contain
$b$).
Every model of $S$ consistent with $\{\dua a,\beta\}$ is consistent
with $\delta$ (clause $c_2$). Moreover, every model consistent with
$\{\dua a,\dua\beta\}$ is consistent with $\gamma$ (clause $c_1$).
Hence, the family
\[
{\A} =\{\{a\},\{\dua a,\beta,\delta\},\{\dua a,\dua\beta,\gamma\}\}
\]
is a cover for $S$ and we set $\rho(S)=\A$.

The theories $S_{\{\dua a,\beta,\delta\}}$ and $S_{\{\dua a,\dua\beta,
\gamma\}}$ contain the 2-clause $\epsilon\vee\lambda$ (it follows from
the fact that $e$ and $\ell$ are different from $a$, $b$, $d$ and $g$).
Thus, $k(S_{\{\dua a,\beta,\delta\}})\geq 1$ and $k(S_{\{\dua a,\dua
\beta,\gamma\}})\geq 1$. It follows,
\[
\Delta(S,S_A)\geq \left\{
            \begin{array}{ll}
              1 & \mbox{if $A=\{ a\}$}\\
              3 + \alpha & \mbox{if $A=\{\dua a,\beta,\delta\},\{\dua a,
          \dua\beta,\gamma\}\mbox{.}$}
            \end{array}
         \right.
\]
The equation (\ref{tau'}) becomes
\[
\tau^{-1}+2\tau^{-3-\alpha}=1
\]
and $\tau_S'\leq 1\mbox{.}66$.

\noindent
{\bf Comment.} From now on we can assume that every clause in $T(a)-
\{c_1,c_2\}$ contains either $g$ or $d$.

\noindent
{\bf Subcase (iv).} $T(a)=\{c_1,c_2\}$.

Clearly, the collection
\[
{\A'} = \{\{ \dua a,\beta\},\{\dua a,\dua\beta\},\{ a,\beta\},\{
a,\dua\beta\}\}
\]
is a cover for $S$.

Every model of $S$ consistent with $\{\dua a,\beta\}$ is consistent
with $\delta$ to satisfy $c_2$. Similarly, every model consistent with
$\{\dua a,\dua\beta\}$ is consistent with $\gamma$ to satisfy $c_1$.
Every minimal model $M$ consistent with $\{ a,\beta\}$ is consistent
with $\dua\delta$ (otherwise $M-\{a\}$ is a model of $S$, as $c_1$ and
$c_2$ are the only clauses in $S$ with a positive occurrence of $a$).
Similarly, every minimal model of $S$ consistent with $\{a,\dua\beta
\}$ is consistent with $\dua\gamma$. Hence, the family
\[
{\A} = \{\{ \dua a,\beta,\delta\},\{\dua a,\dua\beta,\gamma\},\{a,
\beta,\dua\delta\},\{ a,\dua\beta,\dua\gamma\}\}
\]
is a cover for $S$ and we define $\rho(S)=\A$. Moreover, for every
$A\in \A$,
\[
\Delta(S,S_A)\geq 3\mbox{.}
\]
The equation (\ref{tau'}) becomes
\[
4\tau^{-3}=1
\]
and $\tau_S'\leq 1\mbox{.}59$.

\noindent
{\bf Comment.} In the remainder of Case 11, we can assume that
no two clauses in $T(a)$ have the same set of atoms. Indeed, let us
assume that $c'_1,c'_2\in T(a)$ and $\At(c'_1)=\At(c'_2)$. Without
loss of generality, we can assume that $c'_1=a\vee\beta'\vee\gamma'$.
Then, it follows that $c'_2=a\vee\dua{\beta'} \vee\dua{\gamma'}$ (the
cases $c_2'=a\vee{\beta'} \vee\dua{\gamma'}$ and $c'_2=a\vee\dua{\beta'}
\vee{\gamma'}$ are covered by Case 11(i)). We now note that in Cases
11(ii)-(iv) we allowed for the possibility that $\gamma=\dua\delta$.
Thus, if $T(a)= \{c'_1,c'_2\}$, then Case 11(iv) applies. So, let us
assume that there is a 3-clause $c_3=a\vee\epsilon \vee\lambda$, such
that $c_3\in T(a)-\{c'_1,c'_2\}$. If $|\{e,\ell\}\cap \{b',g'\}|=0$,
then Case 11(iii) applies. If $|\{e,\ell\}\cap \{b',g'\}|=1$, then Case
11(ii) applies. Finally, if $\{e,\ell\}=\{b',g'\}$, then Case 11(i)
applies to $c'_1$ and $c_3$ or to $c_2'$ and $c_3$.

\noindent
{\bf Subcase (v).} $T(a)-\{c_1,c_2\}$ contains a clause $c_3$ such that
at least one of $\dua\gamma$ and $\dua\delta$ is a literal of $c_3$.

Without loss of generality we assume that $c_3$ contains $\dua\gamma$,
that is, $c_3=a\vee\dua\gamma\vee\epsilon$, for some literal $\epsilon$.
We can assume that $e\not=b$ (otherwise, $\At(c_1)=\At(c_3)$).

Every model of $S$ consistent with $\{\dua a,\dua\beta\}$ is consistent
with $\{\gamma,\epsilon\}$ to satisfy $c_1$ and $c_3$. Moreover, every
model of $S$ consistent with $\{\dua a,\beta\}$ is consistent with
$\delta$ to satisfy $c_2$. Hence, the family
\[
{\A} = \{\{ a\},\{\dua a,\dua\beta,\gamma,\epsilon\},\{ \dua a,
\beta,\delta\}\}
\]
is a cover for $S$ and we set $\rho(S)=\A$. Thus,
\[
\Delta(S,S_A)\geq \left\{
            \begin{array}{ll}
              1 & \mbox{if $A=\{ a\}$}\\
              4 & \mbox{if $A=\{\dua a,\dua\beta,\gamma,\epsilon\}$}\\
              3 & \mbox{if $A=\{ \dua a,\beta,\delta\}\mbox{.}$}
            \end{array}
         \right.
\]
The equation (\ref{tau'}) becomes
\[
\tau^{-1}+\tau^{-4}+\tau^{-3}=1
\]
and $\tau_S'\leq 1\mbox{.}62$.

\noindent
{\bf Subcase (vi).} No 3-clause in $T(a)-\{c_1,c_2\}$ contains
$\dua\gamma$ or $\dua\delta$.

We recall that by the comment after Case 11(iii), every clause
$T(a)-\{c_1,c_2\}$ contains $\gamma$ or $\delta$.

Let us now define $c_3=a\vee \gamma \vee\delta$. We assume first that
$T(a)-\{c_1,c_2\}=\{ c_3\}$.

Clearly, the collection
\[
{\A'} = \{\{ a,\delta\},\{ a,\dua\delta\},\{\dua a,\gamma\},\{
\dua a,\dua\gamma\}\}
\]
is a cover for $S$.

Every minimal model $M$ of $S$ consistent with $\{a,\delta\}$ is
consistent with $\{\dua\gamma,\dua\beta\}$. Otherwise, $M$ would be
consistent with $\delta$ and $\gamma$ or with $\delta$ and $\beta$.
Consequently, $M-\{a\}$ would satisfy all three clauses $c_1,c_2,c_3$
of $T(a)$, and since $a$ does not belong to any other clause in $S$,
$M-\{ a\}$ would be a model of $S$, contrary to the minimality of $M$.

Moreover, every model of $S$ consistent with $\{\dua a,\dua\gamma\}$ is
consistent with $\{\beta,\delta\}$ to satisfy $c_1$ and $c_3$. Hence,
the family
\[
{\A} = \{\{ a,\delta,\dua\gamma,\dua\beta\},\{ a,\dua\delta\},
\{\dua a,\gamma\},\{\dua a,\dua\gamma,\beta,\delta\}\}
\]
is a cover for $S$ and we set $\rho(S)=\A$.

Let us observe that the theory $S_{\{\dua a,\gamma\}}$ contains the
2-clause $\dua\beta\vee\delta$ (indeed, the atoms $b$ and $d$ are
different from $a$ and $g$) and so, $k(S_{\{\dua a,\gamma\}})\geq 1$.
Thus,
\[
\Delta(S,S_A)\geq \left\{
            \begin{array}{ll}
              4 & \mbox{if $A=\{ a,\delta,\dua\gamma,\dua\beta\},\{
           \dua a,\dua\gamma,\beta,\delta\}$}\\
              2 & \mbox{if $A=\{ a,\dua\delta\}$}\\
              2 + \alpha & \mbox{if $A=\{\dua a,\gamma\}\mbox{.}$}
              \end{array}
         \right.
\]
The equation (\ref{tau'}) becomes
\[
2\tau^{-4}+\tau^{-2}+\tau^{-2-\alpha}=1
\]
and $\tau_S'\leq 1\mbox{.}64$.

It remains to consider the case when $T(a)-\{c_1,c_2\} \not=\{c_3\}$.
If $T(a)-\{c_1,c_2\}=\emptyset$, Case 11(iv) applies. Since $T(a)-\{c_1,
c_2\} \not=\{c_3\}$, there is a clause $c_4\in T(a)-\{c_1,c_2\}$ such
that $c_4\not=c_3$. We can assume that $b\notin
\At(c_4)$ as, otherwise, Case 11(ii) would apply. Moreover, we can assume
that exactly one of $g$ and $d$ belongs to $\At(c_4)$ (if neither $g$
nor $d$ does, Case 11(iii) applies, and if both do, Case 11(v)
applies or $c_4=a\vee\gamma\vee\delta=c_3$, a contradiction with $c_4
\not=c_3$).

Without loss of generality, we can assume that $c_4$ contains $\gamma$
(and so, it does not contain $\delta$). Let $c_4=a\vee\gamma\vee
\epsilon$. Clearly, $\epsilon\not=\delta$. Moreover, $\epsilon\not=\dua
\delta$, as no clause in $T(a)-\{c_1,c_2\}$ contains $\dua\delta$, by
the assumption we adopted in Case 11(vi).

The family
\[
{\A'} = \{\{a,\delta\},\{ a,\dua\delta\},\{\dua a,\gamma\},\{\dua a,
\dua\gamma\}\}
\]
is, trivially, a cover for $S$.

Let us observe that every minimal model $M$ of $S$ consistent with
$\{a,\delta\}$ is consistent with $\dua\gamma$. If not, $M$ would
be consistent with $\gamma$. Since every clause of $T(a)$ contains
$\delta$ or $\gamma$, $M-\{a\}$ would be a model of $T(a)$ and,
consequently, of $S$, contrary to the minimality of $M$.

Moreover, every model of $S$ consistent with $\{ \dua a,\dua\gamma\}$
is consistent with $\{\epsilon,\beta,\delta\}$ to satisfy $c_4$, $c_1$
and $c_2$. Hence, the family
\[
{\A} = \{\{ a,\delta,\dua\gamma\},\{ a,\dua\delta\},\{\dua a,
\gamma\},\{\dua a,\dua\gamma,\epsilon,\beta,\delta\}\}
\]
is a cover for $S$ and we define $\rho(S)=\A$.

We observe that the theory $S_{\{\dua a,\gamma\}}$ contains the
2-clause $\dua\beta\vee\delta$ and so, $k(S_{\{\dua a,\gamma\}})\geq 1$.
Thus,
\[
\Delta(S,S_A)\geq \left\{
            \begin{array}{ll}
              3 & \mbox{if $A=\{ a,\delta,\dua\gamma\}$}\\
              2 & \mbox{if $A=\{ a,\dua\delta\}$}\\
              2 + \alpha & \mbox{if $A=\{\dua a,\gamma\}$}\\
              5 & \mbox{if $A=\{\dua 
a,\dua\gamma,\epsilon,\beta,\delta\}\mbox{.}$}
              \end{array}
         \right.
\]
The the equation (\ref{tau'}) becomes
\[
\tau^{-3}+\tau^{-2}+\tau^{-2-\alpha}+\tau^{-5}=1
\]
and $\tau_S'\leq 1\mbox{.}66$.

\noindent
{\bf Comment.} From now on we will assume that for each atom $a$ no two
clauses of $T(a)$ contain dual literals. We will denote by $\Gamma(a)$
the undirected graph whose vertices are literals that belong to clauses
in $T(a)$, and two literals $\beta$ and $\gamma$ form an edge in
$\Gamma(a)$ if $a\vee\beta\vee\gamma \in S$. We will write such an edge
as $\beta\gamma$. We emphasize that whenever $\beta\gamma$ is an edge of
a graph $\Gamma(a)$, there is a clause $a\vee\beta\vee\gamma$ in $S$.
By Case 3, we will assume in what follows that $\Gamma(a)$ has a nonempty
set of edges.

\noindent
{\bf Case 12.} There is an atom $a$ such that $\Gamma(a)$ has a vertex
of degree at least $5$.

Let $\beta$ be a vertex of degree at least $5$ in $\Gamma(a)$ and let
$\beta_1,\beta_2,\beta_3,\beta_4,\beta_5$ be neighbors of $\beta$ in
$\Gamma(a)$. In particular, it follows that $S$ contains the clauses
$a\vee\beta\vee\beta_i$, $i=1,2,3,4,5$. Moreover, by the most recent
comment, all atoms $b_i$, $i=1,2,3,4,5$, are pairwise distinct.

Clearly, every model of $S$ consistent with $\{\dua a,\dua\beta\}$ is
consistent with $\{\beta_1,\beta_2,\beta_3,\beta_4,\beta_5\}$ (clauses
$a\vee\beta\vee\beta_i$, $i=1,2,3,4,5$). Hence, the family
\[
{\A} =\{\{a\},\{\dua a,\beta\},\{\dua a,\dua\beta,\beta_1,\beta_2,
\beta_3,\beta_4,\beta_5\}\}
\]
is a cover and we set $\rho(S)=\A$. It follows that
\[
\Delta(S,S_A)\geq \left\{
            \begin{array}{ll}
              1 & \mbox{if $A=\{ a\}$}\\
              2 & \mbox{if $A=\{\dua a,\beta\}$}\\
              7 & \mbox{if $A=\{\dua a,\dua\beta,\beta_1,\beta_2,\beta_3,
           \beta_4,\beta_5\}\mbox{.}$}
            \end{array}
         \right.
\]
The equation (\ref{tau'}) becomes
\[
\tau^{-1}+\tau^{-2}+\tau^{-7}=1
\]
and $\tau_S'\leq 1\mbox{.}66$.

\noindent
{\bf Case 13.} There is an atom $a$ such that the maximum degree of
a vertex in $\Gamma(a)$ is $3$ or $4$.

Let $\beta$ be a vertex of maximum degree in $\Gamma(a)$. If the degree
of $\beta$ is $4$ then let $\beta_1,\beta_2,\beta_3,\beta_4$ be its
neighbors. Otherwise let $\beta_1,\beta_2,\beta_3$ be the neighbors of
$\beta$.

\noindent
{\bf Subcase (i).} The degree of $\beta$ is $4$ and there are
at least $5$ edges in $\Gamma(a)$.

Let $\gamma\delta$ be an edge in $\Gamma(a)$ not incident to $\beta$.
Every model of $S$ consistent with $\{\dua a,\dua\beta\}$ is consistent
with $\{\beta_1,\beta_2,\beta_3,\beta_4\}$ (clauses $a\vee \beta\vee
\beta_i$, $i=1,2,3,4$). Hence, the family
\[
{\A} =\{\{a\},\{\dua a,\beta\},\{\dua a,\dua\beta,\beta_1,\beta_2,
\beta_3,\beta_4\}\}
\]
is a cover for $S$ and we set $\rho(S)=\A$.

We observe that $\gamma$ and $\delta$ are vertices in $\Gamma(a)$ and,
consequently, $a\not=g,d$. Next, we note that since the edge $\gamma
\delta$ is not incident to $\beta$, the atoms $g,d$ and $b$ are pairwise
distinct.
Thus, the theory $S_{\{\dua a,\beta\}}$ contains the 2-clause $\gamma
\vee \delta$ and
so, $k(S_{\{\dua a,\beta\}})\geq 1$. It follows that
\[
\Delta(S,S_A)\geq \left\{
            \begin{array}{ll}
              1 & \mbox{if $A=\{ a\}$}\\
              2 + \alpha & \mbox{if $A=\{\dua a,\beta\}$}\\
              6 & \mbox{if $A=\{\dua a,\dua\beta,\beta_1,\beta_2,
          \beta_3,\beta_4\}\mbox{.}$}
            \end{array}
         \right.
\]
The equation (\ref{tau'}) becomes
\[
\tau^{-1}+\tau^{-2-\alpha}+\tau^{-6}=1
\]
and $\tau_S'\leq 1\mbox{.}64$.

\noindent
{\bf Subcase (ii).} All edges in $\Gamma(a)$ are incident to $\beta$.

Every model consistent with $\{\dua a,\dua\beta\}$ is consistent with
$\{\beta_1,\beta_2,\beta_3\}$ (clauses $a\vee\beta\vee\beta_i$, $i=1,2,
3$). Moreover every minimal model $M$ of $S$, consistent with $a$, is
consistent with $\dua\beta$, too. Otherwise, $M$ would be consistent
with $\beta$. That would imply that $M-\{a\}$ is a model of all clauses
in $T(a)$ and, consequently, of $S$, contrary to the minimality of $M$.
Hence, the family
\[
{\A} =\{\{a,\dua\beta\},\{\dua a,\beta\},\{\dua a,\dua\beta,
\beta_1,\beta_2,\beta_3\}\}
\]
is a cover for $S$ and we set $\rho(S)=\A$.
Thus,
\[
\Delta(S,S_A)\geq \left\{
            \begin{array}{ll}
              2 & \mbox{if $A=\{ a,\dua\beta\},\{\dua a,\beta\}$}\\
              5 & \mbox{if $A=\{\dua 
a,\dua\beta,\beta_1,\beta_2,\beta_3\}\mbox{.}$}
            \end{array}
         \right.
\]
The equation (\ref{tau'}) becomes
\[
2\tau^{-2}+\tau^{-5}=1
\]
and $\tau_S'\leq 1\mbox{.}52$.

\noindent
{\bf Comment.}
In the remainder of Case 13, we can assume that the degree of $\beta$
is 3 and that the graph $\Gamma(a)$ has at least 4 edges. If the degree
of $\beta$ was 4 and $\Gamma(a)$ had 5 or more edges, Case 13(i) would
apply. If the degree of $\beta$ was 4 and $\Gamma(a)$ had 4 edges, or if
the degree of $\beta$ was 3 and $\Gamma(a)$ had 3 edges, they all would
be incident to $\beta$ and Case 13(ii) would apply.

\noindent
{\bf Subcase (iii).} The degree of $\beta$ in $\Gamma(a)$ is $3$ and
$\Gamma(a)$ contains at least $5$ edges.

It follows that $\Gamma(a)$ contains two edges, say $\gamma_1\delta_1$
and $\gamma_2\delta_2$, that are not incident to $\beta$.

We will assume first that these two edges are independent. It is easy to
see that
every model of $S$ consistent with $\{\dua a,\dua\beta\}$ is consistent
with $\{\beta_1,\beta_2,\beta_3\}$ (clauses $a\vee\beta\vee\beta_i$,
$i=1,2,3$). Hence, the family
\[
{\A} =\{\{a\},\{\dua a,\beta\},\{\dua a,\dua\beta,\beta_1,\beta_2,
\beta_3\}\}
\]
is a cover for $S$ and we set $\rho(S)=\A$.

The theory $S_{\{\dua a,\beta\}}$ contains two 2-clauses
$\gamma_1\vee \delta_1$ and $\gamma_2\vee\delta_2$ with disjoint sets of
atoms. Consequently,
$k(S_{\{\dua a,\beta\}}) \geq 2$. Thus,
\[
\Delta(S,S_A)\geq \left\{
            \begin{array}{ll}
              1 & \mbox{if $A=\{ a\}$}\\
              2 + 2\alpha & \mbox{if $A=\{\dua a,\beta\}$}\\
              5 & \mbox{if $A=\{\dua a,\dua\beta,\beta_1,\beta_2,
          \beta_3\}\mbox{.}$}
            \end{array}
         \right.
\]
The equation (\ref{tau'}) becomes
\[ \tau^{-1}+\tau^{-2-2\alpha}+\tau^{-5}=1, \]
and $\tau_S'\leq 1\mbox{.}65$.

Next, we will assume that the two edges $\gamma_i\delta_i$, $i=1,2$, are
not independent in $\Gamma(a)$. Without loss of generality we may assume
that $\delta_1=\delta_2=\delta$.

Every model of $S$ consistent with $\{\dua a,\beta,\dua\delta\}$ is
consistent with $\{\gamma_1,\gamma_2\}$ (clauses $a\vee\gamma_1\vee
\delta$ and $a\vee\gamma_2\vee\delta$). Moreover, every model of $S$
consistent with $\{\dua a,\dua\beta\}$ is consistent with $\{\beta_1,
\beta_2,\beta_3\}$ (clauses $a\vee\beta\vee\beta_i$, $i=1,2,3$). Hence,
the family
\[
{\A} =\{\{a\},\{\dua a,\beta,\delta\},\{\dua a,\beta,\dua\delta,
\gamma_1,\gamma_2\},\{\dua a,\dua\beta,\beta_1,\beta_2,\beta_3\}\}
\]
is a cover for $S$ and we set $\rho(S)=\A$.
Thus,
\[
\Delta(S,S_A)\geq \left\{
            \begin{array}{ll}
              1 & \mbox{if $A=\{ a\}$}\\
              3 & \mbox{if $A=\{\dua a,\beta,\delta\}$}\\
              5 & \mbox{if $A=\{\dua a,\beta,\dua\delta,\gamma_1,
          \gamma_2\},\{\dua a,\dua\beta,\beta_1,\beta_2,\beta_3\}\mbox{.}$}
            \end{array}
         \right.
\]
The equation (\ref{tau'}) becomes
\[
\tau^{-1}+\tau^{-3}+2\tau^{-5}=1
\]
and $\tau_S'\leq 1\mbox{.}65$.

\noindent
{\bf Subcase (iv).} The degree of $\beta$ in $\Gamma(a)$ is $3$
and $\Gamma(a)$ contains exactly 4 edges.

Three of the edges of $\Gamma(a)$ are incident to $\beta$. The fourth
one is not. Let us denote by $\gamma\delta$ the edge in $\Gamma(a)$ that
is not incident to $\beta$.

Every model of $S$ consistent with $\{\dua a,\dua\beta\}$ is consistent
with
$\{\beta_1,\beta_2,\beta_3\}$ (clauses $a\vee\beta\vee\beta_i$, $i=1,
2,3$). Moreover, every minimal model $M$ of $S$ consistent with $\{a,
\beta\}$ is consistent with $\{\dua\gamma,\dua\delta\}$. Otherwise, $M$
would be consistent with at least one of $\gamma$ and $\delta$ and
$M-\{a\}$ would a model of $T(a)$ and so, of $S$, contrary to the
minimality of $M$. Hence, the family
\[
{\A} =\{\{a,\dua\beta\},\{a,\beta,\dua\gamma,\dua\delta\},\{\dua a,
\beta\}, \{\dua a,\dua\beta,\beta_1,\beta_2,\beta_3\}\}
\]
is a cover for $S$ and we define $\rho(S)=\A$.

The theory $S_{\{\dua a,\beta\}}$ contains the 2-clause $\gamma\vee
\delta$ and so, $k(S_{\{\dua a,\beta\}})\geq 1$. Thus,
\[
\Delta(S,S_A)\geq \left\{
            \begin{array}{ll}
              2 & \mbox{if $A=\{a,\dua\beta\}$}\\
              4 & \mbox{if $A=\{a,\beta,\dua\gamma,\dua\delta\}$}\\
              2 + \alpha & \mbox{if $A=\{\dua a,\beta\}$}\\
              5 & \mbox{if $A=\{\dua 
a,\dua\beta,\beta_1,\beta_2,\beta_3\}\mbox{.}$}
            \end{array}
         \right.
\]
The equation (\ref{tau'}) becomes
\[
\tau^{-2}+\tau^{-4}+\tau^{-2-\alpha}+\tau^{-5}=1
\]
and $\tau_S'\leq 1\mbox{.}60$.

\noindent
{\bf Comment.} From now on, we can assume that for every atom $a$, every
vertex in the graph $\Gamma(a)$ has degree 1 or 2.

\noindent {\bf Case 14.} There is an atom $a$ such that
$\Gamma(a)$ contains at least $4$ independent edges.

Let $\gamma_1\delta_1$, $\gamma_2\delta_2$,$ \gamma_3\delta_3$,
$\gamma_4\delta_4$ be independent edges in $\Gamma(a)$.
In this case we set $\rho(S)=\A$, where
\[
{\A} =\{\{a\},\{\dua a\}\}
\]
($\A$ is trivially complete).

The theory $S_{\{\dua a\}}$ contains four 2-clauses $\gamma_i\vee
\delta_i$, $i=1,2,3,4$, with pairwise different atoms and so, $k(S_{
\{\dua a\}})\geq 4$.
Thus,
\[
\Delta(S,S_A)\geq \left\{
            \begin{array}{ll}
              1 & \mbox{if $A=\{a\}$}\\
              1 + 4\alpha & \mbox{if $A=\{\dua a\}\mbox{.}$}
            \end{array}
         \right.
\]
The equation (\ref{tau'}) becomes
\[
\tau^{-1}+\tau^{-1-4\alpha}=1
\]
and $\tau_S'=1\mbox{.}6701\mbox{..}\ $.

\noindent
{\bf Case 15.} There is an atom $a$ such that $\Gamma(a)$ has at least
$5$ edges.

\noindent
{\bf Subcase (i).} There is a pair of nonadjacent vertices of degree
$2$ in $\Gamma(a)$.

Let $\beta$ and $\gamma$ be two nonadjacent vertices of degree $2$ in
$\Gamma(a)$. We denote by $\beta_1$ and $\beta_2$ the neighbors of
$\beta$ and by $\gamma_1$ and $\gamma_2$ the neighbors of $\gamma$.

Every model of $S$ consistent with $\{\dua a,\dua\beta\}$ is consistent
with $\{\beta_1,\beta_2\}$ (clauses $a\vee\beta\vee\beta_1$ and
$a\vee\beta\vee\beta_2$). Moreover, every model of $S$ consistent with
$\{\dua a,\beta,\dua\gamma\}$ is consistent with $\{\gamma_1,\gamma_2\}$
(clauses $a\vee\gamma\vee\gamma_1$ and $a\vee\gamma\vee\gamma_2$).
Hence, the family
\[
{\A} =\{\{a\},\{\dua a,\dua\beta,\beta_1,\beta_2\},\{\dua a,\beta,
\gamma\},\{\dua a,\beta,\dua\gamma,\gamma_1,\gamma_2\}\}
\]
is a cover for $S$ and we set $\rho(S)=\A$.

Since the maximum degree of a vertex in $\Gamma(a)$ is $2$ and
$\Gamma(a)$ has at least $5$ edges, there is an edge, say
$\delta\epsilon$, in $\Gamma(a)$ whose endvertices are different from
$\beta,\beta_1$ and $\beta_2$. Thus, the theory $S_{\{\dua a,\dua\beta,
\beta_1,\beta_2\}}$ contains the 2-clause $\delta\vee\epsilon$ and so,
$k(S_{\{\dua a,\dua\beta,\beta_1,\beta_2\}})\geq 1$. Similarly, there
is an edge, say $\lambda\varphi$, in $\Gamma(a)$ whose endvertices are
different from $\beta$ and $\gamma$. Hence, the theory $S_{\{\dua a,
\beta,\gamma\}}$ contains the 2-clause $\lambda\vee\varphi$ and so,
$k(S_{\{\dua a,\beta,\gamma\}})\geq 1$.

Thus,
\[
\Delta(S,S_A)\geq \left\{
            \begin{array}{ll}
              1 & \mbox{if $A=\{a\}$}\\
              4 + \alpha & \mbox{if $A=\{\dua 
a,\dua\beta,\beta_1,\beta_2\}$}\\
              3 + \alpha & \mbox{if $A=\{\dua a,\beta,\gamma\}$}\\
              5 & \mbox{if $A=\{\dua 
a,\beta,\dua\gamma,\gamma_1,\gamma_2\}\mbox{.}$}
            \end{array}
         \right.
\]
The equation (\ref{tau'}) becomes
\[
\tau^{-1}+\tau^{-4-\alpha}+\tau^{-3-\alpha}+\tau^{-5}=1
\]
and $\tau_S'\leq 1\mbox{.}66$.

\noindent
{\bf Subcase (ii).} There are no two nonadjacent vertices of degree
$2$ in $\Gamma(a)$.

If there are no vertices of degree $2$ in $\Gamma(a)$ then $\Gamma(a)$
contains at least $5$ independent edges and Case 14 applies. Thus,
let $\beta$ be a vertex of degree $2$ in $\Gamma(a)$ and let $\beta_1$
and $\beta_2$ be the neighbors of $\beta$.

Every model of $S$ consistent with $\{\dua a,\dua\beta\}$ is consistent
with $\{\beta_1,\beta_2\}$ (clauses $a\vee\beta\vee\beta_1$ and
$a\vee\beta\vee\beta_2$). Hence, the family
\[
{\A} =\{\{a\},\{\dua a,\beta\},\{\dua a,\dua\beta,\beta_1,
\beta_2\}\}
\]
is a cover for $S$ and we define $\rho(S)=\A$.

We claim that the $3$ edges in $\Gamma(a)$ that are not incident to
$\beta$ are independent. Indeed, let us suppose it is not so. Then some
two of these edges have a common vertex, say $\gamma$. The degree of
$\gamma$ in $\Gamma(a)$ is $2$ and none of the edges incident to $\gamma$
is incident to $\beta$, contrary to the assumption we adopt in this
subcase.

It follows that the theory $S_{\{\dua a,\beta\}}$ contains three
2-clauses with pairwise disjoint sets of atoms. Consequently,
$k(S_{\{\dua a,\beta\}}) \geq 3$. Thus,
\[
\Delta(S,S_A)\geq \left\{
            \begin{array}{ll}
              1 & \mbox{if $A=\{a\}$}\\
              2 + 3\alpha & \mbox{if $A=\{\dua a,\beta\}$}\\
              4 & \mbox{if $A=\{\dua a,\dua\beta,\beta_1,\beta_2\}\mbox{.}$}
            \end{array}
         \right.
\]
The equation (\ref{tau'}) becomes
\[
\tau^{-1}+\tau^{-2-3\alpha}+\tau^{-4}=1
\]
and $\tau_S'\leq 1\mbox{.}67$.

\noindent
{\bf Case 16.} There is an atom $a$ such that $\Gamma(a)$ has exactly
$4$ edges and there are two nonadjacent vertices of degree $2$ in
$\Gamma(a)$.

Let $\beta$ and $\gamma$ be two nonadjacent vertices of degree $2$ in
$\Gamma(a)$. We denote by $\beta_1$ and $\beta_2$ the neighbors of
$\beta$ and by $\gamma_1$ and $\gamma_2$ the neighbors of $\gamma$.

Clearly, the collection
\[
{\A'} = \{\{a,\dua\beta\},\{a,\beta\},\{\dua a,\dua\beta\},
\{\dua a,\beta,\gamma\},\{\dua a,\beta,\dua\gamma\}\}
\]
is a cover for $S$.

Every model of $S$ consistent with $\{\dua a,\dua\beta\}$ is consistent
with $\{\beta_1,\beta_2\}$ (clauses $a\vee\beta\vee\beta_1$ and $a\vee
\beta\vee\beta_2$). Moreover, every model of $S$ consistent with $\{\dua
a,\beta,\dua\gamma\}$ is consistent with $\{\gamma_1,\gamma_2\}$
(clauses $a\vee\gamma\vee\gamma_1$ and $a\vee\gamma\vee\gamma_2$).
Finally, every minimal model $M$ of $S$ consistent with $\{a,\beta\}$ is
consistent with $\{\dua\gamma\}$ as, otherwise, $M-\{a\}$ would be a
model of $S$ contrary to the minimality of $M$. Hence, the family
\[
{\A} =\{\{a,\dua\beta\},\{a,\beta,\dua\gamma\},\{\dua a,\dua\beta,
\beta_1,\beta_2\},\{\dua a,\beta,\gamma\},\{\dua a,\beta,\dua\gamma,
\gamma_1,\gamma_2\}\}
\]
is a cover for $S$ and we set $\rho(S)=\A$. As $\beta$ and $\gamma$ are
nonadjacent in $\Gamma(a)$, the vertices $\beta,\gamma,\gamma_1,
\gamma_2$ are pairwise different. Thus,
\[
\Delta(S,S_A)\geq \left\{
            \begin{array}{ll}
              2 & \mbox{if $A=\{a,\dua\beta\}$}\\
              3 & \mbox{if $A=\{a,\beta,\dua\gamma\},\{\dua 
a,\beta,\gamma\}$}\\
              4 & \mbox{if $A=\{\dua a,\dua\beta,\beta_1,\beta_2\}$}\\
              5 & \mbox{if $A=\{\dua 
a,\beta,\dua\gamma,\gamma_1,\gamma_2\}\mbox{.}$}
            \end{array}
         \right.
\]
The equation (\ref{tau'}) becomes
\[
\tau^{-2}+2\tau^{-3}+\tau^{-4}+\tau^{-5}=1
\]
and $\tau_S'\leq 1\mbox{.}67$.

\noindent
{\bf Case 17.} There is an atom $a$ such that $\Gamma(a)$ has exactly
$4$ edges and there is exactly one vertex of degree $2$ in $\Gamma(a)$.

Let $\beta$ be the vertex of degree 2 in $\Gamma(a)$, let $\beta_1$ and
$\beta_2$ be the neighbors of $\beta$ in $\Gamma(a)$ and let
$\gamma\delta$, $\epsilon\lambda$ be the two isolated edges
in $\Gamma(a)$.

Every model of $S$ consistent with $\{\dua a,\dua\beta\}$ is consistent
with $\{\beta_1,\beta_2\}$ (clauses $a\vee \beta\vee \beta_1$ and $a\vee
\beta\vee \beta_2$). Hence, the family
\[
{\A} =\{\{a\},\{\dua a,\beta\},\{\dua a,\dua\beta,\beta_1,\beta_2\}\}
\]
is a cover for $S$ and we set $\rho(S)=\A$.

Both theories $S_{\{\dua a,\beta\}}$ and $S_{\{\dua a,\dua\beta,\beta_1,
\beta_2\}}$ contain two 2-clauses $\gamma\vee\delta$ and
$\epsilon\vee\lambda$, whose sets of atoms are pairwise disjoint.
Thus, k($S_{\{\dua a,\beta\}})\geq 2$ and $k(S_{\{\dua a,\dua\beta,
\beta_1,\beta_2\}})\geq 2$. Consequently,
\[
\Delta(S,S_A)\geq \left\{
            \begin{array}{ll}
              1 & \mbox{if $A=\{a\}$}\\
              2 + 2\alpha & \mbox{if $A=\{\dua a,\beta\}$}\\
              4 + 2\alpha & \mbox{if $A=\{\dua 
a,\dua\beta,\beta_1,\beta_2\}\mbox{.}$}
            \end{array}
         \right.
\]
The equation (\ref{tau'}) becomes
\[
\tau^{-1}+\tau^{-2-2\alpha}+\tau^{-4-2\alpha}=1
\]
and $\tau_S'\leq 1\mbox{.}67$.

\noindent
{\bf Case 18.} There is an atom $a$ such that $\Gamma(a)$ contains a
vertex of degree 2 and has exactly $3$ edges.

Let $\beta$ be a vertex of degree $2$ in $\Gamma(a)$. We denote by
$\beta_1$ and $\beta_2$ the neighbors of $\beta$. Let $\gamma\gamma_1$
be the edge in $\Gamma(a)$ which is not incident to $\beta$.

Clearly, the family
\[
{\A'} = \{\{a,\dua\beta\},\{a,\beta\},\{\dua a,\dua\beta\},\{\dua a,
\beta,\gamma\},\{\dua a,\beta,\dua\gamma\}\}
\]
is a cover for $S$.

Every model of $S$ consistent with $\{\dua a,\dua\beta\}$ is consistent
with $\{\beta_1,\beta_2\}$ (clauses $a\vee\beta\vee\beta_1$ and
$a\vee\beta\vee\beta_2$). Moreover, every model of $S$ consistent with
$\{\dua a,\beta,\dua\gamma\}$ is consistent with $\gamma_1$ (clause
$a\vee\gamma\vee\gamma_1$). Finally, every minimal model $M$ of $S$
consistent with $\{a,\beta\}$ is consistent with $\{\dua\gamma,\dua
\gamma_1\}$ as, otherwise, $M-\{a\}$ would be a model of $S$, contrary
to the minimality of $M$. Hence, the family
\[
{\A} =\{\{a,\dua\beta\},\{a,\beta,\dua\gamma,\dua\gamma_1\},\{\dua
a,\dua\beta,\beta_1,\beta_2\},\{\dua a,\beta,\gamma\},\{\dua a,\beta,
\dua\gamma,\gamma_1\}\}
\]
is a cover for $S$ and we set $\rho(S)=\A$. As $\beta$ and $\gamma$ are
nonadjacent in $\Gamma(a)$, the vertices $\beta,\gamma,\gamma_1$ are
pairwise different. Thus,
\[
\Delta(S,S_A)\geq \left\{
            \begin{array}{ll}
              2 & \mbox{if $A=\{a,\dua\beta\}$}\\
              3 & \mbox{if $A=\{\dua a,\beta,\gamma\}$}\\
              4 & \mbox{if $A=\{a,\beta,\dua\gamma,\dua\gamma_1\},\{\dua
           a,\dua\beta,\beta_1,\beta_2\},\{\dua a,\beta,\dua\gamma,
           \gamma_1\}\mbox{.}$}
            \end{array}
         \right.
\]
The equation (\ref{tau'}) becomes
\[
\tau^{-2}+\tau^{-3}+3\tau^{-4}=1
\]
and $\tau_S\leq 1\mbox{.}65$.

\noindent
{\bf Case 19.} There is an atom $a$ such that the graph $\Gamma(a)$
has exactly $2$ edges and they are independent.

We denote by $\beta_1\beta_2$ and $\gamma_1\gamma_2$ the two edges in
$\Gamma(a)$. We define
\[
{\A} = \{\{a,\dua\beta_1,\dua\beta_2\},\{a,\dua\gamma_1,\dua
\gamma_2\},\{\dua a\}\}
\]
Let us assume that $M$ is a minimal model of $S$. If $a\not\in M$ then
$M$ is consistent with $\{\dua a\}$. Therefore let us assume that $a\in
M$. If $\beta_1,\beta_2\not\in M$ then $M$ is consistent with $\{a,
\dua\beta_1,\dua\beta_2\}$. On the other hand, if $\beta_i\in M$,
for some $i=1,2$, then $M$ is consistent with $\{a,\dua\gamma_1,\dua
\gamma_2\}$ (otherwise $M-\{a\}$ would be a model of $S$, contrary to
the minimality of $M$). It follows that the family
${\A}$
is a cover for $S$ and we set $\rho(S)=\A$.

The theory $S_{\{\dua a\}}$ contains two 2-clauses $\beta_1\vee\beta_2$
and $\gamma_1\vee\gamma_2$ with pairwise disjoint sets of atoms.
Consequently, $k(S_{\{\dua a\}})\geq 2$.
Thus,
\[
\Delta(S,S_A)\geq \left\{
            \begin{array}{ll}
              3 & \mbox{if $A=\{a,\dua\beta_1,\dua\beta_2\},\{a,\dua
          \gamma_1,\dua\gamma_2\}$}\\
              1 + 2\alpha & \mbox{if $A=\{\dua a\}\mbox{.}$}
            \end{array}
         \right.
\]
The equation (\ref{tau'}) becomes
\[
2\tau^{-3}+\tau^{-1-2\alpha} =1
\]
and $\tau_S'\leq 1\mbox{.}61$.

\noindent
{\bf Case 20.} The graph $\Gamma(a)$ has either $1$ edge or $2$ adjacent
edges.

Let $\beta$ be any vertex of $\Gamma(a)$, if $\Gamma(a)$ has $1$ edge or
the vertex of degree $2$, if $\Gamma(a)$ has $2$ edges. We denote by
$\gamma$ a neighbor of $\beta$ in $\Gamma(a)$.

Every minimal model $M$ of $S$ consistent with $a$ is consistent
with $\dua\beta$ (otherwise, $M-\{a\}$ would be a model of $S$,
contrary to the minimality of $M$). Hence, the family
\[
{\A} =\{\{\dua a\},\{a,\dua\beta\}\}
\]
is a cover for $S$ and $\rho(S)=\A$.
Thus,
\[
\Delta(S,S_A)\geq \left\{
            \begin{array}{ll}
              1 & \mbox{if $A=\{\dua a\}$}\\
              2 & \mbox{if $A=\{a,\dua\beta\}\mbox{.}$}
            \end{array}
         \right.
\]
The equation (\ref{tau'}) becomes
\[
\tau^{-1}+\tau^{-2}=1
\]
and $\tau_S'\leq 1\mbox{.}62$.

\noindent
{\bf Comment.} As we already noted just before Case 14, we can now
assume that for every atom $a$, the maximum degree of a vertex in
$\Gamma(a)$ is at most 2. We can also assume that for every atom $a$,
$\Gamma(a)$ has at most 4 edges (otherwise, Case 15 applies).

Let us assume that for some atom $a$, $\Gamma(a)$ has at most three
edges. If $\Gamma(a)$ has 1 or 2 edges, either Case 19 or Case 20
applies. Thus, we can assume that $\Gamma(a)$ has 3 edges. If
$\Gamma(a)$ has a vertex of degree 2, Case 18 applies. It follows that
from now on we can assume that if $\Gamma(a)$ has three edges, then
it is isomorphic to the graph $3P_1$ that consists of three independent
edges.

Let us now consider an atom $a$ such that $\Gamma(a)$ contains exactly
four edges. It is easy to see that the only situation not covered by
Cases 16 and 17 is when $\Gamma(a)$ contains at least 2 vertices of
degree 2, all of them adjacent. Thus, we can assume that $\Gamma(a)$
is isomorphic to the graph whose components are a triangle and a
single isolated edge (denoted by $C_3\cup P_1$) or the graph whose
components are a $3$-edge path and a single isolated edge (denoted
by $P_3\cup P_1$).

\noindent
{\bf Case 21.} For every atom $a$ occurring in $S$, $\Gamma(a)$ is
isomorphic to $C_3\cup P_1$, $P_3\cup P_1$ or $3P_1$. By the comment
above, the assumption we adopt here covers all the situations not
covered by Cases 1-20.

\noindent
{\bf Subcase (i).} For some atom $a$, there is a clause in $S$
containing the literal $\dua a$.

Let $\dua a\vee\delta\vee\epsilon$ be a clause in $S$ with the literal
$\dua a$.

Let us suppose first that $\Gamma(a)$ is $P_3\cup P_1$ or $3P_1$. Then
$\Gamma(a)$ contains 3 independent edges, say $\beta_1\gamma_1$,
$\beta_2\gamma_2$, $\beta_3\gamma_3$. We set
$\rho(S)=\A$, where
\[
{\A} =\{\{a\},\{\dua a\}\}
\]
(it is clearly a cover for $S$).

The theory $S_{\{a\}}$ contains a 2-clause $\delta\vee\epsilon$ and so,
$k(S_{\{a\}})\geq 1$. Moreover, the theory $S_{\{\dua a\}}$ contains
three 2-clauses $\beta_1\vee\gamma_1$, $\beta_2\vee\gamma_2$ and
$\beta_3\vee\gamma_3$ with pairwise disjoint sets of atoms.
Consequently, $k(S_{\{\dua a\}})\geq 3$. Thus,
\[
\Delta(S,S_A)\geq \left\{
            \begin{array}{ll}
              1 + \alpha & \mbox{if $A=\{a\}$}\\
              1 + 3\alpha & \mbox{if $A=\{\dua a\}\mbox{.}$}
            \end{array}
         \right.
\]
The equation (\ref{tau'}) becomes
\[
\tau^{-1-\alpha}+\tau^{-1-3\alpha}=1
\]
and $\tau_S'\leq 1\mbox{.}66$.

Let us suppose next that $\Gamma(a)$ is the graph $C_3\cup P_2$. Let
$\beta$ be a vertex of degree 2 in $\Gamma(a)$, let $\beta_1$, $\beta_2$
be the neighbors of $\beta$ in $\Gamma(a)$, and let $\gamma_1\gamma_2$
be the isolated edge in $\Gamma(a)$.

Every model of $S$ consistent with $\{\dua a,\dua\beta\}$ is consistent
with $\{\beta_1,\beta_2\}$ (clauses $a\vee\beta\vee\beta_1$ and $a\vee
\beta\vee\beta_2$). Hence, the family
\[
{\A} =\{\{a\},\{\dua a,\beta\},\{\dua 
a,\dua\beta,\beta_1,\beta_2\}\}\mbox{.}
\]
is a cover for $S$ and we set $\rho(S)=\A$.

The theory $S_{\{a\}}$ contains a 2-clause $\delta\vee\epsilon$ and so,
$k(S_{\{a\}})\geq 1$. Moreover, the theory $S_{\{\dua a,\beta\}}$
contains two 2-clauses $\beta_1\vee\beta_2$ and $\gamma_1\vee\gamma_2$
with pairwise disjoint sets of atoms. Consequently, $k(S_{\{\dua a,\beta
\}})\geq 2$. Finally, the theory $S_{\{\dua a,\dua\beta,\beta_1,\beta_2
\}}$ contains the 2-clause $\gamma_1\vee\gamma_2$ and so, $k(S_{\{\dua
a,\dua\beta,\beta_1,\beta_2\}})\geq 1$.
Thus,
\[
\Delta(S,S_A)\geq \left\{
            \begin{array}{ll}
              1 + \alpha & \mbox{if $A=\{a\}$}\\
              2 + 2\alpha & \mbox{if $A=\{\dua a,\beta\}$}\\
              4 + \alpha & \mbox{if $A=\{\dua 
a,\dua\beta,\beta_1,\beta_2\}\mbox{.}$}
            \end{array}
         \right.
\]
The equation (\ref{tau'}) becomes
\[
\tau^{-1-\alpha}+\tau^{-2-2\alpha}+\tau^{-4-\alpha}=1
\]
and $\tau_S'\leq 1\mbox{.}63$.

\noindent
{\bf Subcase (ii).} All occurrences of every atom in $S$ are positive.

Let us suppose first that there is an atom $a$ such that $\Gamma(a)$ is
isomorphic to $P_3\cup P_1$. Let $d,b,c,e$ be the consecutive vertices
of the path $P_3$ in $\Gamma(a)$ and let $f,g$ be the vertices of the
isolated edge in $\Gamma(a)$.

We will consider the graph $\Gamma(b)$. Clearly, it contains the edges
$ad$ and $ac$ so it is not isomorphic to $3P_1$. The graph $\Gamma(b)$
is not isomorphic to $C_3\cup P_1$, either. Let us suppose that it is.
Then, the edge $cd$ must be an edge of $\Gamma(b)$. Thus, $S$ contains
the clause $b\vee c\vee d$. Consequently, the graph $\Gamma(d)$ contains
the edges $ab$ and $bc$. The pair $ac$ is not an edge of $\Gamma(d)$
because $a$ belongs to 4 clauses only and $a\vee c\vee d$ is not one of
them. For the same reason $\Gamma(d)$ does not contain any edge of the
form $ah$, where $h\not=b$. Hence the graph $\Gamma(d)$ is isomorphic
to $P_3\cup P_1$ and there is an edge $ch$ in $\Gamma(d)$, for some $h
\not=a,b$. Thus, $S$ contains the clause $c\vee d\vee h$, where $h\not
=a,b$. Hence the following clauses belong to the theory $S$: $a\vee c
\vee e$, $a\vee b\vee c$, $b\vee c\vee d$ and $c\vee d\vee h$. All of
them belong to $T(c)$. Consequently, the pairs $ae$, $ab$, $bd$
and $dh$ belong to the graph $\Gamma(c)$. It follows that the graph
$\Gamma(c)$ is connected, a contradiction with the assumptions we
adopted in Case 21\mbox{.}

It follows that $\Gamma(b)$ is isomorphic to $P_3\cup P_1$. Clearly, $da$
and $ac$ are edges of $\Gamma(b)$. Let $b_1b_2$, $b_3b_4$ be the
remaining two edges of $\Gamma(b)$. Obviously, the edges $b_1b_2$ and
$b_3b_4$ are independent and $b_1,b_2,b_3,b_4\not= a$.

The graph $\Gamma(c)$ contains the edges $ab$ and $ae$. Hence,
$\Gamma (c)$ is isomorphic either to $C_3\cup P_1$ or to $P_3\cup
P_1$. In both cases there is an edge, say $c_1c_2$, in $\Gamma(c)$
with endvertices different from $a,b,e$.

Clearly, the family
\[
{\A'} = \{\{a,b,c \},\{a,b,\dua c\},\{ a,\dua b\},\{\dua a,b\},
\{\dua a,\dua b\}\}
\]
is a cover for $S$.

Every model of $S$ consistent with $\{\dua a,\dua b\}$ is consistent
with $\{c,d\}$ to satisfy the clauses $a\vee b\vee c$ and $a\vee b\vee
d$. Moreover, every minimal model $M$ of $S$ consistent with $\{ a,b,
c\}$ is consistent with $\{\dua f,\dua g\}$. Otherwise $M-\{ a\}$ would
be a model of $T(a)$ and, consequently, of $S$, contrary to the
minimality of $M$. Hence, the family
\[
{\A} = \{\{a,b,c,\dua f,\dua g\},\{a,b,\dua c\},\{ a,\dua b\},
\{\dua a,b\},\{\dua a,\dua b,c,d\}\}
\]
is a cover for $S$ and we set $\rho(S)=\A$.

The theory $S_{\{a,b,\dua c\}}$ contains the 2-clause $c_1\vee c_2$.
Since $c_1,c_2\not= a,b,c$, $k(S_{\{a,b,\dua c\}})\geq 1$. The theory
$S_{\{ a,\dua b\}}$ contains two 2-clauses $b_1\vee b_2$ and $b_3\vee
b_4$ with pairwise different atoms. Thus, $k(S_{\{ a,\dua b\}})\geq 2$.
The theory $S_{\{\dua a,b\}}$ contains two 2-clauses $c\vee e$ and $f
\vee g$ with pairwise different atoms, so $k(S_{\{\dua a,b\}})\geq 2$.
Finally, the theory $S_{\{\dua a,\dua b,c,d\}}$ contains the 2-clause
$f\vee g$, so $k(S_{\{\dua a,\dua b,c,d\}})\geq 1$.

Thus,
\[
\Delta(S,S_A)\geq \left\{
            \begin{array}{ll}
              5 & \mbox{if $A=\{a,b,c,\dua f,\dua g\}$}\\
              3 + \alpha & \mbox{if $A=\{a,b,\dua c\}$}\\
              2 + 2\alpha & \mbox{if $A=\{ a,\dua b\},\{\dua a,b\}$}\\
              4 + \alpha & \mbox{if $A=\{\dua a,\dua b,c,d\}\mbox{.}$}
            \end{array}
         \right.
\]
The equation (\ref{tau'}) becomes
\[
\tau^{-5}+\tau^{-3-\alpha}+2\tau^{-2-2\alpha}+\tau^{-4-\alpha}=1
\]
and $\tau_S\leq 1\mbox{.}66$.

Let us suppose now that there is an atom $a=a_1$ such that $\Gamma(a_1)$
is isomorphic to $C_3\cup P_1$ and that for no atom $a'$, $\Gamma(a')$
is isomorphic to $P_3\cup P_1$. Let $a_2,a_3,a_4$ be the vertices of
degree 2 in $\Gamma(a_1)$ and let $b_1,c_1$ be the vertices of degree 1.
Clearly, $b_1,c_1\not\in \{ a_1,a_2,a_3,a_4\}$. Since the graph
$\Gamma(a_2)$ has a vertex of degree 2 (the edges $a_1a_3$ and $a_1a_4$
belong to $\Gamma(a_2)$), the graph $\Gamma(a_2)$ is isomorphic to $C_3
\cup P_1$ and $a_3a_4$ is one of its edges. It follows that $a_2\vee
a_3\vee
a_4$ is a clause in $S$. For the same reason $\Gamma(a_3)$ and $\Gamma
(a_4)$ are isomorphic to $C_3\cup P_1$. Let, for $i=2,3,4$, $b_i,c_i$ be
the vertices of degree 1 in $\Gamma(a_i)$. Clearly, $b_i,c_i\not\in \{
a_1,a_2,a_3,a_4\}$, for $i=2,3,4$.

If for some $i\not=j$, $\{b_i,c_i\}=\{b_j,c_j\}$, say $b_i=b_j$ and
$c_i=c_j$, then pairs $a_jc_i$, $c_ia_i$ are edges in $\Gamma(b_i)$ and
so, the degree of $c_i$ in $\Gamma(b_i)$ is $2$. Hence $\Gamma(b_i)$ is
isomorphic to $C_3\cup P_1$ and $a_ia_j$ is an edge in $\Gamma(b_i)$, a
contradiction, as $b_i\vee a_i\vee a_j$ is not a clause in $S$ (it
follows from the fact that $b_ia_j$
is not an edge in $\Gamma(a_i)$).

Let us assume now that $\{b_1,c_1\}$, $\{b_2,c_2\}$,
$\{b_3,c_3\}$, $\{ b_4,c_4\}$ are pairwise different. Suppose each
pair of the sets $\{b_1,c_1 \}$, $\{b_2,c_2\}$, $\{b_3,c_3\}$,
$\{b_4,c_4\}$ has a common element. Then there is an element, say
$b_1$, which belongs to all four sets. Thus, $\Gamma(b_1)$
contains the edges $a_1c_1$, $a_2c_2$, $a_3c_3$, $a_4c_4$. Hence,
$\Gamma(b_1)$ is isomorphic to $C_3\cup P_1$. Since
$a_1,a_2,a_3,a_4$ are pairwise different and
$c_i\not\in\{a_1,a_2,a_3, a_4\}$, for $i=1,2,3,4$, we get
$c_1=c_2=c_3=c_4$ ($\Gamma(b_1)$ has 5 vertices as it is
isomorphic to $C_3\cup P_1$). This is a contradiction because we
proved that $\Gamma(b_1)$ is isomorphic to a 4-edge star. Hence,
some two of the sets $\{b_1,c_1\}$, $\{b_2,c_2\}$, $\{b_3,c_3\}$,
$\{b_4,c_4\}$ are disjoint. We assume without loss of generality
that $\{b_1,c_1\}$ and $\{b_2,c_2\}$ are disjoint.

Every model consistent with $\{\dua a_1,\dua a_2\}$ is consistent with
$\{a_3,a_4\}$ to satisfy the clauses $a_1\vee a_2\vee a_3$ and $a_1\vee
a_2\vee a_4$. Hence, the family
\[
{\A} = \{\{a_1\},\{\dua a_1,a_2\},\{\dua a_1,\dua a_2,a_3,a_4\}\}
\]
is a cover for $S$ and we define $\rho(S)=\A$.

The theory $S_{\{\dua a_1,a_2\}}$ contains two 2-clauses with disjoint
sets of atoms: $a_3\vee a_4$ (obtained from the 3-clause $a_1\vee a_3
\vee a_4$ in $S$) and $b_1\vee c_1$ (obtained from the 3-clause $a_1
\vee b_1\vee c_1$ in $S$). Hence $k(S_{\{\dua a_1,a_2\}})\geq 2$. The
theory $S_{\{\dua a_1,\dua a_2,a_3,a_4\}}$ also contains two 2-clauses
with disjoint sets of atoms: $b_1\vee c_1$ (obtained from the 3-clause
$a_1\vee b_1\vee c_1$ in $S$) and $b_2\vee c_2$ (obtained from the
3-clause $a_2\vee b_2\vee c_2$ in $S$). Hence $k(S_{\{\dua a_1,\dua
a_2,a_3,a_4\}})\geq 2$.

Thus,
\[
\Delta(S,S_A)\geq \left\{
            \begin{array}{ll}
              1 & \mbox{if $A=\{a_1\}$}\\
              2 + 2\alpha & \mbox{if $A=\{\dua a_1,a_2\}$}\\
              4 + 2\alpha & \mbox{if $A=\{\dua a_1,\dua 
a_2,a_3,a_4\}\mbox{.}$}
            \end{array}
         \right.
\]
The equation (\ref{tau'}) becomes
\[
\tau^{-1}+\tau^{-2-2\alpha}+\tau^{-4-2\alpha}=1
\]
and $\tau_S'\leq 1\mbox{.}67$.

It remains to consider the case when for every atom $a$ in $S$ the graph
$\Gamma(a)$ is isomorphic to $3P_1$. Let $b_1c_1$, $b_2c_2$ and $b_3c_3$
be the edges of $\Gamma(a)$. We observe that the collection
\[
{\A} = \{\{\dua a\},\{ a,\dua b_1,\dua c_1\},\{ a,\dua b_2,\dua c_2
\},\{ a,\dua b_3,\dua c_3\}\}
\]
is a cover for $S$ and we define $\rho(S)=\A$. Indeed, if $M$ is a
minimal model of $S$ such that $a\in M$ and, for each $i=1,2,3$, $b_i
\in M$ or $c_i\in M$, then $M-\{ a\}$ would be a model of $S$, contrary
to the minimality of $M$.

The theory $S_{\{\dua a\}}$ contains three 2-clauses $b_1\vee
c_1$, $b_2\vee c_2$ and $b_3\vee c_3$ with disjoint sets of atoms.
Hence $k(S_{\{\dua a\}})\geq 3$. Since, for every $i=1,2,3$,
$\Gamma(b_i)$ consists of 3 independent edges one of which is
$ac_i$, the theory $S_{\{ a,\dua b_i,\dua c_i\}}$ contains two
2-clauses with disjoint sets of atoms different from $a$ and
$c_i$. Hence $k(S_{\{ a,\dua b_i, \dua c_i\}})\geq 2$.

Thus,
\[
\Delta(S,S_A)\geq \left\{
            \begin{array}{ll}
              1 + 3\alpha & \mbox{if $A=\{\dua a\}$}\\
              3 + 2\alpha & \mbox{if $A=\{ a,\dua b_1,\dua c_1\},\{ a,
          \dua b_2,\dua c_2\},\{ a,\dua b_3,\dua c_3\}\mbox{.}$}
            \end{array}
         \right.
\]
The equation (\ref{tau'}) becomes
\[
\tau^{-1-3\alpha}+3\tau^{-3-2\alpha}=1
\]
and $\tau_S'\leq 1\mbox{.}66$.

\noindent
{\bf Comment.} There are no other cases to consider. Since the function
$\rho$ we described is splitting and for each $S$ $\tau_S'\leq 
1\mbox{.}6701\mbox{..}\
$, the Lemma \ref{mainth} follows.
\label{lastpage}
\end{document}